\documentclass[reprint,preprintnumbers,amsmath,amssymb,aps,nofootinbib,superscriptaddress]{revtex4-2}

\usepackage[utf8]{inputenc}  
\usepackage{graphicx}  
\usepackage[caption=false]{subfig}  
\usepackage{xcolor}  
\usepackage{amsmath}  
\usepackage{amsfonts}  
\usepackage{latexsym}  
\usepackage{amssymb,bbm}  
\usepackage{bm}  
\usepackage[colorlinks=true,allcolors=blue]{hyperref}  
\usepackage{lipsum}  
\usepackage{newunicodechar}
\newunicodechar{′}{\ensuremath{'} }
\usepackage{setspace}
\usepackage{braket}
\usepackage{musixtex}
\usepackage{siunitx}
\usepackage[version=4]{mhchem}

\newcommand{\app}{\approx}

\newcommand{\Cs}{{}^{13}\R{C}}

\newcommand{\Hs}{{}^{1}\R{H}}

\newcommand{\beq}{\begin{equation}}
\newcommand{\eeq}{\end{equation}}
                  
\newcommand{\benum}{\begin{enumerate}}
\newcommand{\eenum}{\end{enumerate}}
                    
\newcommand{\bit}{\begin{itemize}}
\newcommand{\eit}{\end{itemize}}
\newcommand{\xhat}{\hat{\T{x}}}

\newcommand{\bea}{\begin{eqnarray}}
\newcommand{\eea}{\end{eqnarray}}

\newcommand{\bev}{\begin{verbatim}}
\newcommand{\eev}{\end{verbatim}}

\newcommand{\noi}{\noindent}

\newcommand{\T}[1]{\textbf{#1}}

\newcommand{\R}[1]{\textrm{#1}}

\newcommand{\zfl}[1]{\protect\label{fig:#1}}
\newcommand{\zfr}[1]{\figurename\,\ref{fig:#1}}

\renewcommand{\ket}[1]{\left\vert{#1}\right\rangle}
\renewcommand{\bra}[1]{\left\langle{#1}\right\vert}

\newcommand{\ba}{\left\{ \begin{array}{lr}}
\newcommand{\ea}{\end{array}\right.}

\newcommand{\blist}[1]{
 \begin{list}{#1}
 \begin{align}
	 arrow
 \end{align}
 $\checkmark\star
  { \setlength{\itemsep}{3pt}
     \setlength{\parsep}{2pt}
     \setlength{\topsep}{3pt}
     \setlength{\partopsep}{0pt}
     \setlength{\leftmargin}{1em}
     \setlength{\labelwidth}{1em}
     \setlength{\labelsep}{0.5em} } }
\newcommand{\elist}{
  \end{list}  }

\DeclareMathSymbol{\vartheta}{\mathalpha}{letters}{"12}
\DeclareMathSymbol{\theta}{\mathalpha}{letters}{"23}
\DeclareMathSymbol{\phi}{\mathalpha}{letters}{"27}
\DeclareMathSymbol{\varphi}{\mathalpha}{letters}{"1E}

\newcommand{\bef}
{
\begin{figure}[htbp]
\centering
}

\newcommand{\eef}{\end{figure}}

\renewcommand{\figurename}{Fig.} 

\DeclareRobustCommand{\SIsecref}[1]{\hyperref[#1]{SI Sec.~\ref*{#1}}}
\DeclareRobustCommand{\SIsecrange}[2]{SI Secs.~\hyperref[#1]{\ref*{#1}}--\hyperref[#2]{\ref*{#2}}}
\DeclareRobustCommand{\mainfigref}[2][]{\hyperref[fig:#2]{Fig.~\ref*{fig:#2}#1}}
\DeclareRobustCommand{\mainfigrange}[2]{Figs.~\hyperref[fig:#1]{\ref*{fig:#1}}--\hyperref[fig:#2]{\ref*{fig:#2}}}

\newcommand{\affA}{Department of Chemistry, University of California, Berkeley, Berkeley, CA 94720, USA.}
\newcommand{\affB}{Chemical Sciences Division,  Lawrence Berkeley National Laboratory,  Berkeley, CA 94720, USA.}
\newcommand{\affC}{Kenneth S. Pitzer Center for Theoretical Chemistry, University of California, Berkeley, CA 94720, USA.}
\newcommand{\affD}{CIFAR Azrieli Global Scholars Program, 661 University Ave, Toronto, ON M5G 1M1, Canada.}

\begin{document}
\title{DFT-assisted natural abundance $\Cs$ zero-field NMR via optical magnetometry}
\author{Blake Andrews}\thanks{These authors contributed equally to this work}\affiliation{\affA}\affiliation{\affB}
\author{Xiao Liu}\thanks{These authors contributed equally to this work}\affiliation{\affA}\affiliation{\affC}
\author{Raphael Zumbrunn}\affiliation{\affA}
\author{Calvin Lee}\affiliation{\affA}
\author{Sahand Adibnia}\affiliation{\affA}\affiliation{\affC}
\author{Emanuel Druga}\affiliation{\affA}
\author{Martin Head-Gordon}\affiliation{\affA}\affiliation{\affB}\affiliation{\affC}
\author{Ashok Ajoy}\email{ashokaj@berkeley.edu}
\affiliation{\affA}\affiliation{\affB}\affiliation{\affD}

\begin{abstract}
Zero-field (ZF) nuclear magnetic resonance (NMR) spectroscopy probes scalar 
J-couplings between nuclei while dispensing with large homogeneous magnetic fields, enabling low-cost and geometrically flexible detection, including through conductive enclosures. Despite these advantages, its broader use for chemical analysis has been limited by sensitivity and by the difficulty of predicting the dense spectral multiplets that arise at zero field. Here we demonstrate natural-abundance (1.1\%) $\Cs$ ZF spectroscopy on off-the-shelf liquids using a compact commercial ${}^{87}$Rb magnetometer for the first time, without hyperpolarization or special sample preparation. Instrumental advances yield improved sensitivity, ${\le}$250-mHz linewidths and $>$week-long stability, enabling isotopomer-resolved fingerprint spectra across a 13-molecule library, including the ability to discern rare (0.0121\%) doubly $\Cs$-labelled species. In parallel, we demonstrate vibrationally corrected density-functional theory (DFT) based prediction of ZF NMR spectra for chemically diverse molecules with few-hertz accuracy. Comparing experiment with these calculations renders residual deviations as chemically informative, reporting on hydrogen bonding, hydration and ion pairing at high ionic strength. Together, these results contribute towards DFT-assisted ZF NMR as a general platform for field-constraint-free molecular identification and for extracting transient solution-state structure from responsive J-coupling observables.
\end{abstract}

\maketitle

\noi Nuclear magnetic resonance (NMR) spectroscopy is a central tool for determining molecular structure, but conventional NMR relies on strong magnetic fields and high-frequency inductive detection~\cite{slichterPrinciplesMagneticResonance,ernstPrinciplesNuclearMagnetic1990}. Chemical shifts appear as small parts-per-million variations on top of a much larger Larmor frequency, so high-resolution instruments require magnets with exceptional spatiotemporal homogeneity at the \textless 10~ppb level~\cite{shapiraSpatialEncodingAcquisition2004} together with field-lock strategies~\cite{vanzijlUseDeuteriumNucleus1987}. These requirements increase cost and confine high-field NMR to bulky, facility-scale infrastructure, motivating complementary spectroscopic platforms that retain rich chemical information while relaxing the need for stringent field control.

\begin{figure*}[t]
  \centering
  {\includegraphics[width=0.97\textwidth]{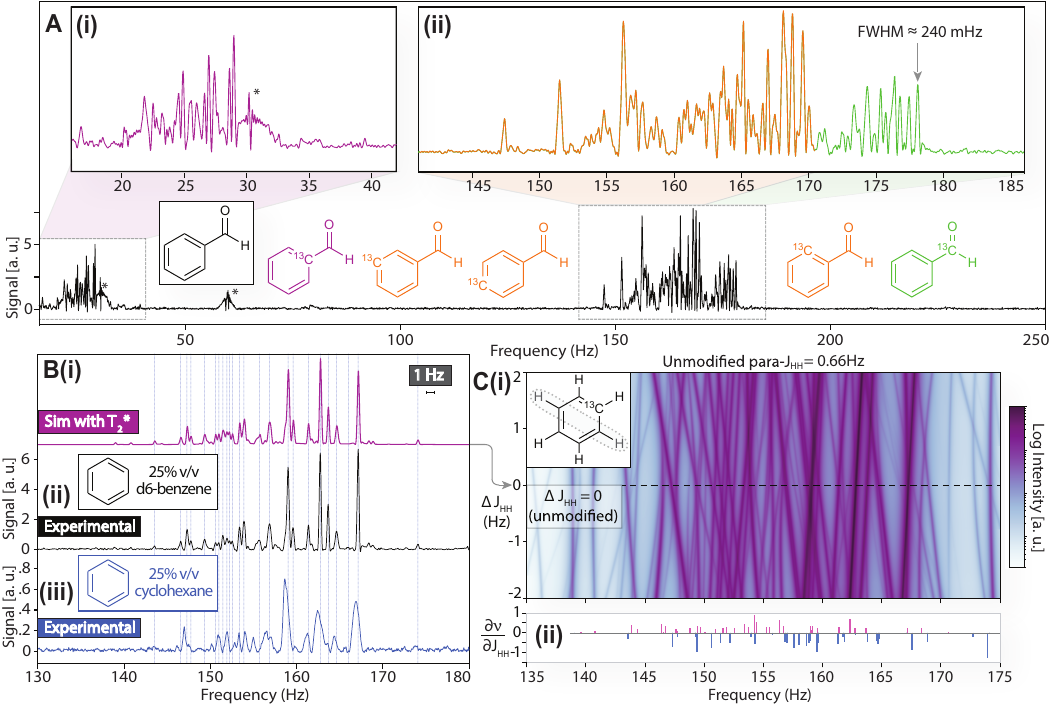}}
  \caption{\T{Isotopomer-resolved ZF NMR at natural abundance $\Cs$.} (A) ZF NMR spectrum of neat, NA benzaldehyde (9,247 scans) serving as a chemical fingerprint. Insets highlight low- and high-frequency windows; colored structures indicate the dominant $\Cs$ isotopomers. The quaternary site produces the lowest-frequency manifold (purple), aromatic substitution sites cluster near the same band as benzene (orange), and the aldehyde substitution appears at the highest frequency (green), consistent with previous enriched reports~\cite{blanchardHighResolutionZeroFieldNMR2013b}; FWHM $\approx$240~mHz is marked. (B) Benchmark experiment on NA benzene in two solvents (25\% v/v): (i) simulation using tabulated $J$ couplings for neat benzene (purple line), with experiment-measured $T_2^*$ broadening applied (Gaussian + exponential, see \SIsecref{subsec:td_sim_proc}), (ii) spectrum in C\textsubscript{6}D\textsubscript{6} (10,161 scans, black line), and (iii) spectrum in cyclohexane (10,231 scans, blue line), showing solvent-dependent line shifts (vertical lines to guide eye). (C) Simulated ZF NMR of benzene responding to perturbations of the para-$J_{\mathrm{HH}}$ (smallest) coupling: (i) stacked spectra for $\Delta J_{\mathrm{HH}}=\pm 2$~Hz about 0.66~Hz (true value, $\Delta J_{\mathrm{HH}}$=0), and (ii) line-dependent slopes ($\partial \nu/\partial J_{\mathrm{HH}}$) for each peak $\nu$, frequency axis shared with (i). Asterisks denote technical noise. Samples were prepared under ambient atmosphere (no oxygen degassing) with standard spectral processing (\SIsecref{subsec:proc_workflow}).
}
\zfl{mfig1}
\end{figure*}

Zero-field (ZF) NMR~\cite{zaxZeroFieldNMR1985,weitekampZeroFieldNuclearMagnetic1983a} offers one such route. Here detection is performed with compact alkali-vapor optically pumped magnetometers (OPMs) in the absence of an applied magnetic field (\textless 10~nT), following pre-polarization in a separate magnet~\cite{taylerInvitedReviewArticle2017}, which can be significantly inhomogeneous, and therefore comparatively inexpensive~\cite{TAYLER2017143}. Rather than chemical shifts, high resolution ZF spectra resolve scalar $J$-coupling networks that report on molecular topology and connectivity in a manner complementary to high-field NMR~\cite{theisChemicalAnalysisUsing2013,blanchardHighResolutionZeroFieldNMR2013b}. Because the observed frequencies are low, small absolute changes in $J$ correspond to exceptionally large fractional shifts, which often exceed $10^3$~ppm, making ZF spectra highly responsive to chemical environment. The zero-field region can be made large and homogeneous with little effort, enabling scalable and potentially parallel detection with high resolution~\cite{andrewsSensitiveMultichannelZeroto2025}. ZF NMR also offers capabilities that are difficult to access at high field: it tolerates substantial susceptibility variation and disorder, and its low-frequency detection can enable spectroscopy through conductive or metallic enclosures~\cite{buruevaChemicalReactionMonitoring2020,bodenstedtFastfieldcyclingUltralowfieldNuclear2021}.

Broader adoption has nevertheless been limited by two long-standing challenges: sensitivity and spectral interpretation~\cite{blanchardHighResolutionZeroFieldNMR2013b}. OPM readout remains less sensitive than high-field inductive detection and, in the absence of hyperpolarization~\cite{theisParahydrogenEnhancedZeroFieldNuclear2011}, ZF studies of organic molecules have relied on isotopically enriched samples---typically bearing one or more $\Cs$ labels (natural abundance 1.1\%)~\cite{blanchardZeroUltralowFieldNMR2016,blanchardZeroUltralowFieldNuclear2020}. Such compounds are often costly, inaccessible, or require dedicated synthesis. Many experiments have also required conditioning steps such as oxygen removal by freeze--pump--thaw cycling~\cite{blanchardZeroUltralowFieldNuclear2020,blanchardHighResolutionZeroFieldNMR2013b}. ZF NMR has therefore remained focused on fundamental studies on a relatively small set of model systems~\cite{wuSearchAxionlikeDark2019a, eillsDirectlyObservableZeemaninsensitive2026, barskiyZerofieldNuclearMagnetic2019b,picazo-frutosZerofieldJspectroscopyQuadrupolar2024b}, rather than becoming a general tool for chemical analysis.

The second difficulty lies in the structure of ZF spectra. Unlike the simpler peak patterns often encountered at high field, ZF spectra can contain a profusion of transitions: evolution under untruncated scalar $J$ couplings allows many coherences, and the number of observable lines grows rapidly with molecular size~\cite{butlerMultipletsZeroMagnetic2013}. These transitions encode the underlying coupling network, but they are rarely intuitive without computational support. First-principles prediction of $J$ couplings, and therefore of complete spectra, is also demanding because even small coupling errors can produce large discrepancies in the predicted multiplet pattern, hindering the construction of reliable reference libraries~\cite{mandzhievaZeroFieldNMRMilliteslaSLIC2025a,helgakerQuantumChemicalCalculationNMR2008}.

In this work, we address both challenges by combining long-term-stable natural-abundance (NA) ZF NMR with predictive quantum chemistry. Instrumental refinements provide improved sensitivity and ${\le}$250-mHz linewidths with week-long measurement stability, which enable isotopomer-resolved spectra across a library of off-the-shelf molecules (see Figs.~\ref{fig:na_lib_full_1} and ~\ref{fig:na_lib_full_2}). In parallel, vibrationally corrected vacuum-state DFT predicts complete ZF multiplets from molecular structure with few-hertz accuracy (\zfr{mfig4}), providing a practical route to forward simulation and assignment of dense spectra. When chemically induced changes in $J$ exceed the remaining prediction error, the deviations between experiment and isolated-molecule calculations themselves become informative, linking ZF observables to hydrogen bonding, hydration and ion pairing in solution (\zfr{mfig5}).

\begin{figure*}[t]
  \centering
  {\includegraphics[width=0.97\textwidth]{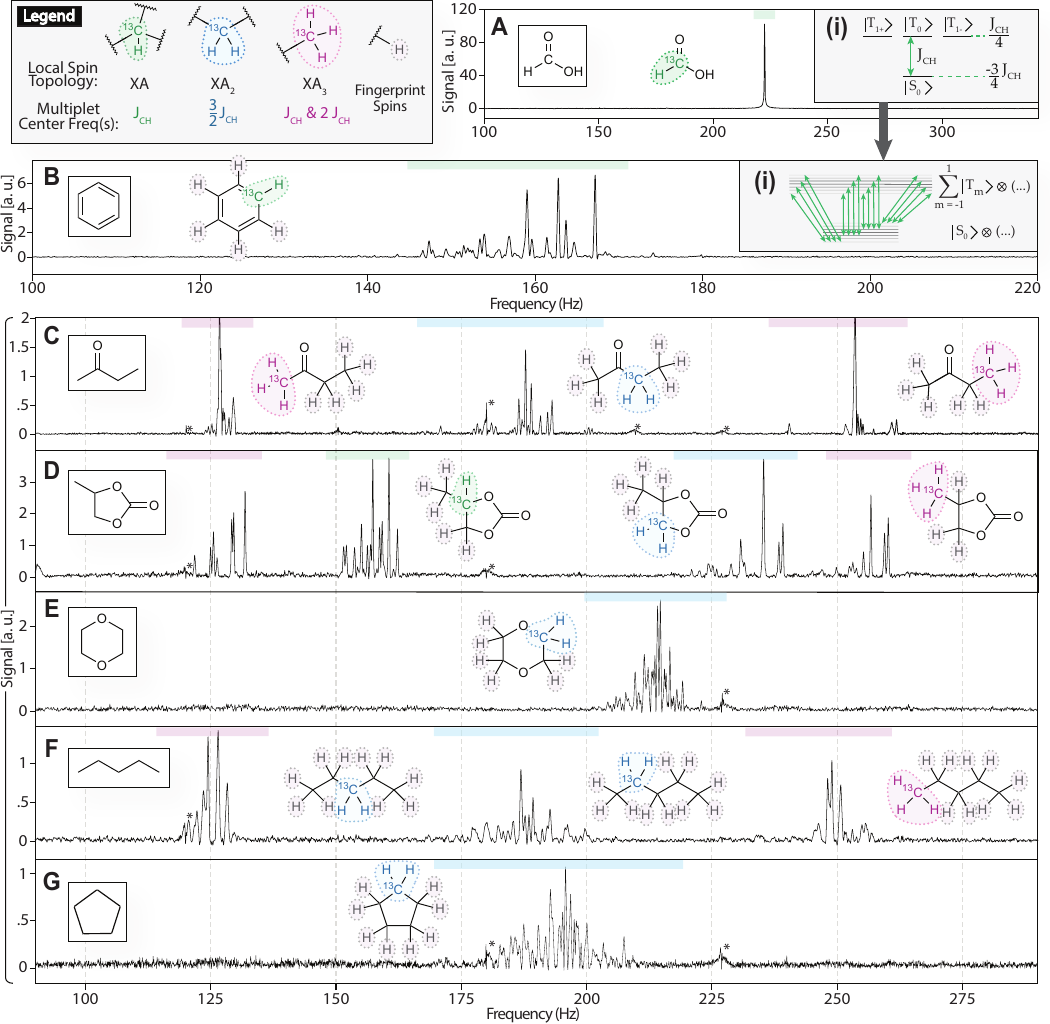}}
  \caption{\T{Natural-abundance ZF NMR library}. (A--G) Representative isotopomer-resolved spectra acquired without isotopic enrichment (complete library in Figs.~\ref{fig:na_lib_full_1} and~\ref{fig:na_lib_full_2}), ordered by increasing complexity. Molecular insets assign features to individual singly $\Cs$-substituted isotopomers; colored bubbles indicate the local $\Cs$H$_n$ motif (XA$_n$) that sets the dominant multiplet center(s) through its corresponding $J_{CH}$; colored overbars mark expected center frequencies from local topologies: XA (CH, green) near ${}^{1}$J$_{\textrm{CH}}$, XA$_2$ (CH$_2$, blue) near $(3/2)\times {}^{1}$J$_{\textrm{CH}}$, and XA$_3$ (CH$_3$, purple) near ${}^{1}$J$_{\textrm{CH}}$ and $2\times {}^{1}$J$_{\textrm{CH}}$ (see legend); the remaining couplings determine the internal fine structure. (A) Formic acid (373 scans) (i) single peak from XA singlet--triplet manifold. (B) Benzene (25\% v/v in C\textsubscript{6}D\textsubscript{6}, 10,161 scans) (i) many lines from perturbed XA manifold. (C) Methyl ethyl ketone (6,876 scans). (D) Propylene carbonate (5,410 scans). (E) 1,4-Dioxane (8,486 scans). (F) \textit{n}-Pentane (15,714 scans). (G) Cyclopentane (14,113 scans). Asterisks denote technical noise. Samples were prepared as neat liquids (excluding B) under ambient atmosphere (no oxygen degassing) with standard spectral processing (\SIsecref{subsec:proc_workflow}).}
\zfl{mfig2}
\end{figure*}

\begin{figure*}[t]
  \centering
  {\includegraphics[width=0.98\textwidth]{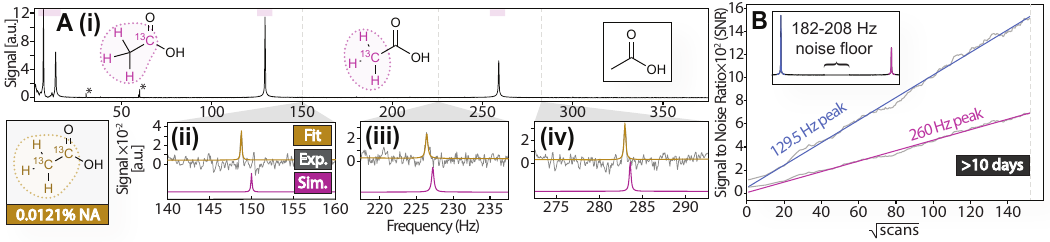}}
  \caption{\T{Stability-enabled detection of rare doubly $\Cs$-labeled acetic acid.} (A) Spectrum of NA acetic acid (23,215 scans), (i) resolving the two singly $\Cs$-labeled isotopomers (sharing XA$_3$ topology, with peaks at 129.6 and 6.7~Hz plus second harmonics) and---after extended averaging---resonances from the doubly $\Cs$-labeled species (0.0121\% probability; dashed markers). (ii-iv) Insets show expanded regions with experiment (grey), Lorentzian fits (yellow), and simulations from conventional $\Cs$ NMR couplings (purple, see \SIsecref{sec:jcc_acetic}). (B) Signal-to-noise ratio of the 129.6 and 260~Hz lines versus $\sqrt{N_{\rm scans}}$, exhibiting linear scaling for more than 10~days (noise estimated from the 182--208~Hz window).}
\zfl{mfig3}
\end{figure*}

\begin{figure*}[t]
  \centering
  {\includegraphics[width=0.99\textwidth]{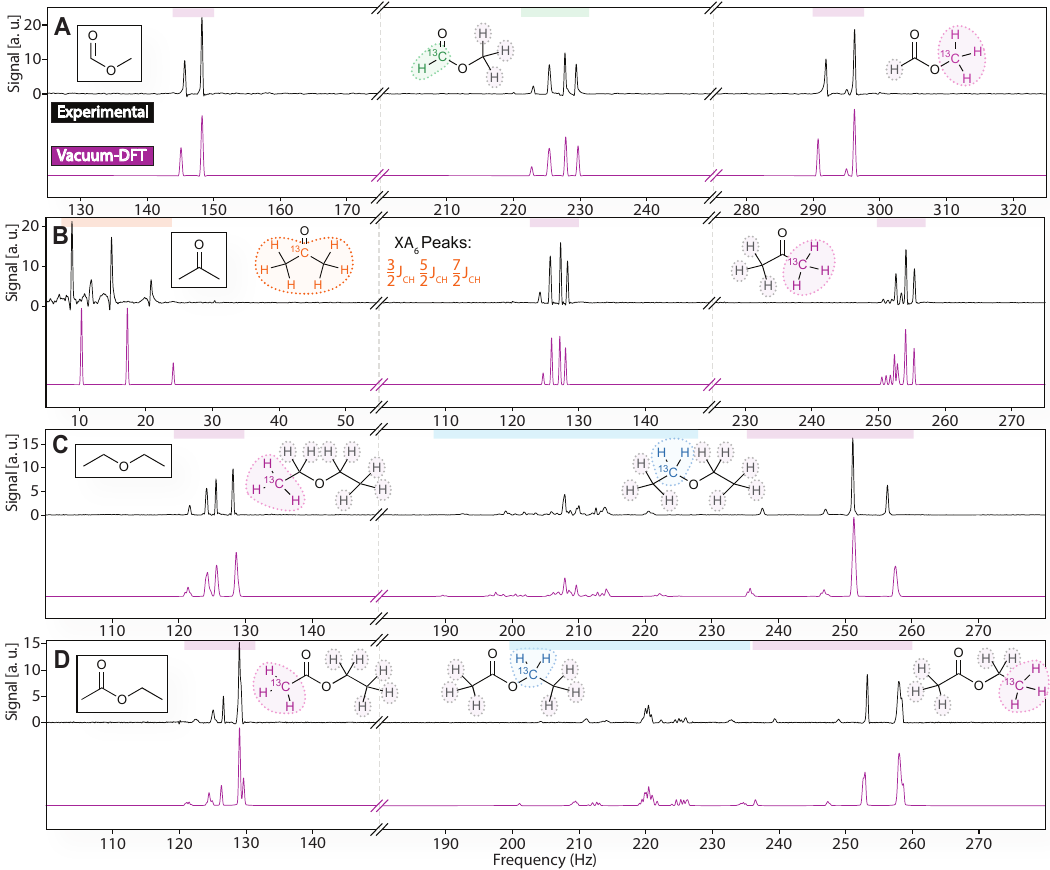}}
  \caption{\T{Vacuum-state \textit{ab initio} prediction of natural-abundance ZF NMR spectra.} Experimental spectra (black) compared with simulations (purple) computed from DFT-predicted $J$ couplings for isolated molecules (``vacuum-DFT''; vibrational corrections included, SI Secs.~\hyperref[subsec:vib_corr]{\ref*{subsec:vib_corr}},\hyperref[subsec:d_comp_workflow]{\ref*{subsec:d_comp_workflow}}). (A) Methyl formate (752 scans), (B) Acetone (3,920 scans; orange: XA$_6$ manifold from the symmetric $\Cs$ isotopomer with six equivalent ${}^{1}$H's), (C) Diethyl ether (10,000 scans), and (D) Ethyl acetate (9,357 scans). Samples were prepared as neat liquids under ambient atmosphere (no oxygen degassing) with standard spectral processing (\SIsecref{subsec:proc_workflow}); simulations are broadened (Gaussian + exponential) and shifted only by a uniform adjustment of the dominant $^{1}J_{\mathrm{CH}}$ for visualization (see \SIsecref{subsec:td_sim_proc}).
}
\zfl{mfig4}
\end{figure*}

\begin{figure*}[t]
  \centering
  {\includegraphics[width=0.99\textwidth]{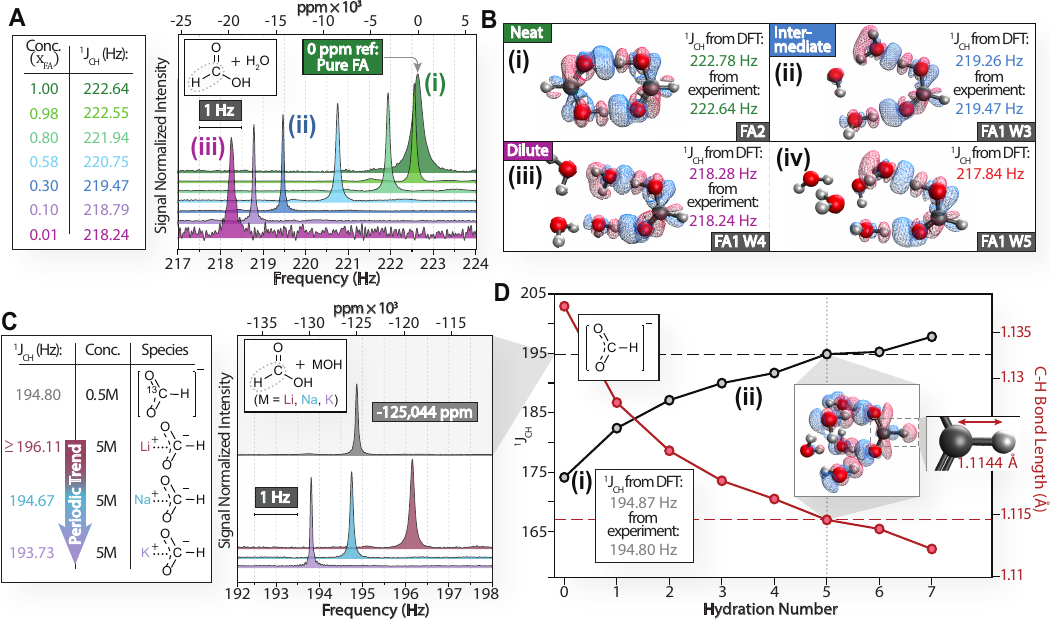}}
  \caption{\T{Solvation- and counterion-dependent $^{1}J_{\mathrm{CH}}$ shifts in aqueous formic acid and formate.} (A) ZF NMR spectra of NA formic acid in water across a dilution series (mole fraction $x_{\mathrm{FA}}$ 1.00--0.01), showing a $\sim -20\,000~\mathrm{ppm}$ shift of the ${}^{1}J_{\mathrm{CH}}$ resonance compared to neat formic acid reference; peak positions were inferred from Lorentzian fits. (B) Global-minimum structures of explicit-solvent clusters used in DFT calculations; blue and red meshes denote net electron-density gain and loss relative to an isolated FA molecule (details in \SIsecref{subsec:e_density_diff}), illustrating how local H-bond networks tune ${}^{1}J_{\mathrm{CH}}$ by redistributing density around the formyl C–H bond and reproduces the observed trend. (C) Upon addition of base, formate's shift in ${}^{1}J_{\mathrm{CH}}$ is $\sim -125\,000~\mathrm{ppm}$ compared to neat formic acid reference, and concentrated (5M) formate solutions in the presence of alkali hydroxides (MOH, M = Li, Na, K) show systematic, ion-specific shifts of ${}^{1}J_{\mathrm{CH}}$, consistent with contact ion pairing (details in \SIsecref{subsec:conc_formate_cip}). (D) Modeled hydration-number dependence of ${}^{1}J_{\mathrm{CH}}$ (black) and predicted C--H bond length (\AA{}, red) at the global minimum structure; dashed lines indicate the measured ${}^{1}J_{\mathrm{CH}}$ value (from the 0.5~M formate peak in (C)) intersecting at the inferred hydration number=5, yielding a predicted bond length of 1.1144~\AA{} within the validated solvation model. Sample preparation is described in SI Secs.~\hyperref[sec:fa_dilutions]{\ref*{sec:fa_dilutions}}--\hyperref[sec:formate_compisition]{\ref*{sec:formate_compisition}}, scan numbers for spectra in (A) and (C) are listed in Fig.~\ref{fig:acq_params} caption, and DFT details are given in \SIsecref{sec:qc_j}. The 0.5~M formate sample in panel C was ${}^{13}$C enriched only as a time-saving measure.
}  
\zfl{mfig5}
\end{figure*}

\vspace{0.5em}
\noi{\normalsize\bfseries System and Principle \par}
\noi
The apparatus used here builds on Ref.~\cite{andrewsSensitiveMultichannelZeroto2025} and is summarized in SI Secs.~\hyperref[subsec:instrument_scheme]{\ref*{subsec:instrument_scheme}}--\hyperref[subsec:NMRduino]{\ref*{subsec:NMRduino}} (Figs.~\ref{fig:apparatus} and \ref{fig:nmrduino}). Samples are pre-polarized at 9.4~T in an inexpensive, inhomogeneous magnet, shuttled in $\sim$1~s into a shielded zero-field region, pulsed, and detected with a commercial ${}^{87}$Rb magnetometer (Quspin~\cite{shahQZFMGen3}; ${\app}$40~fT~Hz$^{-1/2}$ in a Twinleaf MS2 magnetic shield) positioned $\sim$2~mm from a standard GC--MS vial. A low-cost NMRduino provides spectrometer control~\cite{taylerNMRduinoModularOpenSource2024}. All analytes were used off the shelf at natural-abundance ${}^{13}$C; sample compositions and dilution protocols are given in \SIsecref{sec:sample_prep}.

Assuming a single ${}^{13}$C label, the ZF spin Hamiltonian~\cite{barskiyZeroUltralowfieldNuclear2025} is
\begin{equation}
\hat{H}_\mathrm{ZF}
= \sum_{i\in \mathrm{H}} J^{\mathrm{i}}_{\mathrm{CH}}\,\hat{\mathbf{I}}_i \cdot \hat{\mathbf{S}}
+ \sum_{i<j\in \mathrm{H}} J^{\mathrm{ij}}_{\mathrm{HH}}\,\hat{\mathbf{I}}_i \cdot \hat{\mathbf{I}}_{j} ,
\end{equation}
where $\hat{\mathbf{I}}_i$ and $\hat{\mathbf{S}}$ denote the $i$th ${}^{1}$H and ${}^{13}$C spin operators, and $J^{\mathrm{i}}_{\mathrm{CH}}$ and $J^{\mathrm{ij}}_{\mathrm{HH}}$ are the heteronuclear and homonuclear couplings. At least one heteronucleus is therefore required for an observable spectrum. In the ideal XA$_n$ limit (for example, $\Cs$$^{1}$H$_n$), eigenstates are labelled by the total angular momentum $\hat{\mathbf{F}} = \sum_{n\in \mathrm{H}} \hat{\mathbf{I}}_n + \hat{\mathbf{S}}$ and its projection $M_F$; allowed ZF transitions satisfy $\Delta F = 0,\pm 1$ and $\Delta M_F = 0$~\cite{butlerMultipletsZeroMagnetic2013}. More weakly coupled spins then split these manifolds into the dense multiplets observed experimentally~\cite{theisChemicalAnalysisUsing2013}.

The detected observable is the longitudinal magnetization,
$
\hat{M}_z =
\sum_{i\in \mathrm{H}} \gamma_\mathrm{H} \hat{I}_{z,i}
+ \gamma_\mathrm{C}\hat{S}_{z},
$
with $\gamma_\mathrm{H}$ and $\gamma_\mathrm{C}$ as the gyromagnetic ratios. After adiabatic transfer from the pre-polarizing magnet to zero field, the high-temperature density operator is $\hat{\rho}_0 \propto \hat{M}_x$, where $\xhat$ lies along the shuttle axis. A broadband DC ${}^{1}$H $\pi/2$ pulse then initiates evolution under $\hat{H}_\mathrm{ZF}$, and the measured signal is
$
S(t) = \mathrm{Tr}\!\left\{\hat{M}_z\, e^{-i\hat{H}_\mathrm{ZF} t}\, \hat{\rho}_0\, e^{+i\hat{H}_\mathrm{ZF} t}\right\},
$
a ZF free-induction decay whose Fourier transform yields the spectrum. Peak positions depend on both coupling magnitudes and spin topology: one-bond ${}^{1}J_{\mathrm{CH}}$ couplings are typically $\mathcal{O}(10^2)$~Hz and increase with carbon $s$ character (sp$^3$ $<$ sp$^2$ $<$ sp), whereas geminal and vicinal couplings report on local geometry and dihedral angle through Karplus-type behaviour~\cite{hadadNMRSpinCouplingsSaccharides2017b,sternhellCorrelationInterprotonSpin1969}. Selected long-range pathways remain measurable up to $\sim$10~Hz~\cite{barfieldLongrangeProtonSpinspin1969,sternhellCorrelationInterprotonSpin1969}, and sub-hertz couplings can also be resolved~\cite{wilzewskiMethodMeasurementSpinSpin2017a}. Nonequivalent ${}^{13}$C substitution sites therefore generate distinct isotopomer spectra that complement chemical shifts by encoding connectivity in numerous peaks in a highly overconstrained way.

\vspace{0.5em}
\noi{\normalsize\bfseries ZF NMR with isotopomer resolution and $J$ sensitivity \par}
\noi Natural-abundance benzaldehyde (\zfr{mfig1}A) illustrates the basic idea. Its spectrum is a weighted sum of five distinct $\Cs$-labelled isotopomers (colored spectra). Insets \zfr{mfig1}A(i,ii) highlight the low- and high-frequency windows: the quaternary phenyl site gives the lowest-frequency manifold, the aromatic CH sites cluster near the same band as benzene, and the aldehyde carbon gives the highest-frequency features, consistent with enriched-sample studies~\cite{blanchardHighResolutionZeroFieldNMR2013b}. The fine splittings resolved in \zfr{mfig1}A(ii) achieve $\approx$240~mHz frequency resolution (marked). Even a one-dimensional ZF spectrum therefore provides a rich, isotopomer-specific $J$-coupling fingerprint.

Benzene provides a stringent benchmark for ZF NMR spectroscopy because its couplings are independently known~\cite{blanchardHighResolutionZeroFieldNMR2013b,wilzewskiMethodMeasurementSpinSpin2017a}. Panels \zfr{mfig1}B(i--iii) show, respectively, the corresponding simulation broadened with measured $T_2^{\ast}$ values (SI Secs.~\hyperref[subsec:td_sim]{\ref*{subsec:td_sim}}--\hyperref[subsec:td_sim_proc]{\ref*{subsec:td_sim_proc}}), the spectrum of NA benzene diluted in \ce{C6D6}, and the spectrum in cyclohexane. In the latter, many lines shift by 100--500~mHz (dashed lines guide the eye) and broaden measurably, indicating that both the couplings and relaxation depend on the chemical environment. When directly compared (SI \hyperref[subsec:prev_sota]{\ref*{subsec:prev_sota}}), the SNR of \zfr{mfig1}B(i) exceeds the previous benzene benchmark by $\sim 3\times$~\cite{blanchardHighResolutionZeroFieldNMR2013b}, despite our use of a compact commercial OPM with worse sensitivity (see discussion in \SIsecref{sec:sensitivity_budget}). The gain comes from apparatus refinements that reduce noise, stronger pre-polarization, larger sample volume and, crucially, excellent long-term stability (SI Secs.~\hyperref[subsec:com_vs_home_built]{\ref*{subsec:com_vs_home_built}}--~\hyperref[subsec:sens_stability]{\ref*{subsec:sens_stability}}), as also illustrated in \zfr{mfig3}B.

\zfr{mfig1}C illustrates the spectral dependence of even weak couplings at zero field. Starting from the simulated benzene spectrum in \zfr{mfig1}B(ii), \zfr{mfig1}C(i) shows the expected ZF spectra as the para-$J_{\mathrm{HH}}$ coupling is varied about its true value of 0.66~Hz by $\pm 2$~Hz. Even this weakest interaction, which is difficult to measure directly at high field and does not directly involve the $\Cs$ spin, produces conspicuous rearrangements of the multiplet pattern, so even small changes remain measurable in ZF NMR. The line-by-line response slopes in \zfr{mfig1}C(ii) span $\sim -1.3$ to 0.9~Hz/Hz, showing that different peaks shift at markedly different rates~\cite{bodenstedtOpticallyDetectedNuclear2024a}. Rather than changing a single $J$, the solvent dependence between \zfr{mfig1}B(ii,iii) reflects coordinated changes in several couplings to shift peaks. More broadly, \zfr{mfig1}C shows that ZF spectra are highly overconstrained: many line positions respond together, allowing subtle environmental perturbations to be inferred from the full multiplet response rather than from a single resonance.

\vspace{0.5em}
\noi{\normalsize\bfseries Stability-enabled natural-abundance ZF NMR library \par}
\noi \zfr{mfig2} compiles a subset of the first experimental natural-abundance ZF NMR library. A complete spectral library is provided in Figs.~\ref{fig:na_lib_full_1} and \ref{fig:na_lib_full_2}, with acquisition and processing details in \SIsecref{sec:full_na_lib_para} and Figs.~\ref{fig:acq_params}--\ref{fig:phase_params}. Representative spectra in \zfr{mfig2}A--G progress from a simple XA system to increasingly complex coupling networks. Formic acid (\zfr{mfig2}A) provides the minimal XA case with a single peak, benzene (\zfr{mfig2}B) embeds the same local motif in an aromatic ring with additional splitting (compare \zfr{mfig2}A(i) to \zfr{mfig2}B(i)). Methyl ethyl ketone (\zfr{mfig2}C) and propylene carbonate (\zfr{mfig2}D) introduce multiple inequivalent $\Cs$ sites, 1,4-dioxane (\zfr{mfig2}E) and \textit{n}-pentane (\zfr{mfig2}F) illustrate nonaromatic ring versus chain topology, and cyclopentane (\zfr{mfig2}G) gives the densest manifold as its fixed dihedral angles enhance couplings across an 11-spin network.

To a first approximation, each isotopomer's multiplet centre is set by its dominant heteronuclear coupling ${}^{m}J_{\mathrm{CH}}$ (typically $m=1$), and the number of directly bonded protons to $\Cs$ defines the local XA$_n$ topology. Thus XA, XA$_2$, and XA$_3$ motifs give centres near ${}^{1}J_{\mathrm{CH}}$, $(3/2){}^{1}J_{\mathrm{CH}}$, and ${}^{1}J_{\mathrm{CH}}$ together with $2{}^{1}J_{\mathrm{CH}}$, respectively~\cite{theisChemicalAnalysisUsing2013}. Throughout \zfr{mfig2}, the green, blue, and purple bubbles denote XA, XA$_2$, and XA$_3$, within each $\Cs$ isotopomer respectively, and the matching colored overbars mark the associated spectral regions following the legend. The same logic organizes \zfr{mfig2}C--G; it also rationalizes the XA$_6$ manifold of the symmetric acetone isotopomer in \zfr{mfig4}B. In this sense, ZF isotopomer spectra play a role analogous to DEPT $\Cs$ at high field, but with much richer internal structure~\cite{doddrelldistortionless1982}.

\zfr{mfig3}A,B return to sensitivity and stability. \zfr{mfig3}A(i) clearly shows two single $\Cs$ isotopomers of acetic acid following \zfr{mfig2} convention. As an illustration of detection limits, the lower panels of \zfr{mfig3}A zoom into the regions marked by vertical dashed lines in \zfr{mfig3}A(i), where weak resonances from the rare (0.0121\%) doubly $^{13}$C-labeled species appear (\zfr{mfig3}A(ii--iv)); their assignments are confirmed by conventional ${}^{13}$C data and explicit simulation (\SIsecref{sec:jcc_acetic} and Fig.~\ref{fig:Jcc}). With dilution, we further show in \SIsecref{subsec:LOD} that a 100~mM sample of NA formic acid remains detectable. To illustrate instrument stability, \zfr{mfig3}B shows the SNR scaling of the 129.5 and 260~Hz lines in \zfr{mfig3}A(i). We observe linear growth with $\sqrt{N_{\rm scans}}$ for more than 10~days, representing, to our knowledge, the best reported ZF NMR SNR stability, showing that environmental perturbations can be tightly controlled.

ZF NMR also intrinsically suppresses water signals because \ce{H2O} lacks a heteronuclear probe pair such as ${}^{13}$C--${}^{1}$H. Accordingly, the ZF spectrum of acetic acid in water remains essentially unchanged from the neat case (\SIsecref{subsec:water_supp} and Fig.~\ref{fig:H2Osup}), unlike the corresponding high-field ${}^{1}$H spectrum. The aqueous measurements in \zfr{mfig5} exploit this same advantage and point toward chemically complex, salt-rich samples that are difficult to access by conventional means~\cite{optical_constants_water,the_snr_of_nmr,cryogenically_cooled_probes}.

\vspace{0.5em}
\noi{\normalsize\bfseries Vibrational corrections are key for \textit{ab initio} ZF NMR spectral prediction\par}
\noi
We next turn to the second challenge: predicting ZF spectra accurately enough to interpret dense multiplets. Because ZF patterns are highly structured, the method places unusually strict demands on quantum-chemical $J$ prediction. In principle, electronic-structure calculations coupled to spin-dynamics simulation can predict complete spectra \textit{ab initio}~\cite{hogbenSpinachSoftwareLibrary2011b,mandzhievaZeroFieldNMRMilliteslaSLIC2025a}, but the measured couplings are rovibrationally averaged, so neglecting vibrational contributions can introduce substantial systematic errors~\cite{ruden2003vibrational}.

Motivated by this point, we combine non-empirical hybrid functional PBE0-based~\cite{adamo1999toward} electronic $J$ calculations with explicit vibrational corrections (SI Secs.~\hyperref[subsec:vib_corr]{\ref*{subsec:vib_corr}},\hyperref[subsec:d_comp_workflow]{\ref*{subsec:d_comp_workflow}}) and propagate the full coupling network through spin dynamics (SI Secs.~\hyperref[subsec:td_sim]{\ref*{subsec:td_sim}}--\hyperref[subsec:td_sim_proc]{\ref*{subsec:td_sim_proc}}). The resulting vacuum-state calculations (purple) reproduce the experimental fingerprints (black) in \zfr{mfig4}A--D for methyl formate, acetone, diethyl ether, and ethyl acetate with few-hertz accuracy, while retaining the smaller couplings that determine the internal multiplet structure. For display, we apply only a small uniform shift of the dominant ${}^{1}J_{\mathrm{CH}}$ where noted (\SIsecref{subsec:td_sim_proc} and Table~\ref{table:shifts}); the fine structure itself is not tuned. The remaining discrepancies need not be random: the shifts in benzene's spectrum across solvents in \zfr{mfig1}B(ii,iii) and across the formic-acid series in \zfr{mfig5}A point to solution-phase effects beyond an isolated-solute model. Even so, isolated-molecule DFT already appears accurate enough for molecular identification, conformational analysis in solution-phase systems~\cite{shimizu2015zero}, and construction of \textit{in silico} ZF reference libraries amenable to data-driven spectral assignment~\cite{chowdhuryEvaluationMachineLearning2021,mccarthyMoleculeIdentificationRotational2020,horaiMassBankPublicRepository2010}. 

\vspace{0.5em}
\noi{\normalsize\bfseries DFT-assisted ZF NMR provides new chemical contrast: a case study on formic acid and formate \par}
\noi
Residual deviations between experiment and vacuum-state DFT can themselves be treated as observables. Using formic acid and formate as a model system, \zfr{mfig5} shows that multi-hertz changes in ${}^{1}J_{\mathrm{CH}}$ encode hydrogen bonding, hydration, and ion pairing when environmental shifts exceed prediction error. Because ZF resonances occur at only $\sim 10^2$~Hz, these changes correspond to effective shifts of $10^3$--$10^5$~ppm, giving an unusually large contrast scale for NMR.

For neutral formic acid, \zfr{mfig5}A shows pronounced concentration-dependent shifts across an aqueous dilution series (mole fraction $x_{\mathrm{FA}}$ 1.00--0.01). The structures in \zfr{mfig5}B(i--iii) correspond to the experimental points in \zfr{mfig5}A(i--iii), as the mole fraction roughly matches the stoichiometry of each cluster; there, blue/red meshes denote electron-density gain/loss upon the formation of hydrogen bonds relative to one isolate FA molecule (details in \SIsecref{subsec:e_density_diff}). Neat FA is well modeled as a hydrogen-bonded cyclic dimer (\zfr{mfig5}B(i))~\cite{scheiner1997hydrogen,chelli2005structure}; upon dilution, explicit hydration reorganizes the carboxyl $\pi$ system and the formyl C--H bond, as visualized by the electron-density-difference plots in \zfr{mfig5}B and in more detail in Figs.~\ref{fig:FA1_EDD} and \ref{fig:FA-_EDD}. 
Relative to the vacuum-state value (228.35~Hz; SI Secs.~\hyperref[subsec:vib_corr]{\ref*{subsec:vib_corr}}--\hyperref[subsec:d_comp_workflow]{\ref*{subsec:d_comp_workflow}}), hydrogen bonding lowers ${}^{1}J_{\mathrm{CH}}$ by about 6~Hz in neat FA and by a further $\sim$4.5~Hz upon dilution. The calculations faithfully track this trend as the averaged recurring local solvation motifs match the physical system~\cite{giovannini2020molecular}.

Deprotonation produces an even larger response. The 0.5~M formate resonance in \zfr{mfig5}C moves by $\sim\!-125\,000$~ppm ($\sim$28Hz) relative to neat FA, reflecting a major electron reorganization of the carboxyl $\pi$ system. In \zfr{mfig5}D, the black curve gives calculated ${}^{1}J_{\mathrm{CH}}$ versus hydration number, and the red curve gives the associated C--H bond length; matching the experimental value from \zfr{mfig5}C to the calculated trend confirms an average hydration number of $\sim 5$, consistent with previous experiments and simulations~\cite{rahman2012hydration,leung2004ab,rudolph2022raman}. The corresponding bond length is 1.1144~\AA{}. At 5~M, \zfr{mfig5}C also resolves counterion-specific shifts for Li$^{+}$, Na$^{+}$, and K$^{+}$, and the same periodic trend is reproduced by the contact-ion-pair calculations in \SIsecref{subsec:conc_formate_cip} and Fig.~\ref{fig:CIP}. These concentrated aqueous samples can be measured without special adaptation, whereas conventional NMR and IR often struggle with water background or salt loading~\cite{optical_constants_water,the_snr_of_nmr,cryogenically_cooled_probes}. Relative to the few-ppm concentration-dependent shifts reported for high-field sodium formate~\cite{trapp2025electrolyte}, the ZF readout resolves much larger and more chemically specific $J$ shifts.

\vspace{0.5em}
\noi{\normalsize\bfseries Conclusions and Outlook \par}
\noi
Taken together, these results suggest that ZF NMR can provide quantitative access to scalar-coupling networks at natural abundance and, when combined with DFT, may offer chemical insight beyond molecular identification.

Vibrationally corrected DFT together with spin-dynamics simulation may already provide useful forward predictions of complete ZF spectra for chemical assignment and reference-library construction. For larger molecules, additional computational acceleration, including quantum processor and AI-assisted approaches, could broaden the reach of such predictions. Equally important, residual deviations from vacuum-state calculations need not be treated simply as error: they can report on changes in local electronic structure caused by hydrogen bonding, hydration and ion pairing. In \zfr{mfig5}, these $J$ shifts follow the progression from cyclic dimers to hydrated motifs, distinguish counterion-specific contact ion pairing at high salt, and correlate with C--H bond length within an explicit-solvent model.

This suggests that DFT-assisted ZF NMR could offer a useful window onto noncovalent interactions. Because such interactions underlie processes such as protein--ligand recognition and $\pi$--$\pi$ stacking~\cite{meyer2003interactions}; the method could ultimately help identify dominant interaction classes for candidate small-molecule drugs~\cite{zhou2012specific}. Explicit modeling of a ligand within its binding environment may reveal measurable perturbations to the ZF spectral fingerprint upon binding. Complementarily, site-specific isotopic labeling of amino-acid residues in proteins could, in principle, report on local noncovalent interactions and hence on local structure, including binding sites.

More broadly, high resolution zero field may make it possible to observe extremely weak $J$ couplings between diastereotopic hydrogens many bonds away from a chiral centre, providing a handle on conformational bias and stereochemical assignment in complex natural products. The ability to acquire high-resolution spectra in highly disordered, conductive and/or high-ionic-strength media (\zfr{mfig5}C), without field-homogeneity constraints such as magnetic-susceptibility broadening, also points to new measurements on ionic liquids, concentrated electrolytes and related materials that can be challenging for high-field NMR.

There is also substantial headroom for improving sensitivity. State-of-the-art potassium-OPMs optimized for ZF NMR achieve a 1.2 fT Hz$^{-1/2}$ noise floor~\cite{HONG2025200170}—an order of magnitude better than used here—potentially enabling permanent-magnet prepolarization. Compact superconducting approaches could also raise prepolarization fields when homogeneity is not required~\cite{chenWatchSized12Tesla2023,leeConstructionTestResult2022}; flux concentrators and rapid shuttling could offer additional gains (see \SIsecref{subsec:sens_stability}). Together, these advances could make natural-abundance ZF NMR practical for routine chemical analysis or real-time, deployable, \textit{in situ} monitoring of industrial chemical processes.

\providecommand{\beginmethods}{}
\beginmethods
\section*{Methods}

    \vspace{0.5em}
    \noi{\normalsize\bfseries Natural-abundance samples and solution series\par}
    \noi
        All analytes were measured at the natural abundance of $^{13}$C unless otherwise noted. Liquids were used off-the-shelf, loaded under ambient atmosphere into \SI{2}{\milli\liter} GC--MS vials compatible with the shuttle geometry, and measured without freeze--pump--thaw degassing or other special conditioning. Samples span neat liquids, dilute benzene samples in \ce{C6D6} or cyclohexane, and aqueous formic-acid/formate solutions. Formic-acid dilution series were prepared volumetrically with micropipettes at fixed total volume, and concentrated formate solutions were generated by neutralizing formic acid with alkali hydroxides under conditions chosen to favor complete deprotonation and, at high salt loading, contact ion pairing. The \SI{0.5}M formate sample in \zfr{mfig5}C was $^{13}$C enriched only as a time-saving measure; all other experiments were acquired at true natural abundance.

    \vspace{0.5em}
    \noi{\normalsize\bfseries Zero-field NMR apparatus and pulse--acquire workflow\par}
    \noi
        The instrument builds on our multichannel zero-field platform and combines pre-polarization in an inhomogeneous \SI{9.4}{\tesla} magnet with detection in a mu metal magnetic shield (Twinleaf MS2). Samples were polarized for approximately \SI{17.5}{\second}, shuttled to the shield in approximately \SI{0.9}{\second} under a \SI{210}{\micro\tesla} guiding field, and then detected with a Gen 3 dual-axis QuSpin $^{87}$Rb optically pumped magnetometer positioned at a standoff of about \SI{2}{\milli\meter} from the sample vial. After transfer to zero field, a broadband DC $^{1}$H $\pi/2$ pulse initiates evolution under the scalar-coupling Hamiltonian, and the ensuing zero-field free-induction decay was recorded along the magnetometer-sensitive axis. The magnetic shield was degaussed before experiments, and its integrated compensation coils were used to null residual fields to the low-nT regime.

    \vspace{0.5em}
    \noi{\normalsize\bfseries Spectrometer control and acquisition parameters\par}
    \noi
        Shuttling, guiding-field control, pulse timing and digitization were synchronized with an NMRduino spectrometer built around a Teensy 4.1 microcontroller. The sequence file specified the timing of the shuttling, guiding field, the $^{1}$H pulse and the acquisition block for each scan. For all datasets, the digitizer recorded $N_{\mathrm{pts}}=65536$ points, with sampling rates between \SI{4}{\kilo\hertz} and \SI{16.666}{\kilo\hertz} depending on the experiment, corresponding to acquisition windows of \SIrange{3.932}{16.384}{\second} per scan. Samples and shielding hardware were mechanically isolated to suppress motion artifacts, and all electronics were powered from an uninterruptible power supply to reduce line-noise pickup during the long averaging intervals required for natural-abundance work.

    \vspace{0.5em}
    \noi{\normalsize\bfseries Spectral processing and sensitivity analysis\par}
    \noi
        Displayed spectra were generated from scan-averaged time-domain traces using the following workflow. First, a smoothened time-domain trace was estimated with Savitzky--Golay filtering and subtracted to remove low-frequency magnetometer drift and residual relaxation contributions from the overwhelmingly abundant $^{13}$C-free isotopomers. The traces were then edge-trimmed to suppress boundary artefacts, optionally zero-filled for visualization, apodized with an exponential window, and Fourier transformed. Final spectra were phase-corrected with zeroth- and first-order adjustments, followed by an asymmetric least-squares baseline correction in the frequency domain. Signal-to-noise ratios were determined from the dominant peaks relative to noise-only spectral windows processed in the same way.

    \vspace{0.5em}
    \noi{\normalsize\bfseries Spin-dynamics simulations and isotopomer-resolved spectral reconstruction\par}
    \noi
        Time-domain zero-field simulations were carried out with the Spinach library. Each isotopomer was represented by its list of NMR-active nuclei together with an isotropic scalar-coupling network, and the density matrix was propagated under the zero-field Hamiltonian with detection through the longitudinal magnetization operator along the optically pumped magnetometer axis. To reproduce experimental line shapes, the simulated free-induction decays were multiplied post hoc by combined Gaussian and exponential envelopes chosen to match the measured broadening. For coupling-response calculations such as \zfr{mfig1}C, transition frequencies and amplitudes were evaluated directly in the Hamiltonian eigenbasis, and numerical derivatives $\partial \nu/\partial J$ calculated by repeating after a small perturbation of a selected coupling is applied. These two complementary simulation approaches were used throughout for parameter response visualization, validation of the double-$^{13}$C acetic-acid spectrum, and comparison with vacuum-state DFT predictions.

    \vspace{0.5em}
    \noi{\normalsize\bfseries Quantum-chemical prediction of scalar couplings\par}
    \noi
        J coupling calculations were performed in ORCA 6.0.1~\cite{neese2025software}. All calculations apply the RIJCOSX approximation~\cite{helmich2021improved}, with the auxiliary basis automatically generated on the fly~\cite{stoychev2017automatic}. For isolated molecules, candidate conformers were first identified with GOAT global optimization~\cite{de2025goat} at GFN2-xTB~\cite{bannwarth2019gfn2} level and then reoptimized at the PBE0-D3~\cite{adamo1999toward,grimme2010consistent}/aug-cc-pCVTZ level~\cite{dunning1989gaussian,kendall1992electron,woon1995gaussian}; harmonic frequency calculations were performed to verify that the final structures were true minima. Electronic contributions to the scalar couplings were computed at the PBE0-D3/pcJ-3~\cite{jensen2006basis} level. Second-order vibrational perturbation theory (VPT2)~\cite{franke2021vpt2}-type vibrational corrections were then added explicitly by evaluating cubic force constants at the PBE0-D3/aug-cc-pCVTZ level, and numerical derivatives of the electronic $J$ couplings with respect to normal modes at the PBE0-D3/pcJ-3 level. The choices of $J_{elec}$ and $J_{vib}$ computational levels are motivated by rigorous analysis and benchmarking, to be presented in a future publication. For molecules with multiple relevant conformers, final couplings were Boltzmann averaged at \SI{298}{\kelvin} using electronic energies refined at the RHF:RCCSD(T)-F12~\cite{pavovsevic2014geminal}/cc-pVDZ-F12~\cite{peterson2008systematically} level together with the DFT zero-point energies. This workflow furnished the vacuum-state coupling networks used to predict the spectra in \zfr{mfig4}.

    \vspace{0.5em}
    \noi{\normalsize\bfseries Explicit-solvent hydration and ion-pair models\par}
    \noi
        To interpret the chemically induced $J$ shifts in formic acid and formate, we complemented the vacuum-state calculations with explicit-solvent cluster models. Neutral formic-acid microhydrates were built from literature-inspired starting structures and reoptimized through the same geometry/vibrational workflow used for the isolated molecules. Hydrated formate anions and hydrated alkali contact-ion pairs were generated by combining AutoSolvator-produced solvent shells with GOAT global optimization with GFN2-xTB, followed by full geometry optimization and coupling calculations (details in SI Sec.\hyperref[subsec:d_comp_workflow]{\ref*{subsec:d_comp_workflow}}). A conductor-like polarizable continuum (C-PCM)~\cite{cossi2003energies} model was included for the hydrated anion and contact-ion-pair calculations. Because the pcJ-3 basis set is not available in ORCA for Li and K, pc-3~\cite{jensen2007polarization, jensen2012polarization} was used for those atoms in the electronic $J$ calculations, and aug-pc-2~\cite{jensen2012polarization} replaced aug-cc-pCVTZ on K during the geometry and vibrational steps. The resulting $^{1}J_{\mathrm{CH}}$ values were compared directly with the dilution-series and counterion-dependent resonances in \zfr{mfig5}A,C.

    \vspace{0.5em}
    \noi{\normalsize\bfseries Electron-density analysis of the $J$-shift observable\par}
    \noi
        To visualize the origin of the measured $J$ shifts, we carried out absolutely localized molecular orbital energy decomposition analysis (ALMO-EDA)~\cite{horn2016probing} in Q-Chem 6.3~\cite{epifanovsky2021software} for representative hydrated formic-acid and formate clusters at $\omega$B97X-D3~\cite{chai2008systematic, chai2008long, lin2013long}/def2-TZVPD~\cite{weigend2005balanced, rappoport2010property} level. The solute and water cluster were treated as separate fragments, allowing construction of frozen, polarized and fully relaxed electronic states. All density-difference plots were evaluated at the same isovalue (0.002 a.u.), enabling direct comparison across the neat, intermediate-dilution, dilute and ion-paired structures discussed in the electron density plots between the frozen and fully interacting states in \zfr{mfig5}B,D. Details of a more thorough breakdown of the electron density change from polarization and charge transfer effects are analyzed in SI Sec.\hyperref[subsec:e_density_diff]{\ref*{subsec:e_density_diff}}.

\vspace{0.5em}
\noi{\normalsize\bfseries Contributions\par}
\noi BA constructed the apparatus with ED, performed experiments, data analysis, and simulations. XL developed the quantum chemistry calculation workflow, performed the calculations (with assistance from SA) and analysis, and helped design experiments with BA. RZ and CL refined apparatus, performed experiments and simulations. MHG supervised quantum chemistry calculations. AA supervised experiments. BA, XL, and AA wrote the manuscript.

\vspace{0.5em}
\noi{\normalsize\bfseries Acknowledgements\par}
\noi We gratefully thank Prof. Evan Williams for the magnet used in this work. We acknowledge the labs of R. Sarpong, J. Long, and P. Arnold for lab chemicals used in this study. We additionally acknowledge Michael C. D. Tayler for helpful conversations and access to NMRduino hardware. We acknowledge helpful and stimulating conversations with J. Kyle Polack and Keith Fritzsching at Sandia National Lab. This work was supported in part by the U.S. Department of Energy National Nuclear Security Administration through the LB26-ML-Rb ZULF NMR-PD3Ra project. This work was supported by NSF Partnerships for Innovation (PFI) (2141083) and NSF Early-concept Grants for Exploratory Research (EAGER) (2231634). The computational work was supported by the Director, Office of Basic Energy Sciences, Chemical Sciences, Geosciences, and Biosciences Division of the U.S.  Department of Energy, under Contract No. DE-AC02-05CH11231.

\vspace{-5mm}

\bibliographystyle{apsrev4-1}
\bibliography{zulf_na_exp}

@string{ pnas = {Proc. Nat. Acad. Sc.} }

@article{cryogenically_cooled_probes,
  title = {Cryogenically Cooled Probes---a Leap in {{NMR}} Technology},
  author = {Kovacs, Helena and Moskau, Detlef and Spraul, Manfred},
  year = 2005,
  month = may,
  journal = {Progress in Nuclear Magnetic Resonance Spectroscopy},
  volume = {46},
  number = {2},
  pages = {131--155},
  issn = {0079-6565},
  doi = {10.1016/j.pnmrs.2005.03.001},
  urldate = {2026-02-20}
}

@article{the_snr_of_nmr,
  title = {The Signal-to-Noise Ratio of the Nuclear Magnetic Resonance Experiment},
  author = {Hoult, D. I and Richards, R. E},
  year = 1976,
  month = oct,
  journal = {Journal of Magnetic Resonance (1969)},
  volume = {24},
  number = {1},
  pages = {71--85},
  issn = {0022-2364},
  doi = {10.1016/0022-2364(76)90233-X},
  urldate = {2026-02-20}
}

@article{optical_constants_water,
  title = {Optical {{Constants}} of {{Water}} in the 200-Nm to 200-Microm {{Wavelength Region}}},
  author = {Hale, G. M. and Querry, M. R.},
  year = 1973,
  month = mar,
  journal = {Applied Optics},
  volume = {12},
  number = {3},
  pages = {555--563},
  issn = {1559-128X},
  doi = {10.1364/AO.12.000555},
  langid = {english},
  pmid = {20125343}
}

@article{doddrelldistortionless1982,
  title = {Distortionless Enhancement of {{NMR}} Signals by Polarization Transfer},
  author = {Doddrell, D. M and Pegg, D. T and Bendall, M. R},
  year = 1982,
  month = jun,
  journal = {Journal of Magnetic Resonance (1969)},
  volume = {48},
  number = {2},
  pages = {323--327},
  issn = {0022-2364},
  doi = {10.1016/0022-2364(82)90286-4},
  urldate = {2026-02-20}
}

@article{andrewsSensitiveMultichannelZeroto2025,
  title = {Sensitive Multichannel Zero-to Ultralow-Field {{NMR}} with Atomic Magnetometer Arrays},
  author = {Andrews, Blake and Lai, Matthew and Wang, Zhen and Kato, Norihisa and Tayler, Michael C D and Druga, Emanuel and Ajoy, Ashok},
  editor = {Thompson, Levi},
  year = 2025,
  month = jun,
  journal = {PNAS Nexus},
  volume = {4},
  number = {6},
  pages = {pgaf187},
  issn = {2752-6542},
  doi = {10.1093/pnasnexus/pgaf187},
  urldate = {2026-02-20},
  copyright = {https://creativecommons.org/licenses/by-nc/4.0/},
  langid = {english}
}

@article{barfieldLongrangeProtonSpinspin1969,
  title = {Long-Range Proton Spin-Spin Coupling},
  author = {Barfield, Michael and Chakrabarti, Bireswar},
  year = 1969,
  month = dec,
  journal = {Chemical Reviews},
  volume = {69},
  number = {6},
  pages = {757--778},
  issn = {0009-2665, 1520-6890},
  doi = {10.1021/cr60262a001},
  urldate = {2026-02-19},
  langid = {english}
}

@article{barskiyZerofieldNuclearMagnetic2019b,
  title = {Zero-Field Nuclear Magnetic Resonance of Chemically Exchanging Systems},
  author = {Barskiy, Danila A. and Tayler, Michael C. D. and {Marco-Rius}, Irene and Kurhanewicz, John and Vigneron, Daniel B. and Cikrikci, Sevil and Aydogdu, Ayca and Reh, Moritz and Pravdivtsev, Andrey N. and H{\"o}vener, Jan-Bernd and Blanchard, John W. and Wu, Teng and Budker, Dmitry and Pines, Alexander},
  year = 2019,
  month = jul,
  journal = {Nature Communications},
  volume = {10},
  number = {1},
  pages = {3002},
  issn = {2041-1723},
  doi = {10.1038/s41467-019-10787-9},
  urldate = {2026-02-18},
  langid = {english}
}

@article{barskiyZeroUltralowfieldNuclear2025,
  title = {Zero- to Ultralow-Field Nuclear Magnetic Resonance},
  author = {Barskiy, Danila A. and Blanchard, John W. and Budker, Dmitry and Eills, James and Pustelny, Szymon and Sheberstov, Kirill F. and Tayler, Michael C.D. and Trabesinger, Andreas H.},
  year = 2025,
  month = aug,
  journal = {Progress in Nuclear Magnetic Resonance Spectroscopy},
  volume = {148--149},
  pages = {101558},
  issn = {00796565},
  doi = {10.1016/j.pnmrs.2025.101558},
  urldate = {2026-02-18},
  langid = {english}
}

@article{blanchardHighResolutionZeroFieldNMR2013b,
  title = {High-{{Resolution Zero-Field NMR}} \textbackslash{{emphJ}} -{{Spectroscopy}} of {{Aromatic Compounds}}},
  author = {Blanchard, John W. and Ledbetter, Micah P. and Theis, Thomas and Butler, Mark C. and Budker, Dmitry and Pines, Alexander},
  year = 2013,
  month = mar,
  journal = {Journal of the American Chemical Society},
  volume = {135},
  number = {9},
  pages = {3607--3612},
  issn = {0002-7863, 1520-5126},
  doi = {10.1021/ja312239v},
  urldate = {2024-04-24}
}

@incollection{blanchardZeroUltralowFieldNMR2016,
  title = {Zero- to {{Ultralow-Field NMR}}},
  booktitle = {{{eMagRes}}},
  author = {Blanchard, John W. and Budker, Dmitry},
  editor = {Harris, Robin K. and Wasylishen, Roderick L.},
  year = 2016,
  month = sep,
  pages = {1395--1410},
  publisher = {John Wiley \& Sons, Ltd},
  address = {Chichester, UK},
  doi = {10.1002/9780470034590.emrstm1369},
  urldate = {2024-06-05},
  isbn = {978-0-470-03459-0 978-0-470-05821-3}
}

@article{blanchardZeroUltralowFieldNuclear2020,
  title = {Zero- to {{Ultralow-Field Nuclear Magnetic Resonance J-spectroscopy}} with {{Commercial Atomic Magnetometers}}},
  author = {Blanchard, John W. and Wu, Teng and Eills, James and Hu, Yinan and Budker, Dmitry},
  year = 2020,
  journal = {Journal of Magnetic Resonance},
  volume = {314},
  pages = {106723},
  issn = {1090-7807},
  doi = {10.1016/j.jmr.2020.106723}
}

@article{bodenstedtFastFieldCyclingUltralowFieldNuclear2021,
  title = {Fast-{{Field-Cycling Ultralow-Field Nuclear Magnetic Relaxation Dispersion}}},
  author = {Bodenstedt, Sven and Mitchell, Morgan W. and Tayler, Michael C. D.},
  year = 2021,
  month = jun,
  journal = {Nature Communications},
  volume = {12},
  number = {1},
  pages = {4041},
  issn = {2041-1723},
  doi = {10.1038/s41467-021-24248-9},
  urldate = {2024-06-18}
}

@phdthesis{bodenstedtOpticallyDetectedNuclear2024a,
  title = {Optically Detected Nuclear Magnetic Resonance above and Far below Earth's Magnetic Field: Spin Dynamics and Relaxation in Unconventional Regimes},
  shorttitle = {Optically Detected Nuclear Magnetic Resonance above and Far below Earth's Magnetic Field},
  author = {Bodenstedt, Sven and Mitchell, Morgan W. and Tayler, Michael C. D.},
  year = 2024,
  month = feb,
  eprint = {2117/406472},
  eprinttype = {hdl},
  doi = {10.5821/dissertation-2117-406472},
  urldate = {2026-02-19},
  copyright = {http://creativecommons.org/licenses/by-nc-sa/4.0/},
  langid = {english},
  school = {Universitat Polit\`ecnica de Catalunya}
}

@article{botoMeasuringFunctionalConnectivity2021a,
  title = {Measuring {{Functional Connectivity}} with {{Wearable MEG}}},
  author = {Boto, Elena and Hill, Ryan M. and Rea, Molly and Holmes, Niall and Seedat, Zelekha A. and Leggett, James and Shah, Vishal and Osborne, James and Bowtell, Richard and Brookes, Matthew J.},
  year = 2021,
  month = apr,
  journal = {NeuroImage},
  volume = {230},
  pages = {117815},
  issn = {10538119},
  doi = {10.1016/j.neuroimage.2021.117815},
  urldate = {2024-05-13}
}

@article{buruevaChemicalReactionMonitoring2020,
  title = {Chemical {{Reaction Monitoring Using Zero-Field Nuclear Magnetic Resonance Enables Study}} of {{Heterogeneous Samples}} in {{Metal Containers}}},
  author = {Burueva, Dudari B. and Eills, James and Blanchard, John W. and Garcon, Antoine and {Picazo-Frutos}, Rom{\'a}n and Kovtunov, Kirill V. and Koptyug, Igor V. and Budker, Dmitry},
  year = 2020,
  month = sep,
  journal = {Angewandte Chemie International Edition},
  volume = {59},
  number = {39},
  pages = {17026--17032},
  issn = {1433-7851, 1521-3773},
  doi = {10.1002/anie.202006266},
  urldate = {2024-04-23}
}

@article{butlerMultipletsZeroMagnetic2013,
  title = {Multiplets at Zero Magnetic Field: {{The}} Geometry of Zero-Field {{NMR}}},
  shorttitle = {Multiplets at Zero Magnetic Field},
  author = {Butler, Mark C. and Ledbetter, Micah P. and Theis, Thomas and Blanchard, John W. and Budker, Dmitry and Pines, Alexander},
  year = 2013,
  month = may,
  journal = {The Journal of Chemical Physics},
  volume = {138},
  number = {18},
  pages = {184202},
  issn = {0021-9606},
  doi = {10.1063/1.4803144},
  urldate = {2026-02-18}
}

@article{chenWatchSized12Tesla2023,
  title = {Watch-{{Sized}} 12 {{Tesla All-High-Temperature-Superconducting Magnet}}},
  author = {Chen, Pin-Hui and Gao, Chukun and Alaniva, Nicholas and Bj{\"o}rgvinsd{\'o}ttir, Sn{\ae}d{\'i}s and Pagonakis, Ioannis Gr. and Urban, Michael A. and D{\"a}pp, Alexander and Gunzenhauser, Ronny and Barnes, Alexander B.},
  year = 2023,
  month = dec,
  journal = {Journal of Magnetic Resonance},
  volume = {357},
  pages = {107588},
  issn = {10907807},
  doi = {10.1016/j.jmr.2023.107588},
  urldate = {2024-05-13}
}

@article{chowdhuryEvaluationMachineLearning2021,
  title = {Evaluation of {{Machine Learning Methods}} for {{Classification}} of {{Rotational Absorption Spectra}} for {{Gases}} in the 220--330 {{GHz Range}}},
  author = {Chowdhury, M. Arshad Zahangir and Rice, Timothy E. and Oehlschlaeger, Matthew A.},
  year = 2021,
  month = feb,
  journal = {Applied Physics B},
  volume = {127},
  number = {3},
  pages = {34},
  issn = {1432-0649},
  doi = {10.1007/s00340-021-07582-0},
  urldate = {2024-11-21}
}

@misc{eillsDirectlyObservableZeemaninsensitive2026,
  title = {A Directly Observable, {{Zeeman-insensitive}} Nuclear Spin Coherence in Solution},
  author = {Eills, James and Singh, Anushka and Teimoori, Amir-Mahyar and {Marco-Rius}, Irene and Mitchell, Morgan W. and Tayler, Michael C. D.},
  year = 2026,
  month = jan,
  number = {arXiv:2601.07614},
  eprint = {2601.07614},
  primaryclass = {physics},
  publisher = {arXiv},
  doi = {10.48550/arXiv.2601.07614},
  urldate = {2026-02-20},
  archiveprefix = {arXiv}
}

@book{ernstPrinciplesNuclearMagnetic1990,
  title = {Principles of {{Nuclear Magnetic Resonance}} in {{One}} and {{Two Dimensions}}},
  author = {Ernst, Richard R and Bodenhausen, Geoffrey and Wokaun, Alexander},
  year = 1990,
  month = may,
  publisher = {Oxford University PressOxford},
  doi = {10.1093/oso/9780198556473.001.0001},
  urldate = {2024-04-23},
  isbn = {978-0-19-855647-3 978-1-383-02862-1}
}

@incollection{hadadNMRSpinCouplingsSaccharides2017b,
  title = {{{NMR Spin-Couplings}} in {{Saccharides}}: {{Relationships Between Structure}}, {{Conformation}} and the {{Magnitudes}} of {{{\emph{J}}}} {{HH}}, {{{\emph{J}}}} {{CH}} and {{{\emph{J}}}} {{CC Values}}},
  shorttitle = {{{NMR Spin-Couplings}} in {{Saccharides}}},
  booktitle = {{{NMR}} in {{Glycoscience}} and {{Glycotechnology}}},
  author = {Hadad, Matthew J. and Zhang, Wenhui and Turney, Toby and Sernau, Luke and Wang, Xiaocong and Woods, Robert J. and Incandela, Andrew and Surjancev, Ivana and Wang, Amy and Yoon, Mi-Kyung and Coscia, Atticus and Euell, Christopher and Meredith, Reagen and Carmichael, Ian and Serianni, Anthony S.},
  editor = {Kato, Koichi and Peters, Thomas},
  year = 2017,
  month = may,
  pages = {20--100},
  publisher = {The Royal Society of Chemistry},
  doi = {10.1039/9781782623946-00020},
  urldate = {2026-02-19},
  isbn = {978-1-78262-310-6 978-1-78801-128-0 978-1-78262-394-6},
  langid = {english}
}

@article{helgakerQuantumChemicalCalculationNMR2008,
  title = {The {{Quantum-Chemical Calculation}} of {{NMR Indirect Spin}}--{{Spin Coupling Constants}}},
  author = {Helgaker, Trygve and Jaszu{\'n}ski, Micha{\l} and Pecul, Magdalena},
  year = 2008,
  month = nov,
  journal = {Progress in Nuclear Magnetic Resonance Spectroscopy},
  volume = {53},
  number = {4},
  pages = {249--268},
  issn = {00796565},
  doi = {10.1016/j.pnmrs.2008.02.002},
  urldate = {2024-11-21},
  copyright = {https://www.elsevier.com/tdm/userlicense/1.0/}
}

@article{hillOptimisingSensitivityOpticallyPumped2024,
  title = {Optimising the {{Sensitivity}} of {{Optically-Pumped Magnetometer Magnetoencephalography}} to {{Gamma Band Electrophysiological Activity}}},
  author = {Hill, Ryan M. and Schofield, Holly and Boto, Elena and Rier, Lukas and Osborne, James and Doyle, Cody and Worcester, Frank and Hayward, Tyler and Holmes, Niall and Bowtell, Richard and Shah, Vishal and Brookes, Matthew J.},
  year = 2024,
  month = mar,
  journal = {Imaging Neuroscience},
  volume = {2},
  pages = {1--19},
  issn = {2837-6056},
  doi = {10.1162/imag_a_00112},
  urldate = {2024-11-21}
}

@article{hogbenSpinachSoftwareLibrary2011b,
  title = {{\emph{Spinach}} -- {{A}} Software Library for Simulation of Spin Dynamics in Large Spin Systems},
  author = {Hogben, H. J. and Krzystyniak, M. and Charnock, G. T. P. and Hore, P. J. and Kuprov, Ilya},
  year = 2011,
  month = feb,
  journal = {Journal of Magnetic Resonance},
  volume = {208},
  number = {2},
  pages = {179--194},
  issn = {1090-7807},
  doi = {10.1016/j.jmr.2010.11.008},
  urldate = {2026-02-20}
}

@article{horaiMassBankPublicRepository2010,
  title = {{{MassBank}}: {{A Public Repository}} for {{Sharing Mass Spectral Data}} for {{Life Sciences}}},
  shorttitle = {{{MassBank}}},
  author = {Horai, Hisayuki and Arita, Masanori and Kanaya, Shigehiko and Nihei, Yoshito and Ikeda, Tasuku and Suwa, Kazuhiro and Ojima, Yuya and Tanaka, Kenichi and Tanaka, Satoshi and Aoshima, Ken and Oda, Yoshiya and Kakazu, Yuji and Kusano, Miyako and Tohge, Takayuki and Matsuda, Fumio and Sawada, Yuji and Hirai, Masami Yokota and Nakanishi, Hiroki and Ikeda, Kazutaka and Akimoto, Naoshige and Maoka, Takashi and Takahashi, Hiroki and Ara, Takeshi and Sakurai, Nozomu and Suzuki, Hideyuki and Shibata, Daisuke and Neumann, Steffen and Iida, Takashi and Tanaka, Ken and Funatsu, Kimito and Matsuura, Fumito and Soga, Tomoyoshi and Taguchi, Ryo and Saito, Kazuki and Nishioka, Takaaki},
  year = 2010,
  month = jul,
  journal = {Journal of Mass Spectrometry},
  volume = {45},
  number = {7},
  pages = {703--714},
  issn = {1076-5174, 1096-9888},
  doi = {10.1002/jms.1777},
  urldate = {2024-11-21},
  copyright = {http://onlinelibrary.wiley.com/termsAndConditions\#vor}
}

@article{leeConstructionTestResult2022,
  title = {Construction and {{Test Result}} of an {{All-REBCO Conduction-Cooled}} 23.5 {{T Magnet Prototype}} towards a {{Benchtop}} 1 {{GHz NMR Spectroscopy}}},
  author = {Lee, Wooseung and Park, Dongkeun and Bascu{\~n}{\'a}n, Juan and Iwasa, Yukikazu},
  year = 2022,
  month = oct,
  journal = {Superconductor Science and Technology},
  volume = {35},
  number = {10},
  pages = {105007},
  issn = {0953-2048, 1361-6668},
  doi = {10.1088/1361-6668/ac8773},
  urldate = {2024-06-26}
}

@article{mandzhievaZeroFieldNMRMilliteslaSLIC2025a,
  title = {Zero-{{Field NMR}} and {{Millitesla-SLIC Spectra}} for {$>$}200 {{Molecules}} from {{Density Functional Theory}} and {{Spin Dynamics}}},
  author = {Mandzhieva, Iuliia and Theiss, Franziska and He, Xingtao and Ortmeier, Adam and Koirala, Anuja and McBride, Stephen J. and DeVience, Stephen J. and Rosen, Matthew S. and Blum, Volker and Theis, Thomas},
  year = 2025,
  month = jul,
  journal = {Journal of Chemical Information and Modeling},
  volume = {65},
  number = {14},
  pages = {7554--7568},
  issn = {1549-9596, 1549-960X},
  doi = {10.1021/acs.jcim.5c00111},
  urldate = {2026-02-19},
  copyright = {https://doi.org/10.15223/policy-029},
  langid = {english}
}

@article{mccarthyMoleculeIdentificationRotational2020,
  title = {Molecule {{Identification}} with {{Rotational Spectroscopy}} and {{Probabilistic Deep Learning}}},
  author = {McCarthy, Michael and Lee, Kin Long Kelvin},
  year = 2020,
  month = apr,
  journal = {The Journal of Physical Chemistry A},
  volume = {124},
  number = {15},
  pages = {3002--3017},
  issn = {1089-5639, 1520-5215},
  doi = {10.1021/acs.jpca.0c01376},
  urldate = {2024-11-21},
  copyright = {https://doi.org/10.15223/policy-029}
}

@article{picazo-frutosZerofieldJspectroscopyQuadrupolar2024b,
  title = {Zero-Field {{J-spectroscopy}} of Quadrupolar Nuclei},
  author = {{Picazo-Frutos}, Rom{\'a}n and Sheberstov, Kirill F. and Blanchard, John W. and Van Dyke, Erik and Reh, Moritz and Sjoelander, Tobias and Pines, Alexander and Budker, Dmitry and Barskiy, Danila A.},
  year = 2024,
  month = may,
  journal = {Nature Communications},
  volume = {15},
  number = {1},
  pages = {4487},
  issn = {2041-1723},
  doi = {10.1038/s41467-024-48390-2},
  urldate = {2026-02-18},
  langid = {english}
}

@misc{shahQZFMGen3,
  title = {{{QZFM Gen-3}}},
  author = {Shah, Vishal},
  journal = {Quspin},
  urldate = {2024-05-15},
  url = {https://quspin.com/products-qzfm/}
}

@article{meyer2003interactions,
  title={Interactions with aromatic rings in chemical and biological recognition},
  author={Meyer, Emmanuel A and Castellano, Ronald K and Diederich, Fran{\c{c}}ois},
  journal={Angewandte Chemie International Edition},
  volume={42},
  number={11},
  pages={1210--1250},
  year={2003},
  publisher={Wiley Online Library}
}

@article{zhou2012specific,
  title={Specific noncovalent interactions at protein-ligand interface: implications for rational drug design},
  author={Zhou, P and Huang, J and Tian, F},
  journal={Current medicinal chemistry},
  volume={19},
  number={2},
  pages={226--238},
  year={2012},
  publisher={Bentham Science Publishers}
}

@article{shapiraSpatialEncodingAcquisition2004,
  title = {Spatial {{Encoding}} and the {{Acquisition}} of {{High-Resolution NMR Spectra}} in {{Inhomogeneous Magnetic Fields}}},
  author = {Shapira, Boaz and Frydman, Lucio},
  year = 2004,
  month = jun,
  journal = {Journal of the American Chemical Society},
  volume = {126},
  number = {23},
  pages = {7184--7185},
  issn = {0002-7863, 1520-5126},
  doi = {10.1021/ja048859u},
  urldate = {2024-06-06}
}

@article{TAYLER2017143,
title = {Low-cost, pseudo-Halbach dipole magnets for NMR},
journal = {Journal of Magnetic Resonance},
volume = {277},
pages = {143-148},
year = {2017},
issn = {1090-7807},
doi = {https://doi.org/10.1016/j.jmr.2017.03.001},
url = {https://www.sciencedirect.com/science/article/pii/S1090780717300630},
author = {Michael C.D. Tayler and Dimitrios Sakellariou}
}

@article{HONG2025200170,
title = {Femtotesla atomic magnetometer for zero- and ultralow-field nuclear magnetic resonance},
journal = {Magnetic Resonance Letters},
volume = {5},
number = {3},
pages = {200170},
year = {2025},
issn = {2772-5162},
doi = {https://doi.org/10.1016/j.mrl.2024.200170},
url = {https://www.sciencedirect.com/science/article/pii/S2772516224000779},
author = {Taizhou Hong and Yuanhong Wang and Zhenhan Shao and Qing Li and Min Jiang and Xinhua Peng}
}

@book{slichterPrinciplesMagneticResonance,
  title = {Principles of {{Magnetic Resonance}}},
  author = {Slichter, Charles}
}

@article{sternhellCorrelationInterprotonSpin1969,
  title = {Correlation of Interproton Spin--Spin Coupling Constants with Structure},
  author = {Sternhell, S.},
  year = 1969,
  month = jan,
  journal = {Quarterly Reviews, Chemical Society},
  volume = {23},
  number = {2},
  pages = {236--270},
  publisher = {The Royal Society of Chemistry},
  issn = {0009-2681},
  doi = {10.1039/QR9692300236},
  urldate = {2026-02-19},
  langid = {english}
}

@article{taylerInvitedReviewArticle2017,
  title = {Invited {{Review Article}}: {{Instrumentation}} for {{Nuclear Magnetic Resonance}} in {{Zero}} and {{Ultralow Magnetic Field}}},
  shorttitle = {Invited {{Review Article}}},
  author = {Tayler, Michael C. D. and Theis, Thomas and Sjolander, Tobias F. and Blanchard, John W. and Kentner, Arne and Pustelny, Szymon and Pines, Alexander and Budker, Dmitry},
  year = 2017,
  month = sep,
  journal = {Review of Scientific Instruments},
  volume = {88},
  number = {9},
  pages = {091101},
  issn = {0034-6748, 1089-7623},
  doi = {10.1063/1.5003347},
  urldate = {2024-06-05}
}

@article{taylerNMRduinoModularOpenSource2024,
  title = {{{NMRduino}}: {{A Modular}}, {{Open-Source}}, {{Low-Field Magnetic Resonance Platform}}},
  shorttitle = {{{NMRduino}}},
  author = {Tayler, Michael C.D. and Bodenstedt, Sven},
  year = 2024,
  month = may,
  journal = {Journal of Magnetic Resonance},
  volume = {362},
  pages = {107665},
  issn = {10907807},
  doi = {10.1016/j.jmr.2024.107665},
  urldate = {2024-05-14}
}

@article{theisChemicalAnalysisUsing2013,
  title = {Chemical {{Analysis Using J-coupling Multiplets}} in {{Zero-Field NMR}}},
  author = {Theis, Thomas and Blanchard, John W. and Butler, Mark C. and Ledbetter, Micah P. and Budker, Dmitry and Pines, Alexander},
  year = 2013,
  journal = {Chemical Physics Letters},
  volume = {580},
  pages = {160--165},
  issn = {0009-2614},
  doi = {10.1016/j.cplett.2013.06.042}
}

@article{theisParahydrogenEnhancedZeroFieldNuclear2011,
  title = {Parahydrogen-{{Enhanced Zero-Field Nuclear Magnetic Resonance}}},
  author = {Theis, T. and Ganssle, P. and Kervern, G. and Knappe, S. and Kitching, J. and Ledbetter, M. P. and Budker, D. and Pines, A.},
  year = 2011,
  month = jul,
  journal = {Nature Physics},
  volume = {7},
  number = {7},
  pages = {571--575},
  issn = {1745-2473, 1745-2481},
  doi = {10.1038/nphys1986},
  urldate = {2024-04-23},
  copyright = {http://www.springer.com/tdm}
}

@article{vanzijlUseDeuteriumNucleus1987,
  title = {The {{Use}} of {{Deuterium}} as a {{Nucleus}} for {{Locking}}, {{Shimming}}, and {{Measuring NMR}} at {{High Magnetic Fields}}},
  author = {Van Zijl, Peter C.M},
  year = 1987,
  month = nov,
  journal = {Journal of Magnetic Resonance (1969)},
  volume = {75},
  number = {2},
  pages = {335--344},
  issn = {00222364},
  doi = {10.1016/0022-2364(87)90039-4},
  urldate = {2024-06-05},
  copyright = {https://www.elsevier.com/tdm/userlicense/1.0/}
}

@article{weitekampZeroFieldNuclearMagnetic1983a,
  title = {Zero-{{Field Nuclear Magnetic Resonance}}},
  author = {Weitekamp, D. P. and Bielecki, A. and Zax, D. and Zilm, K. and Pines, A.},
  year = 1983,
  month = may,
  journal = {Physical Review Letters},
  volume = {50},
  number = {22},
  pages = {1807--1810},
  issn = {0031-9007},
  doi = {10.1103/PhysRevLett.50.1807},
  urldate = {2026-02-20},
  copyright = {http://link.aps.org/licenses/aps-default-license},
  langid = {english}
}

@article{wilzewskiMethodMeasurementSpinSpin2017a,
  title = {A {{Method}} for {{Measurement}} of {{Spin-Spin Couplings}} with {{Sub-mHz Precision Using Zero-}} to {{Ultralow-Field Nuclear Magnetic Resonance}}},
  author = {Wilzewski, A. and Afach, S. and Blanchard, J.W. and Budker, D.},
  year = 2017,
  month = nov,
  journal = {Journal of Magnetic Resonance},
  volume = {284},
  pages = {66--72},
  issn = {10907807},
  doi = {10.1016/j.jmr.2017.08.016},
  urldate = {2024-04-23}
}

@article{wuSearchAxionlikeDark2019a,
  title = {Search for {{Axionlike Dark Matter}} with a {{Liquid-State Nuclear Spin Comagnetometer}}},
  author = {Wu, Teng and Blanchard, John W. and Centers, Gary P. and Figueroa, Nataniel L. and Garcon, Antoine and Graham, Peter W. and Kimball, Derek F. Jackson and Rajendran, Surjeet and Stadnik, Yevgeny V. and Sushkov, Alexander O. and Wickenbrock, Arne and Budker, Dmitry},
  year = 2019,
  month = may,
  journal = {Physical Review Letters},
  volume = {122},
  number = {19},
  pages = {191302},
  issn = {0031-9007, 1079-7114},
  doi = {10.1103/PhysRevLett.122.191302},
  urldate = {2026-02-19},
  langid = {english}
}

@article{zaxZeroFieldNMR1985,
  title = {Zero Field {{NMR}} and {{NQR}}},
  author = {Zax, D. B. and Bielecki, A. and Zilm, K. W. and Pines, A. and Weitekamp, D. P.},
  year = 1985,
  month = nov,
  journal = {The Journal of Chemical Physics},
  volume = {83},
  number = {10},
  pages = {4877--4905},
  issn = {0021-9606, 1089-7690},
  doi = {10.1063/1.449748},
  urldate = {2026-02-20},
  langid = {english}
}

@article{zhangMeasuringHumanAuditory2024a,
  title = {Measuring {{Human Auditory Evoked Fields}} with a {{Flexible Multi-Channel OPM-Based MEG System}}},
  author = {Zhang, Xin and Chang, Yan and Wang, Hui and Zhang, Yin and Hu, Tao and Feng, Xiao-yu and Zhang, Ming-kang and Yao, Ze-kun and Chen, Chun-qiao and Xu, Jia-yu and Fu, Fang-yue and Guo, Qing-qian and Zhu, Jian-bing and Xie, Hai-qun and Yang, Xiao-dong},
  year = 2024,
  month = apr,
  journal = {Journal of Integrative Neuroscience},
  volume = {23},
  number = {5},
  pages = {93},
  issn = {0219-6352},
  doi = {10.31083/j.jin2305093}
}

@article{ruden2003vibrational,
  title={Vibrational corrections to indirect nuclear spin--spin coupling constants calculated by density-functional theory},
  author={Ruden, Torgeir A and Lutn{\ae}s, Ola B and Helgaker, Trygve and Ruud, Kenneth},
  journal={Journal of Chemical Physics},
  volume={118},
  number={21},
  pages={9572--9581},
  year={2003},
  publisher={American Institute of Physics},
  doi = {https://doi.org/10.1063/1.1569846}
}

@article{adamo1999toward,
  title={Toward reliable density functional methods without adjustable parameters: The PBE0 model},
  author={Adamo, Carlo and Barone, Vincenzo},
  journal={Journal of Chemical Physics},
  volume={110},
  number={13},
  pages={6158--6170},
  year={1999},
  publisher={American Institute of Physics},
  doi = {https://doi.org/10.1063/1.478522}
}

@article{shimizu2015zero,
  title={Zero-field nuclear magnetic resonance spectroscopy of viscous liquids},
  author={Shimizu, Y and Blanchard, JW and Pustelny, Szymon and Saielli, G and Bagno, Alessandro and Ledbetter, MP and Budker, D and Pines, A},
  journal={Journal of Magnetic Resonance},
  volume={250},
  pages={1--6},
  year={2015},
  publisher={Elsevier},
  doi = {https://doi.org/10.1016/j.jmr.2014.10.012}
}

@book{scheiner1997hydrogen,
  title={Hydrogen bonding: a theoretical perspective},
  author={Scheiner, Steve},
  volume={7},
  year={1997},
  publisher={Oxford University Press},
  doi = {https://doi.org/10.1093/oso/9780195090116.001.0001}
}

@article{chelli2005structure,
  title={Structure of liquid formic acid investigated by first principle and classical molecular dynamics simulations},
  author={Chelli, Riccardo and Righini, Roberto and Califano, Salvatore},
  journal={Journal of Physical Chemistry B},
  volume={109},
  number={35},
  pages={17006--17013},
  year={2005},
  publisher={ACS Publications},
  doi = {https://doi.org/10.1021/jp051731u}
}

@article{giovannini2020molecular,
  title={Molecular spectroscopy of aqueous solutions: a theoretical perspective},
  author={Giovannini, Tommaso and Egidi, Franco and Cappelli, Chiara},
  journal={Chemical Society Reviews},
  volume={49},
  number={16},
  pages={5664--5677},
  year={2020},
  publisher={Royal Society of Chemistry},
  doi = {https://doi.org/10.1039/C9CS00464E}
}

@article{rahman2012hydration,
  title={Hydration of formate and acetate ions by dielectric relaxation spectroscopy},
  author={Rahman, Hafiz MA and Hefter, Glenn and Buchner, Richard},
  journal={Journal of Physical Chemistry B},
  volume={116},
  number={1},
  pages={314--323},
  year={2012},
  publisher={ACS Publications},
  doi = {https://doi.org/10.1021/jp207504d}
}

@article{leung2004ab,
  title={Ab initio molecular dynamics study of formate ion hydration},
  author={Leung, Kevin and Rempe, Susan B},
  journal={Journal of the American Chemical Society},
  volume={126},
  number={1},
  pages={344--351},
  year={2004},
  publisher={ACS Publications},
  doi = {https://doi.org/10.1021/ja036267q}
}

@article{rudolph2022raman,
  title={Raman spectroscopic studies on aqueous sodium formate solutions and DFT calculations},
  author={Rudolph, Wolfram W and Irmer, Gert},
  journal={Journal of Solution Chemistry},
  volume={51},
  number={8},
  pages={935--961},
  year={2022},
  publisher={Springer},
  doi = {https://doi.org/10.1007/s10953-022-01170-2}
}

@article{trapp2025electrolyte,
  title={Electrolyte Structure Governs Formate Oxidation at High Concentrations: from Clusters to Currents},
  author={Trapp, Katharina and Kosasang, Soracha and Ingenmey, Johannes and Gomez Vazquez, Dario and Salanne, Mathieu and Lukatskaya, Maria},
  journal={ChemRxiv},
  year={2025},
  publisher={Cambridge University Press},
  doi = {https://doi.org/10.26434/chemrxiv-2025-7j1vd-v2}
}

@article{neese2025software,
  title={Software update: the ORCA program system—version 6.0},
  author={Neese, Frank},
  journal={WIREs Computational Molecular Science},
  volume={15},
  number={2},
  pages={e70019},
  year={2025},
  publisher={Wiley Online Library},
  doi = {https://doi.org/10.1002/wcms.70019}
}

@article{helmich2021improved,
  title={An improved chain of spheres for exchange algorithm},
  author={Helmich-Paris, Benjamin and de Souza, Bernardo and Neese, Frank and Izs{\'a}k, R{\'o}bert},
  journal={Journal of Chemical Physics},
  volume={155},
  number={10},
  pages={104109},
  year={2021},
  publisher={AIP Publishing},
  doi = {https://doi.org/10.1063/5.0058766}
}

@article{stoychev2017automatic,
  title={Automatic generation of auxiliary basis sets},
  author={Stoychev, Georgi L and Auer, Alexander A and Neese, Frank},
  journal={Journal of Chemical Theory and Computation},
  volume={13},
  number={2},
  pages={554--562},
  year={2017},
  publisher={ACS Publications},
  doi = {https://doi.org/10.1021/acs.jctc.6b01041}
}

@article{de2025goat,
  title={GOAT: A global optimization algorithm for molecules and atomic clusters},
  author={de Souza, Bernardo},
  journal={Angewandte Chemie International Edition},
  volume={64},
  number={18},
  pages={e202500393},
  year={2025},
  publisher={Wiley Online Library},
  doi = {https://doi.org/10.1002/anie.202500393}
}

@article{bannwarth2019gfn2,
  title={GFN2-xTB—An accurate and broadly parametrized self-consistent tight-binding quantum chemical method with multipole electrostatics and density-dependent dispersion contributions},
  author={Bannwarth, Christoph and Ehlert, Sebastian and Grimme, Stefan},
  journal={Journal of Chemical Theory and Computation},
  volume={15},
  number={3},
  pages={1652--1671},
  year={2019},
  publisher={ACS Publications},
  doi = {https://doi.org/10.1021/acs.jctc.8b01176}
}

@article{grimme2010consistent,
  title={A consistent and accurate ab initio parametrization of density functional dispersion correction (DFT-D) for the 94 elements H-Pu},
  author={Grimme, Stefan and Antony, Jens and Ehrlich, Stephan and Krieg, Helge},
  journal={Journal of Chemical Physics},
  volume={132},
  number={15},
  pages={154104},
  year={2010},
  publisher={AIP Publishing},
  doi = {https://doi.org/10.1063/1.3382344}
}

@article{dunning1989gaussian,
  title={Gaussian basis sets for use in correlated molecular calculations. I. The atoms boron through neon and hydrogen},
  author={Dunning Jr, Thom H},
  journal={Journal of Chemical Physics},
  volume={90},
  number={2},
  pages={1007--1023},
  year={1989},
  publisher={American Institute of Physics},
  doi = {https://doi.org/10.1063/1.456153}
}

@article{kendall1992electron,
  title={Electron affinities of the first-row atoms revisited. Systematic basis sets and wave functions},
  author={Kendall, Rick A and Dunning Jr, Thom H and Harrison, Robert J},
  journal={Journal of Chemical Physics},
  volume={96},
  number={9},
  pages={6796--6806},
  year={1992},
  publisher={American Institute of Physics},
  doi = {https://doi.org/10.1063/1.462569}
}

@article{woon1995gaussian,
  title={Gaussian basis sets for use in correlated molecular calculations. V. Core-valence basis sets for boron through neon},
  author={Woon, David E and Dunning Jr, Thom H},
  journal={Journal of Chemical Physics},
  volume={103},
  number={11},
  pages={4572--4585},
  year={1995},
  publisher={American Institute of Physics},
  doi = {https://doi.org/10.1063/1.470645}
}

@article{jensen2006basis,
  title={The basis set convergence of spin- spin coupling constants calculated by density functional methods},
  author={Jensen, Frank},
  journal={Journal of Chemical Theory and Computation},
  volume={2},
  number={5},
  pages={1360--1369},
  year={2006},
  publisher={ACS Publications},
  doi = {https://doi.org/10.1021/ct600166u}
}

@article{franke2021vpt2,
  title={How to VPT2: Accurate and intuitive simulations of CH stretching infrared spectra using VPT2+ K with large effective Hamiltonian resonance treatments},
  author={Franke, Peter R and Stanton, John F and Douberly, Gary E},
  journal={Journal of Physical Chemistry A},
  volume={125},
  number={6},
  pages={1301--1324},
  year={2021},
  publisher={ACS Publications},
  doi = {https://doi.org/10.1021/acs.jpca.0c09526}
}

@article{pavovsevic2014geminal,
  title={Geminal-spanning orbitals make explicitly correlated reduced-scaling coupled-cluster methods robust, yet simple},
  author={Pavo{\v{s}}evi{\'c}, Fabijan and Neese, Frank and Valeev, Edward F},
  journal={Journal of chemical physics},
  volume={141},
  number={5},
  pages={054106},
  year={2014},
  publisher={AIP Publishing},
  doi = {https://doi.org/10.1063/1.4890002}
}

@article{peterson2008systematically,
  title={Systematically convergent basis sets for explicitly correlated wavefunctions: The atoms H, He, B--Ne, and Al--Ar},
  author={Peterson, Kirk A and Adler, Thomas B and Werner, Hans-Joachim},
  journal={Journal of Chemical Physics},
  volume={128},
  number={8},
  pages={084102},
  year={2008},
  publisher={AIP Publishing},
  doi = {https://doi.org/10.1063/1.2831537}
}

@article{cossi2003energies,
  title={Energies, structures, and electronic properties of molecules in solution with the C-PCM solvation model},
  author={Cossi, Maurizio and Rega, Nadia and Scalmani, Giovanni and Barone, Vincenzo},
  journal={Journal of Computational Chemistry},
  volume={24},
  number={6},
  pages={669--681},
  year={2003},
  publisher={Wiley Online Library},
  doi = {https://doi.org/10.1002/jcc.10189}
}

@article{jensen2007polarization,
  title={Polarization Consistent Basis Sets. IV: The Elements He, Li, Be, B, Ne, Na, Mg, Al, and Ar},
  author={Jensen, Frank},
  journal={Journal of Physical Chemistry A},
  volume={111},
  number={44},
  pages={11198--11204},
  year={2007},
  publisher={ACS Publications},
  doi = {https://doi.org/10.1021/jp068677h}
}

@article{jensen2012polarization,
  title={Polarization consistent basis sets. VII: The elements K, Ca, Ga, Ge, As, Se, Br, and Kr},
  author={Jensen, Frank},
  journal={Journal of Chemical Physics},
  volume={136},
  number={11},
  pages={114107},
  year={2012},
  publisher={AIP Publishing},
  doi = {https://doi.org/10.1063/1.3690460}
}

@article{horn2016probing,
  title={Probing non-covalent interactions with a second generation energy decomposition analysis using absolutely localized molecular orbitals},
  author={Horn, Paul R and Mao, Yuezhi and Head-Gordon, Martin},
  journal={Physical Chemistry Chemical Physics},
  volume={18},
  number={33},
  pages={23067--23079},
  year={2016},
  publisher={Royal Society of Chemistry},
  doi = {https://doi.org/10.1039/C6CP03784D}
}

@article{epifanovsky2021software,
  title={Software for the frontiers of quantum chemistry: An overview of developments in the Q-Chem 5 package},
  author={Epifanovsky, Evgeny and Gilbert, Andrew TB and Feng, Xintian and Lee, Joonho and Mao, Yuezhi and Mardirossian, Narbe and Pokhilko, Pavel and White, Alec F and Coons, Marc P and Dempwolff, Adrian L and others},
  journal={Journal of Chemical Physics},
  volume={155},
  number={8},
  pages={084801},
  year={2021},
  publisher={AIP Publishing},
  doi = {https://doi.org/10.1063/5.0055522}
}

@article{chai2008systematic,
  title={Systematic optimization of long-range corrected hybrid density functionals},
  author={Chai, Jeng-Da and Head-Gordon, Martin},
  journal={Journal of Chemical Physics},
  volume={128},
  number={8},
  pages={084106},
  year={2008},
  publisher={AIP Publishing},
  doi = {https://doi.org/10.1063/1.2834918}
}

@article{chai2008long,
  title={Long-range corrected hybrid density functionals with damped atom--atom dispersion corrections},
  author={Chai, Jeng-Da and Head-Gordon, Martin},
  journal={Physical Chemistry Chemical Physics},
  volume={10},
  number={44},
  pages={6615--6620},
  year={2008},
  publisher={Royal Society of Chemistry},
  doi = {https://doi.org/10.1039/B810189B}
}

@article{lin2013long,
  title={Long-range corrected hybrid density functionals with improved dispersion corrections},
  author={Lin, You-Sheng and Li, Guan-De and Mao, Shan-Ping and Chai, Jeng-Da},
  journal={Journal of Chemical Theory and Computation},
  volume={9},
  number={1},
  pages={263--272},
  year={2013},
  publisher={ACS Publications},
  doi = {https://doi.org/10.1021/ct300715s}
}

@article{weigend2005balanced,
  title={Balanced basis sets of split valence, triple zeta valence and quadruple zeta valence quality for H to Rn: Design and assessment of accuracy},
  author={Weigend, Florian and Ahlrichs, Reinhart},
  journal={Physical Chemistry Chemical Physics},
  volume={7},
  number={18},
  pages={3297--3305},
  year={2005},
  publisher={Royal Society of Chemistry},
  doi = {https://doi.org/10.1039/B508541A}
}

@article{rappoport2010property,
  title={Property-optimized Gaussian basis sets for molecular response calculations},
  author={Rappoport, Dmitrij and Furche, Filipp},
  journal={Journal of Chemical Physics},
  volume={133},
  number={13},
  pages={134105},
  year={2010},
  publisher={AIP Publishing},
  doi = {https://doi.org/10.1063/1.3484283}
}

@book{mao2017advances,
  title={Advances in Density Functional Theory Calculations and Their Analysis},
  author={Mao, Yuezhi},
  year={2017},
  publisher={University of California, Berkeley}
}

@article{valeev2004improving,
  title={Improving on the resolution of the identity in linear R12 ab initio theories},
  author={Valeev, Edward F},
  journal={Chemical Physics Letters},
  volume={395},
  number={4-6},
  pages={190--195},
  year={2004},
  publisher={Elsevier},
  doi = {https://doi.org/10.1016/j.cplett.2004.07.061}
}

@article{weigend2002efficient,
  title={Efficient use of the correlation consistent basis sets in resolution of the identity MP2 calculations},
  author={Weigend, Florian and K{\"o}hn, Andreas and H{\"a}ttig, Christof},
  journal={Journal of Chemical Physics},
  volume={116},
  number={8},
  pages={3175--3183},
  year={2002},
  publisher={American Institute of Physics},
  doi = {https://doi.org/10.1063/1.1445115}
}

@article{li2024hydrated,
  title={Hydrated formic acid clusters and their interaction with electrons},
  author={Li, Kevin and {\v{D}}urana, Jozef and Koc{\'a}bkov{\'a}, Barbora and Pysanenko, Andrij and Yan, Yihui and On{\v{c}}{\'a}k, Milan and F{\'a}rn{\'\i}k, Michal and Lengyel, Jozef},
  journal={ChemPhysChem},
  volume={25},
  number={10},
  pages={e202400071},
  year={2024},
  publisher={Wiley Online Library},
  doi = {https://doi.org/10.1002/cphc.202400071}
}

@article{o2024pair,
  title={To pair or not to pair? Machine-learned explicitly-correlated electronic structure for NaCl in water},
  author={O’Neill, Niamh and Shi, Benjamin X and Fong, Kara and Michaelides, Angelos and Schran, Christoph},
  journal={Journal of Physical Chemistry Letters},
  volume={15},
  number={23},
  pages={6081--6091},
  year={2024},
  publisher={ACS Publications},
  doi = {https://doi.org/10.1021/acs.jpclett.4c01030}
}


%

\setcounter{figure}{0}
\renewcommand{\thefigure}{S\arabic{figure}}
\renewcommand{\theHfigure}{S\arabic{figure}}
\providecommand{\Cs}{^{13}\mathrm{C}}
\providecommand{\Hs}{^{1}\mathrm{H}}
\providecommand{\R}[1]{\mathrm{#1}}
\providecommand{\zfr}[1]{#1}
\providecommand{\SIfigwidth}{0.90\linewidth}
\providecommand{\SIfigwidthNarrow}{0.72\linewidth}

\clearpage
\onecolumngrid

\setcounter{section}{0}
\setcounter{subsection}{0}
\setcounter{figure}{0}
\setcounter{table}{0}
\setcounter{equation}{0}
\renewcommand{\thesection}{\Roman{section}}
\renewcommand{\thesubsection}{\thesection.\Alph{subsection}}
\makeatletter
\renewcommand{\p@subsection}{}
\makeatother
\renewcommand{\thefigure}{S\arabic{figure}}
\renewcommand{\thetable}{S\arabic{table}}
\renewcommand{\theequation}{S\arabic{equation}}

\begin{center}
{\Large \bfseries Supplementary Information}\\[0.5em]
{\large DFT-assisted natural abundance $\Cs$ zero-field NMR}
\end{center}

\vspace{0.5em}

\tableofcontents

\clearpage

\section{Guide to the Supplementary Information}
\label{sec:executive_summary}

This Supporting Information provides more details about the experiments, data analysis and DFT methodology employed in the main paper. We begin with the full experimental spectral library and the processing steps used to generate the displayed spectra, then move to the quantum-chemical and spin-dynamics tools used to interpret those measurements, and finally document the apparatus, benchmarking, and sample-preparation details needed for reproducibility.

\SIsecref{sec:full_na_lib_para} expands the representative spectra in \mainfigref[A--G]{mfig2}, \mainfigref[A]{mfig3} and \mainfigref[A--D]{mfig4} into the complete natural-abundance ZF NMR library. It also records the acquisition and display-processing workflow used to produce the spectra shown in \mainfigref[A--B]{mfig1}, \mainfigref[A--G]{mfig2}, \mainfigref[A]{mfig3}, \mainfigref[A--D]{mfig4}, and \mainfigref[A,C]{mfig5}, so this section serves as the experimental foundation for the rest of the SI.

\SIsecref{sec:qc_j} lays out the quantum-chemical framework used for the DFT-assisted interpretation in \mainfigref[A--D]{mfig4} and \mainfigref[B--D]{mfig5}. It covers vibrational corrections to computed $J$ couplings, electron-density visualizations for shifts in the $J$ coupling values, the overall computational protocol, and the explicit-solvent treatment of concentrated formate solutions. \SIsecref{sec:zf_spin_sim} follows by describing the complementary spin-dynamics tools used throughout the paper, including time-domain Spinach simulations, frequency-domain response calculations, machine-readable simulation inputs, and the double-$^{13}$C acetic-acid analysis that supports \mainfigref[A]{mfig3}. Together, these two sections connect the measured spectra to predictions and interpretations as discussed in the main text.

The remaining sections document the experimental platform and the practical details needed to reproduce the measurements. \SIsecref{sec:instrument_exp} describes the zero-field apparatus, pulse sequence, and NMRduino control workflow that underlie the experiments summarized in the main paper. \SIsecref{sec:sensitivity_budget} benchmarks sensitivity and long-term stability against prior ZF NMR work and includes the 100~mM formic-acid detection-limit comparison, while \SIsecref{sec:sample_prep} lists the off-the-shelf liquids, the aqueous formic-acid dilution series, and the concentrated formate solution compositions used for the datasets in \mainfigref[A--G]{mfig2} and \mainfigref[A,C]{mfig5}. 

\section{Complete natural-abundance spectral library and data processing}
\label{sec:full_na_lib_para}

\begin{figure}[h!]
  \centering
  \includegraphics[width=0.97\textwidth]{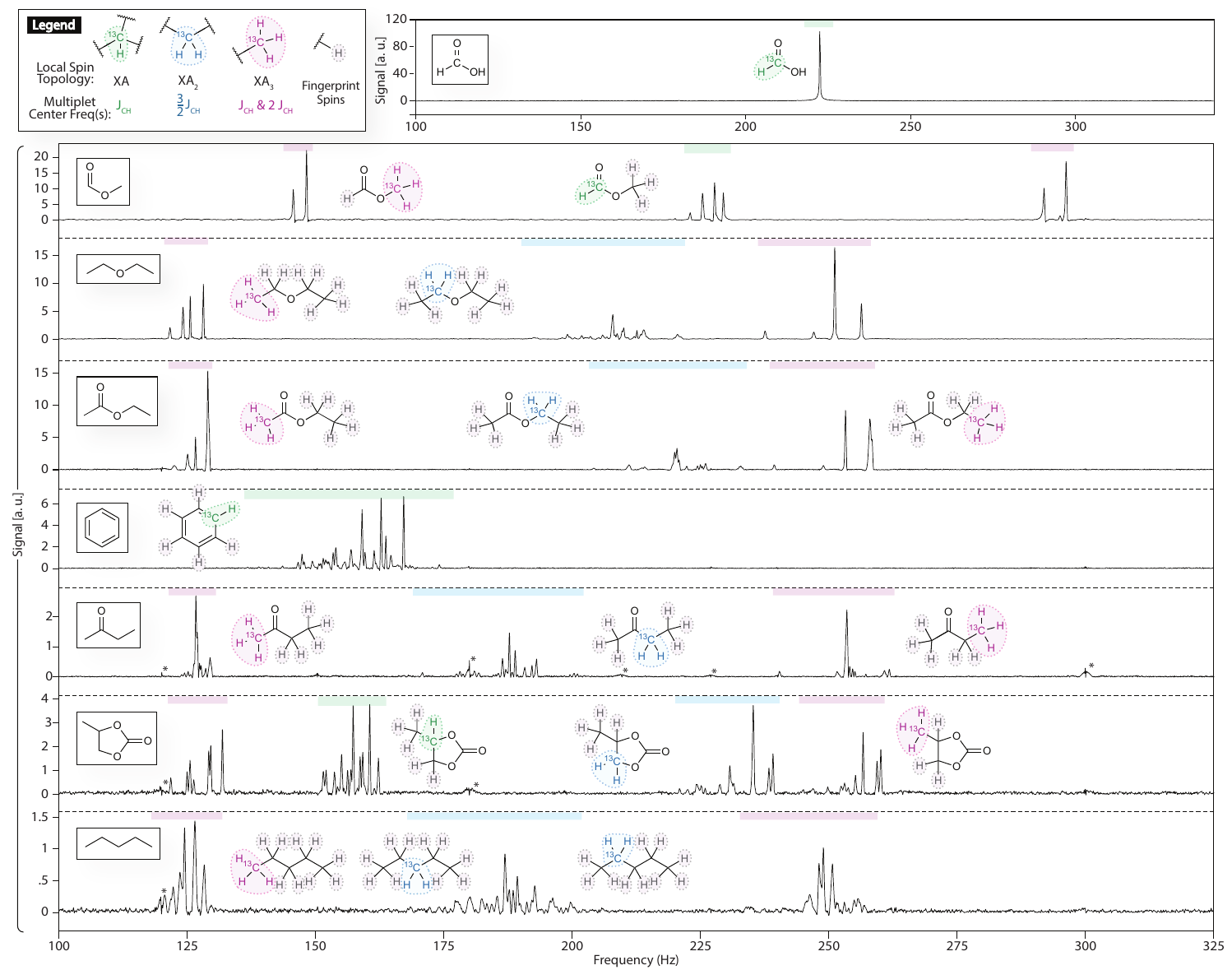}
  \caption{\textbf{Complete natural-abundance ZF NMR library: Part 1.} From top to bottom the spectra are formic acid (see \mainfigref[A]{mfig2}), methyl formate, diethyl ether (see \mainfigref[C]{mfig4}), ethyl acetate (see \mainfigref[D]{mfig4}), benzene (see \mainfigref[B]{mfig1} and \mainfigref[B]{mfig2}), methyl ethyl ketone (see \mainfigref[C]{mfig2}), propylene carbonate (see \mainfigref[D]{mfig2}), and \textit{n}-pentane (see \mainfigref[F]{mfig2}). Colored overbars and molecular insets follow the isotopomer-assignment scheme introduced in \mainfigref{mfig2} (see legend); acquisition and processing parameters are compiled in Figs.~\ref{fig:acq_params}, \ref{fig:proc_params}, and \ref{fig:phase_params}. The remaining library entries are shown in Fig.~\ref{fig:na_lib_full_2}.}
  \label{fig:na_lib_full_1}
\end{figure}

\begin{figure}
  \centering
  \includegraphics[scale=.70]{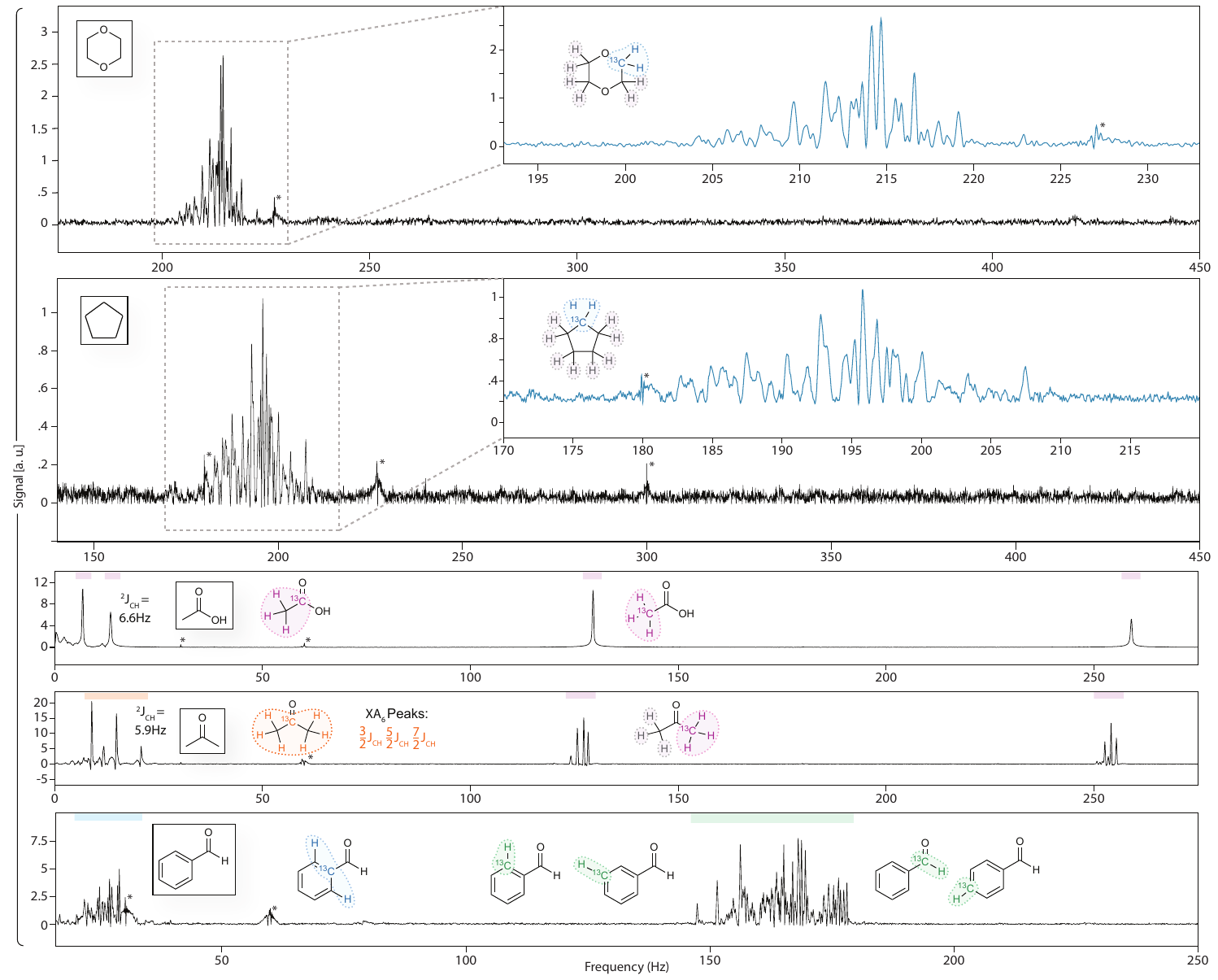}
  \caption{\textbf{Complete natural-abundance ZF NMR library: Part 2.} Continuation of the library in Fig.~\ref{fig:na_lib_full_1}. The top two rows show 1,4-dioxane and cyclopentane (see \mainfigref[E]{mfig2} and \mainfigref[G]{mfig2}) together with expanded insets that highlight spectral complexity. The bottom three rows show acetic acid (linked to the double-$^{13}$C analysis of \mainfigref[A]{mfig3}), acetone (see \mainfigref[B]{mfig4}, highlighting XA$_6$ topology), and benzaldehyde (see \mainfigref[A]{mfig1}), which contain quaternary carbons, where $^2J_{CH}$ is marked for acetic acid and acetone from simple, low frequency XA$_n$ patterns.}
  \label{fig:na_lib_full_2}
\end{figure}

This section expands the representative spectra in \mainfigref[A--G]{mfig2}, \mainfigref[A]{mfig3} and \mainfigref[A--D]{mfig4} into a complete natural-abundance ZF NMR library (Figs.~\ref{fig:na_lib_full_1} and \ref{fig:na_lib_full_2}), including every unique molecule discussed in the main text together with additional zoomed insets where helpful.

In addition, this section compiles (i) acquisition parameters for each spectrum in the library (Fig.~\ref{fig:acq_params}) and (ii) the display-spectrum processing workflow used to convert averaged time-domain data into the spectra shown in \mainfigref[A--B]{mfig1}, \mainfigref[A--G]{mfig2}, \mainfigref[A]{mfig3}, \mainfigref[A--D]{mfig4}, and \mainfigref[A,C]{mfig5}. A Jupyter notebook in the accompanying repository reproduces the processing used throughout this work. The ethyl-acetate example in Figs.~\ref{fig:proc_filter}--\ref{fig:proc_before_after} corresponds to the data compared with vacuum-DFT in \mainfigref[D]{mfig4}.

The experiments corresponding to \mainfigref[B(iii)]{mfig1} and \mainfigref[A,C]{mfig5} were acquired after the magnetic shield (Twinleaf MS2) was replaced with a Twinleaf MS1-LF. The dilute 0.5~M formate measurement in \mainfigref[C]{mfig5} employed $^{13}$C enrichment only as a time-saving measure; all other datasets were acquired at the true natural abundance of $^{13}$C.

\subsection{Acquisition parameters and representative experiment times}
\label{subsec:acq_params_ex}

\begin{figure}
  \centering
  \includegraphics[scale=.75]{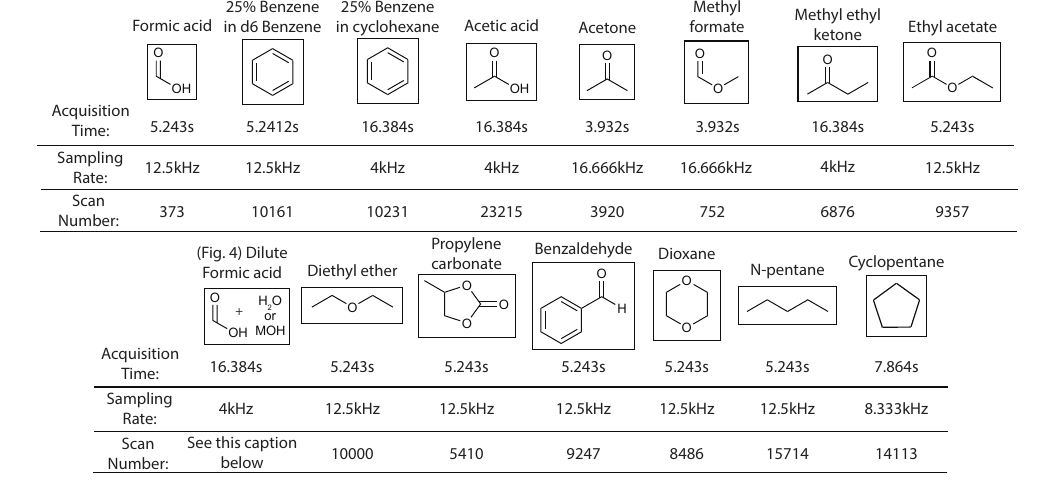}
  \caption{\textbf{Acquisition parameters for the NA ZF NMR library.} The figure is a single two-block table. In each analyte column, the rows list acquisition time per scan, sampling rate, and number of averaged scans, with the skeletal structure as the header. The upper block covers the library and DFT-comparison molecules used in \mainfigref[B]{mfig1}, \mainfigref[A--G]{mfig2}, \mainfigref[A]{mfig3}, and \mainfigref[A--D]{mfig4}; the lower block adds the remaining library entries together with the aqueous formic-acid/formate series of \mainfigref[A,C]{mfig5}. The scan counts used for \mainfigref[A]{mfig5} are $x_\mathrm{FA}$=1.00: 1979, 0.98: 5640, 0.80: 670, 0.58: 736, 0.30: 1342, 0.10: 2218, 0.01: 9852; for \mainfigref[C]{mfig5} they are 0.5~M ($^{13}$C enriched): 181, 5~M Li${^+}$: 3970, 5~M Na${^+}$: 4076, and 5~M K${^+}$: 4038.}
  \label{fig:acq_params}
\end{figure}

Each ZF NMR scan consists of pre-polarization in the fringe field of the 9.4~T magnet, adiabatic shuttling to the shield under a guiding field, a 90$^\circ$ $\Hs$ pulse, and time-domain acquisition with the optically pumped magnetometer (OPM), as summarized in Fig.~\ref{fig:apparatus} and used for the datasets in \mainfigrange{mfig1}{mfig5}.
For all measurements in this work, the digitizer records a fixed number of points ($N_\mathrm{pts}=65536$) at a sampling rate $f_s$, so that the
acquisition time per scan is
\begin{equation}
t_\mathrm{acq}=\frac{N_\mathrm{pts}}{f_s}.
\end{equation}
The sampling rates and acquisition times used for each analyte are summarized in Fig.~\ref{fig:acq_params}.

\paragraph{Example: total elapsed experiment time.}
A useful back-of-the-envelope estimate for the total wall-clock time is obtained by multiplying the number of scans by the per-scan cycle time,
\begin{equation}
t_\mathrm{tot}\approx N_\mathrm{scans}\,\Big(t_\mathrm{pol}+2\times t_\mathrm{shuttle}+t_\mathrm{acq}+t_\mathrm{dead}\Big),
\end{equation}
where $t_\mathrm{pol}$ is the pre-polarization time, $t_\mathrm{shuttle}$ is the shuttling time, and $t_\mathrm{dead}$ is the post-pulse dead time
before the OPM recovers.
Using representative values ($t_\mathrm{pol}\approx 17.5$~s, $t_\mathrm{shuttle}\approx 0.9$~s, $t_\mathrm{dead}\approx 30$~ms),
acetic acid ($N_\mathrm{scans}=23215$, $t_\mathrm{acq}=16.384$~s) corresponds to
\begin{equation}
t_\mathrm{tot}\approx 23215\times(17.5+2\times0.9+16.384+0.03)\mathrm{s}\approx 10\ \mathrm{days}.
\end{equation}
This averaging was performed continuously without spectral degradation~\cite{andrewsSensitiveMultichannelZeroto2025} and with near-ideal behavior, consistent with \mainfigref[B]{mfig3}. This scaling illustrates why long-term instrumental stability (\SIsecref{sec:sensitivity_budget}) is essential for natural-abundance ZF NMR.

\subsection{Display-spectrum processing workflow}
\label{subsec:proc_workflow}
We note that processing applied here is intended for the clearest visualization of experimental spectra for a human reader; in practice, measured data can be optimized against simulated data with $J$'s as adjustable parameters, which can produce accurate $J$'s at the $10^{-2}$--$10^{-4}$~Hz level ~\cite{wilzewskiMethodMeasurementSpinSpin2017a}. Such a scheme could be implemented directly in the time domain, eliminating the need for some processing steps, including Fourier transform.

All spectra shown in the main text are derived from scan-averaged time-domain data using the same processing steps:
\begin{enumerate}
  \item \textbf{Baseline removal / high-pass filtering.} A smooth estimate of the slowly varying baseline is constructed from the raw time trace
  (here, using a Savitzky--Golay polynomial smoothing procedure). Subtracting this baseline suppresses low-frequency components (typically 0--50~Hz)
  that arise from slow magnetometer drift and residual relaxation signals from the overwhelmingly abundant $\Cs$-less isotopomers.
  \item \textbf{Edge truncation.} The smoothed baseline estimate is imperfect near the boundaries of the time trace. A small number of points are removed
  from the beginning and end to avoid boundary artifacts. Points removed from beginning are replaced with null (mean-valued) points so as to not alter the true acquisition start time.
  \item \textbf{Zero-filling (optional).} The trimmed trace is extended by appending zeros to increase the apparent frequency resolution of the discrete
  Fourier transform (DFT). (Zero-filling interpolates the spectrum; it does not add information, but improves peak visualization and fitting.)
  \item \textbf{Apodization.} An exponential window $e^{-Rt}$ is applied to increase signal-to-noise ratio (SNR) at the
  expense of linewidth.
  \item \textbf{Fourier transform.} The frequency-domain spectrum is obtained by Fourier transforming the processed time trace. The real part of the spectrum is plotted and subsequently phased.
\end{enumerate}

\begin{figure}[h!]
  \centering
  \includegraphics[scale=.25]{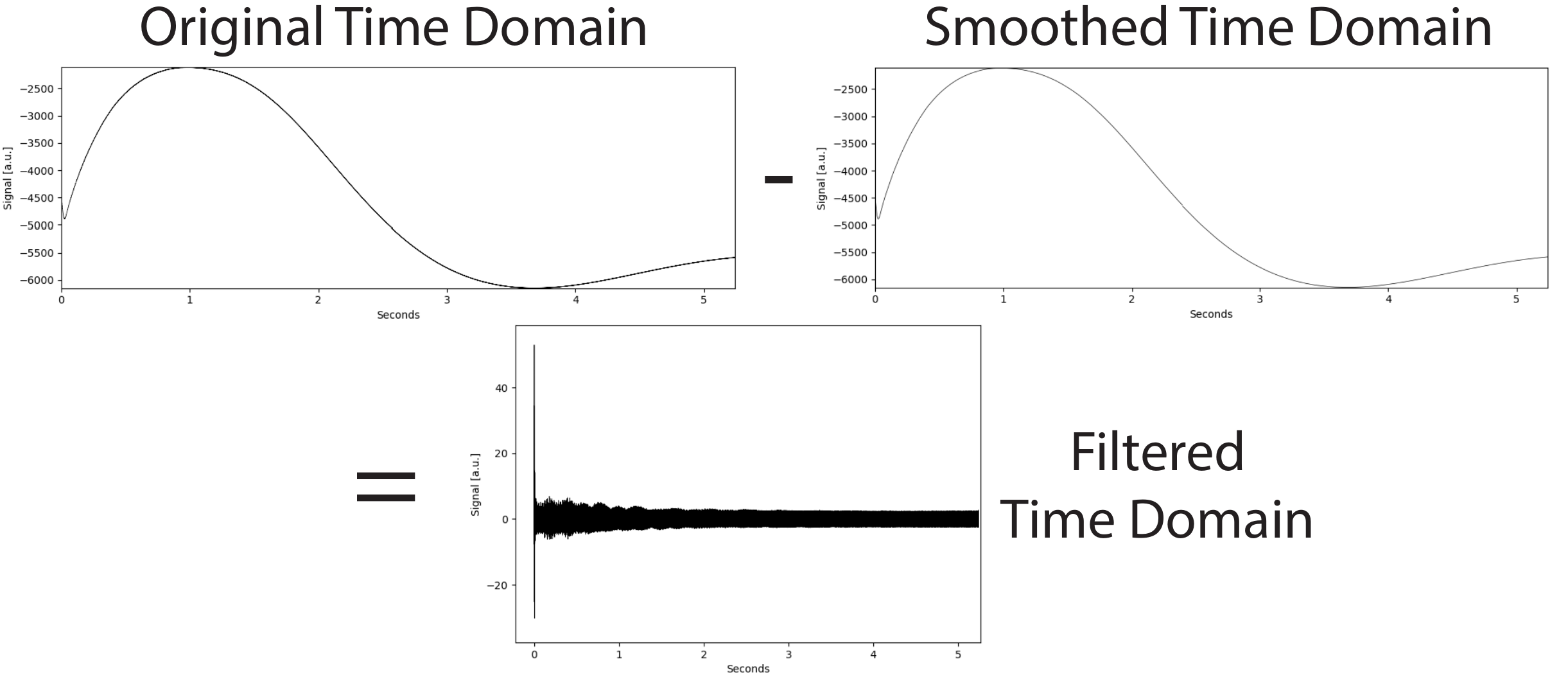}
  \caption{\textbf{Baseline removal by smoothing and subtraction (ethyl acetate example).} Top left: original scan-averaged time-domain signal for the ethyl acetate dataset shown in \mainfigref[D]{mfig4}. Top right: Savitzky--Golay smoothed baseline estimate. Bottom: filtered trace obtained by subtracting the smoothed baseline from the original data, removing the slow drift and low-frequency background before trimming, apodization, and Fourier transformation. The same workflow is used for the processed spectra in \mainfigrange{mfig1}{mfig5}.}
  \label{fig:proc_filter}
\end{figure}

\begin{figure}[h!]
  \centering
  \includegraphics[scale=.175]{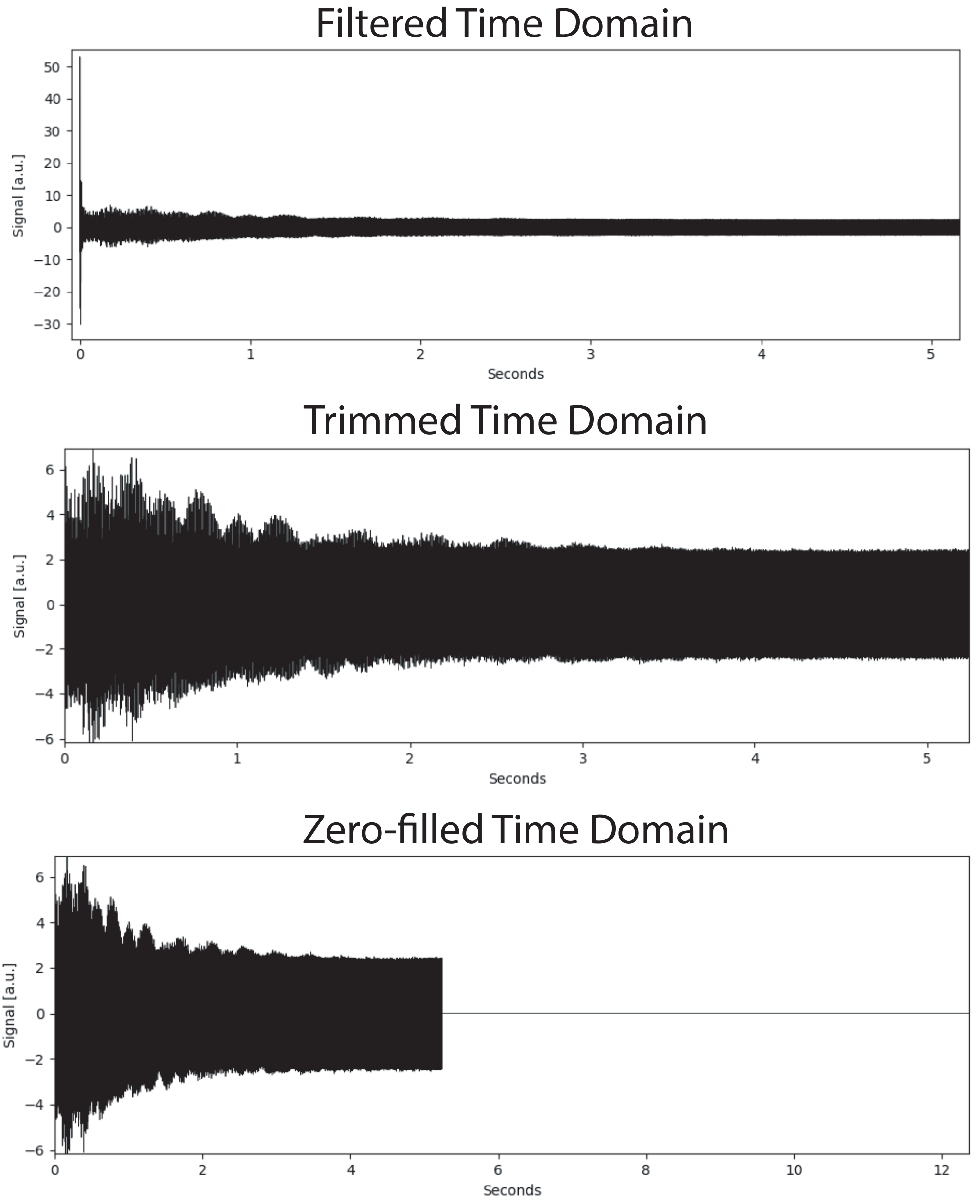}
  \caption{\textbf{Edge trimming and zero-filling (ethyl acetate example).} Top: filtered time-domain trace after baseline subtraction (output of Fig.~\ref{fig:proc_filter}). Middle: edge-trimmed trace after removing boundary regions where the smoothing procedure is least reliable. Bottom: zero-filled trace used for display, which interpolates the discrete Fourier spectrum without adding information. These steps feed into the final spectrum analyzed for ethyl acetate in \mainfigref[D]{mfig4} and, more generally, the spectra in \mainfigrange{mfig1}{mfig5}.}
  \label{fig:proc_trim}
\end{figure}

\begin{figure}[h!]
  \centering
  \includegraphics[scale=.175]{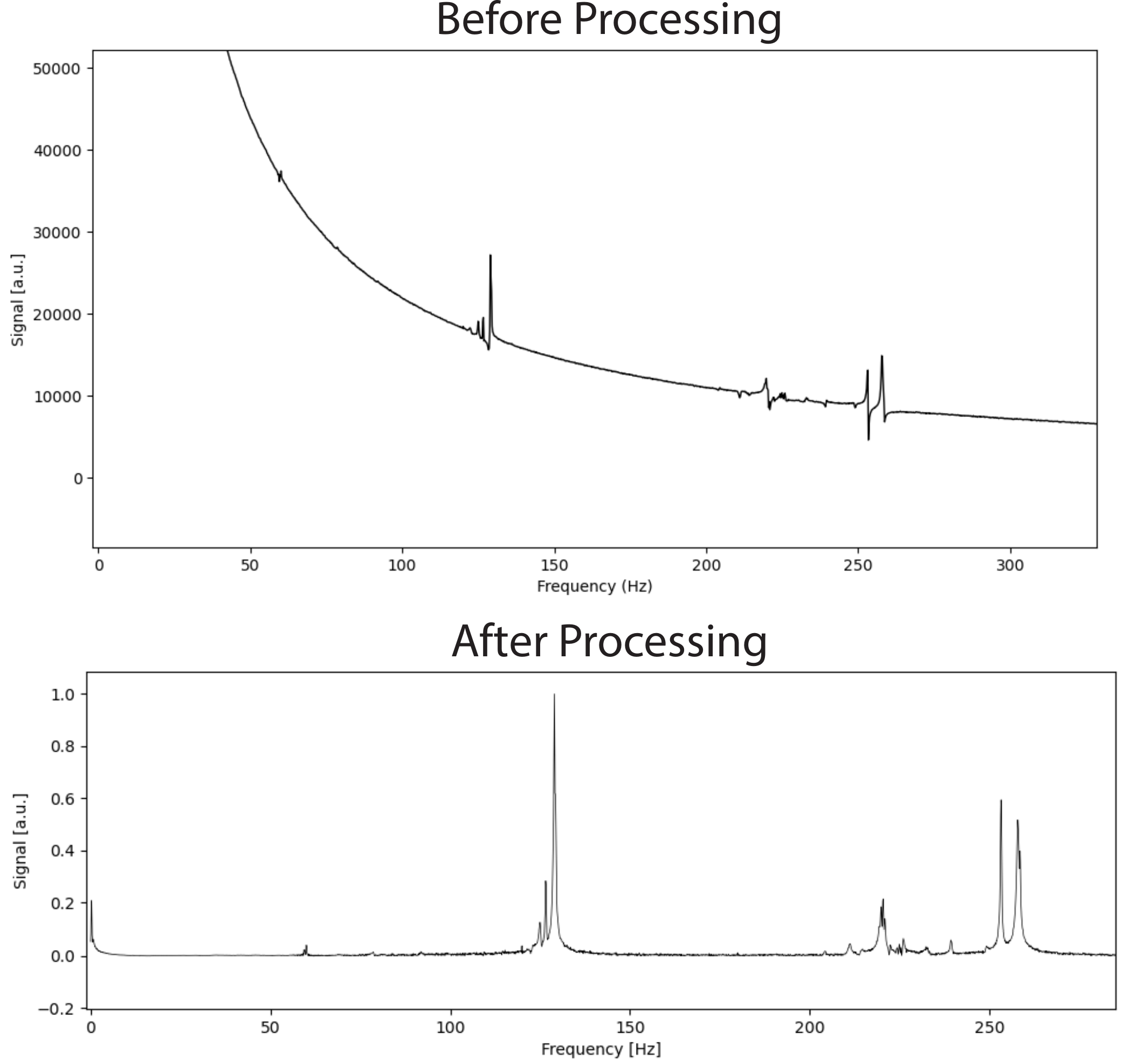}
  \caption{\textbf{Frequency-domain spectrum before and after processing (ethyl acetate).} Top: magnitude (unphased) spectrum formed directly from the raw time-domain data, dominated by a rolling baseline and low-frequency artifacts. Bottom: spectrum after the full processing pipeline of Figs.~\ref{fig:proc_filter} and \ref{fig:proc_trim}, revealing the ZF multiplets compared with vacuum-DFT in \mainfigref[D]{mfig4}. The spectrum shown here is magnitude-only (unphased).}
  \label{fig:proc_before_after}
\end{figure}

\begin{figure}[h!]
  \centering
  \includegraphics[scale=.75]{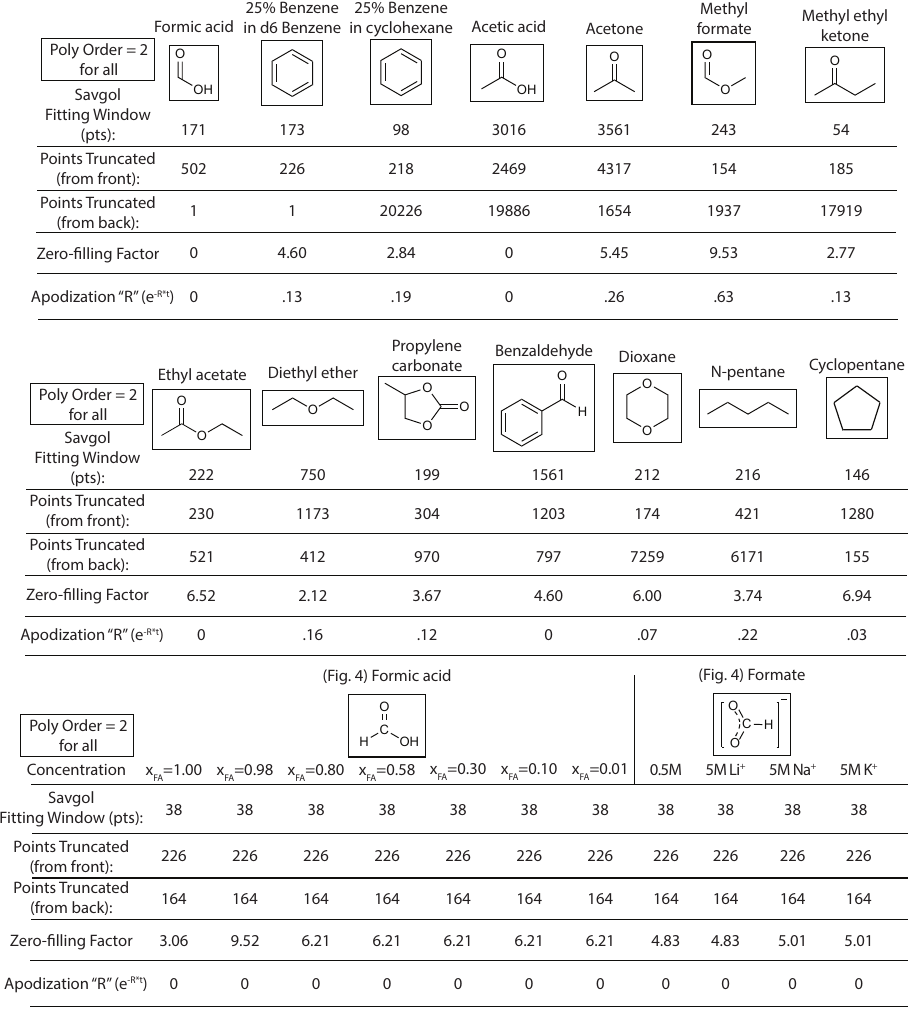}
  \caption{\textbf{Time-domain processing parameters for the NA ZF NMR library.} The upper 2 tables lists parameters for the library of molecules in \mainfigref[B]{mfig1}, \mainfigref[A--G]{mfig2}, \mainfigref[A]{mfig3}, and \mainfigref[A--D]{mfig4}; the lower tables list parameters for the dilution series of \mainfigref[A]{mfig5} and the formate series of \mainfigref[C]{mfig5}. Within each analyte column the entries report the Savitzky--Golay fitting window, the numbers of points truncated from the front and back of the trace, the zero-fill factor, and the exponential apodization rate. The Savitzky--Golay polynomial order is 2 throughout. Together with Figs.~\ref{fig:acq_params} and \ref{fig:phase_params}, these values reproduce the display spectra in the main text. These parameters can be pasted directly into the repository code to reproduce all spectra in the main text.}
  \label{fig:proc_params}
\end{figure}

Figs.~\ref{fig:proc_filter}--\ref{fig:proc_before_after} show representative intermediate steps for ethyl acetate (the spectrum compared with vacuum-DFT in \mainfigref[D]{mfig4}).
Figure~\ref{fig:proc_params} lists the processing parameters (trimming window, truncation points, zero-fill factor, and apodization rate) used for each analyte.

\begin{figure}[h!]
  \centering
  \includegraphics[scale=.75]{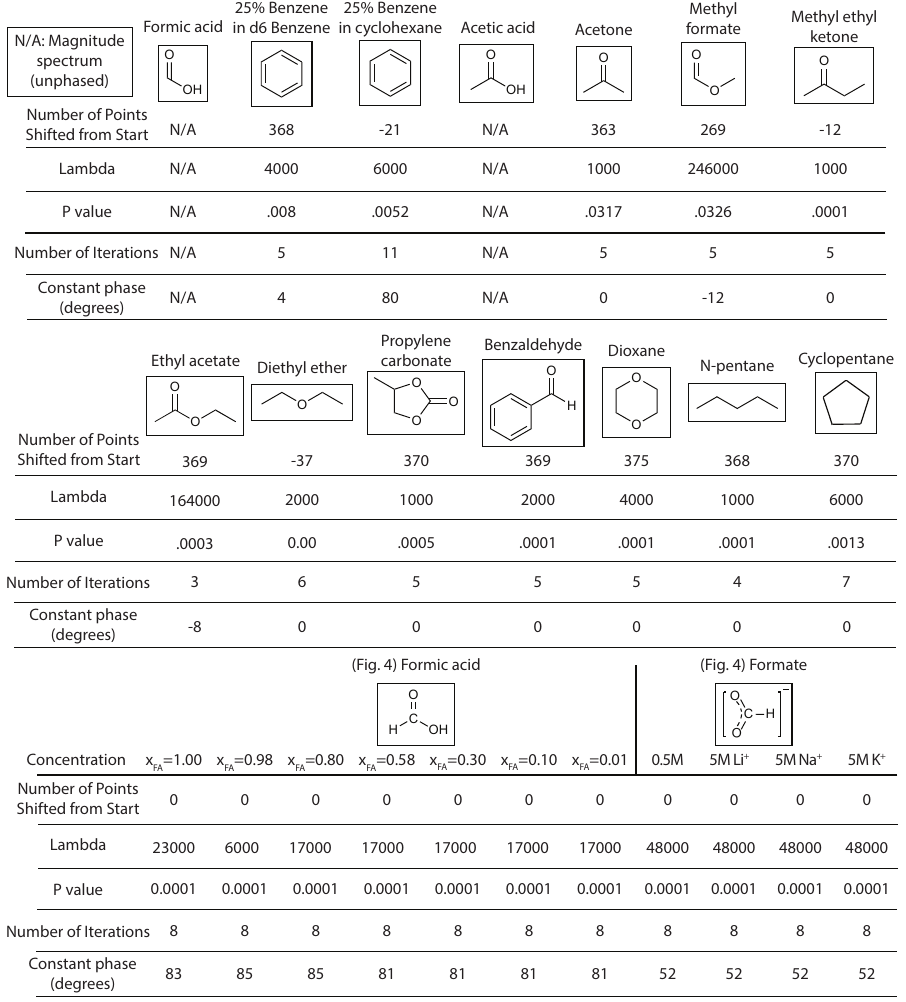}
  \caption{\textbf{Phasing and baseline-correction parameters for the NA ZF NMR library.} The layout matches Fig.~\ref{fig:proc_params}: The upper 2 tables lists parameters for the library of molecules in \mainfigref[B]{mfig1}, \mainfigref[A--G]{mfig2}, \mainfigref[A]{mfig3}, and \mainfigref[A--D]{mfig4}, and the lower table cover the aqueous series in \mainfigref[A,C]{mfig5}. For each analyte the entries give the number of points shifted from the start of the trace (first-order phase adjustment), the asymmetric least-squares parameters $\lambda$ and $p$, the number of iterations, and the constant phase in degrees. N/A denotes spectra displayed as magnitudes without phasing. Together with Figs.~\ref{fig:acq_params} and \ref{fig:proc_params}, these parameters reproduce the main-text spectra, which can be performed with the respository code.}
  \label{fig:phase_params}
\end{figure}

Finally, zeroth- and first-order phase corrections, followed by a baseline correction, are applied in the frequency domain. The zeroth-order correction is implemented by multiplying the spectrum by a constant phase factor $e^{i\phi}$. First-order phase correction is achieved by introducing an effective acquisition delay, implemented by adding or removing (null) points at the beginning of the time-domain signal (denoted "Number of Points Shifted from Start"). Finally, an asymmetric least-squares baseline correction is applied, using the parameters listed in Fig.~\ref{fig:phase_params}. The corresponding function for assymstric least-squares correction is defined separately and implemented in the accompanying jupyter notebook. Likewise, parameters in ~\ref{fig:acq_params} and ~\ref{fig:proc_params} can be inputted directly into repository code to reproduce spectra in the main text.

\section{Quantum-chemical workflow for $J$-coupling prediction}
\label{sec:qc_j}

\subsection{Vibrational corrections to computed $J$ couplings}
\label{subsec:vib_corr}
The main text establishes an \textit{ab initio} workflow for predicting ZF spectra from quantum-chemically calculated $J$ couplings (\mainfigref[A--D]{mfig4} and \mainfigref[B--D]{mfig5}). The physical basis for this workflow is a hierarchy of timescales: electronic motion is fastest, nuclear vibration is slower, and nuclear spin evolution is slower still. In the spirit of the Born-Oppenheimer approximation, the effective ZF nuclear-spin Hamiltonian can be written as follows,

\begin{equation}
    \hat{H}_{\mathrm{ZF}} = \sum_{K,L}^{\mathrm{active}} \langle \mathbf{J}^{K,L} \rangle \, \hat{\mathbf{I}}_{K} \cdot \hat{\mathbf{I}}_{L}
\end{equation}

Here the two nuclear-spin indices $K, L$ of the double summation need to satisfy the total spin angular momentum selection rule (as demonstrated in the System and Principle section in the main text). The $\langle \mathbf{J}^{K,L} \rangle$ matrix encodes electronic structure and nuclear vibrational information indirectly.
\begin{equation}
\begin{aligned}
    \langle \mathbf{J}^{K,L} \rangle
    & = \bra{\Phi_{total}(\boldsymbol{r, R})} \hat{J}^{K, L}(\boldsymbol{r}, \boldsymbol{R}) \ket{\Phi_{total}(\boldsymbol{r, R})} \\
    & = \bra{\Psi_{e}(\boldsymbol{r; R}) \Psi_{n}(\boldsymbol{R})} \hat{J}^{K, L}(\boldsymbol{r}, \boldsymbol{R}) \ket{\Psi_{e}(\boldsymbol{r; R}) \Psi_{n}(\boldsymbol{R})} \\
    & = \boldsymbol{J_{\mathrm{elec}}^{K, L}} + \boldsymbol{J_{\mathrm{vib}}^{K, L}}
\end{aligned}
\end{equation}

Using a second-order vibrational perturbation theory (VPT2)~\cite{franke2021vpt2} expansion of the nuclear vibrational wavefunction, one can obtain the following working equations to calculate the electronic ($J_{\mathrm{elec}}$) and vibrational ($J_{\mathrm{vib}}$) contributions to J coupling~\cite{ruden2003vibrational}. For the electronic contribution $J_{\mathrm{elec}}$: 
\begin{equation}\label{eq:J_elec}
\begin{aligned}
     \boldsymbol{J_{\mathrm{elec}}^{K, L}}
     & = h \frac{ \gamma_{K} }{2 \pi} \frac{ \gamma_{L}  }{2 \pi} \cdot \frac{ d^2 E_{\mathrm{elec}}\left( \textbf{m}_{K}, \textbf{m}_{L} \right) }{ d \boldsymbol{m}_{K} d \boldsymbol{m}_{L} }\bigg|_{\textbf{m}_{K}=\textbf{0}, \textbf{m}_{L}=\textbf{0}} ~~~ 
\end{aligned}
\end{equation}

It is calculated as a second-order response property of the electronic energy's derivative with respect to the 2 nuclear magnetic moments on centers K and L ($\textbf{m}_{K}$ and $\textbf{m}_{L}$), and the prefactor includes product of the gyromagnetic ratio of the two nuclei ($\gamma_{K}$ and $\gamma_{L}$). We will elaborate more about $J_{\mathrm{elec}}$ in the following subsection. For the nuclear vibrational contribution $J_{\mathrm{vib}}$:

\begin{equation}\label{eq:J_vib}
\begin{aligned}
     \boldsymbol{J_{\mathrm{vib}}^{K, L}} & = -\frac{1}{4} \sum_{M}^{ \substack{\text{normal} \\ \text{mode}} } \frac{1}{\omega^2_{M}}  \frac{\partial \boldsymbol{J_{\mathrm{elec}}^{K, L}}}{\partial \textbf{Q}_{M}} \sum_{N}^{ \substack{\text{normal} \\ \text{mode}} } \frac{F_{MNN}}{\omega_{N}}
     + \frac{1}{4} \sum_{M}^{ \substack{\text{normal} \\ \text{mode}} } \frac{1}{\omega^2_{M}}  \frac{\partial^2 \boldsymbol{J_{\mathrm{elec}}^{K, L}}}{\partial \textbf{Q}_{M}^2}
\end{aligned}
\end{equation}

\begin{equation}
    F_{MNN} = \frac{\partial^3 E_{\mathrm{elec}} ( \{ \boldsymbol{Q} \}) }{\partial \textbf{Q}_{M} \partial^2 \textbf{Q}_{N}} \bigg|_{\textbf{Q}_{M}=\textbf{0}, \textbf{Q}_{N}=\textbf{0}} ~~~ 
\end{equation}

Here $\textbf{Q}_{M}$ indicates the normal mode, $\omega_{M}$ indicates the corresponding harmonic vibrational frequency, and $F_{MNN}$ is the semi-diagonal elements of the cubic force constant. In equation \ref{eq:J_vib}, the first term corresponds to anharmonic zero-point vibrational contribution, and the second term corresponds to harmonic zero-point vibrational contribution. Note that we need to calculate the $J_{\mathrm{vib}}$ at the optimized minimum geometry without any imaginary frequency, to ensure the validity of the harmonic approximation in the VPT2 expansion~\cite{ruden2003vibrational}.

In routine NMR property calculations, however, vibrational corrections are often omitted due to the high computational cost of the cubic force constant. For relatively strong coupling interactions (i.e. $^{1}J_{\mathrm{CH}}$, usually above 100 Hz), $J_{\mathrm{vib}}$ is typically much smaller than $J_{\mathrm{elec}}$. However, for relatively weak coupling interactions (multiple bond coupling or through-space coupling, ~10 Hz or even lower), this rule-of-thumb may not hold anymore. As established in previous work, incorporating vibrational contributions can substantially improve the agreement between quantum chemical predictions of J coupling and experiment~\cite{helgakerQuantumChemicalCalculationNMR2008}, even for the relatively strong $^{1}J_{\mathrm{CH}}$. To bring the model closer to the underlying physical system with fortuitous error cancellation removed to the degree possible, we explicitly include vibrational correction in all of the J calculations. The computational details are summarized in the computational protocol below.

\subsection{Electron-density analysis of the $J$-shift observable}
\label{subsec:e_density_diff}
With the second order response formalism (equation \ref{eq:J_elec}), the non-relativistic molecular Hamiltonian $\hat{H}_{\mathrm{e}}$ to calculate $J_{\mathrm{elec}}$ writes as follows,
\begin{equation}
\begin{aligned}
    \hat{H}_{\mathrm{e}}(\boldsymbol{r; R}) = & \ -\frac{1}{2} \sum_{i} \nabla^2_i - \sum_{i} \sum_{K} \frac{Z_K}{r_{iK}} + \sum_{ij} \frac{1}{r_{ij}} + \alpha^2 \sum_{i} \sum_{K} \frac{\boldsymbol{m_{K}} \cdot \boldsymbol{L_{iK}}}{r_{iK}^3} \\
    & + \frac{\alpha^4}{2} \sum_{i} \sum_{KL} \frac{ (\boldsymbol{m_{K}} \cdot \boldsymbol{m_{L}}) (\boldsymbol{r_{iK}} \cdot \boldsymbol{r_{iL}}) - (\boldsymbol{m_{K}} \cdot \boldsymbol{r_{iL}})(\boldsymbol{r_{iK}} \cdot \boldsymbol{m_{L}}) }{r_{iK}^3 r_{iL}^3} \\
    & + \frac{8 \pi \alpha^2}{3} \sum_i \sum_K \delta(r_{iK}) \cdot \boldsymbol{s_{i}} \cdot \boldsymbol{m_{K}} + \alpha^2 \sum_i \sum_{K} \frac{ 3 (\boldsymbol{s}_{i} \cdot \boldsymbol{r_{iK}}) (\boldsymbol{m_{K}} \cdot \boldsymbol{r_{iK}})  - r_{iK}^2 \cdot \boldsymbol{s_{i}} \cdot \boldsymbol{m_{K}} }{ r_{iK}^5 } \\
\end{aligned}
\end{equation}

In most common cases, the dominant contribution to $J_{\mathrm{elec}}$ is the Fermi Contact (FC) interaction. It is very sensitive to changes in electron density in the vicinity of the nuclear center, as reflected in the $\hat{h}^{FC}_{K}$ operator for contributions from nuclear center K,

\begin{equation}
\begin{aligned}
    \hat{h}^{FC}_{K} = \frac{8 \pi \alpha^2}{3} \sum_i \delta(r_{iK}) \cdot \boldsymbol{s_{i}} \cdot \boldsymbol{m_{K}}
\end{aligned}
\end{equation}

Here $\alpha$ is the fine-structure constant, $\boldsymbol{s}_i$ is the electronic-spin operator, and $\boldsymbol{m}_K$ is the nuclear magnetic moment on center $K$. Because of the $\delta(r_{iK})$ dependence, the FC interaction is non-zero only when the position of electron $i$ coincides with nucleus $K$ ($r_{iK}=0$), making $J$ an especially sensitive probe of local electronic structure. The electron-density-difference plots associated with \mainfigref[B]{mfig5} therefore help interpret the $J$ shifts induced by changes in electron density near the formyl C--H bond.

\begin{figure}[ht!]
    \centering
    \includegraphics[width=0.85\linewidth]{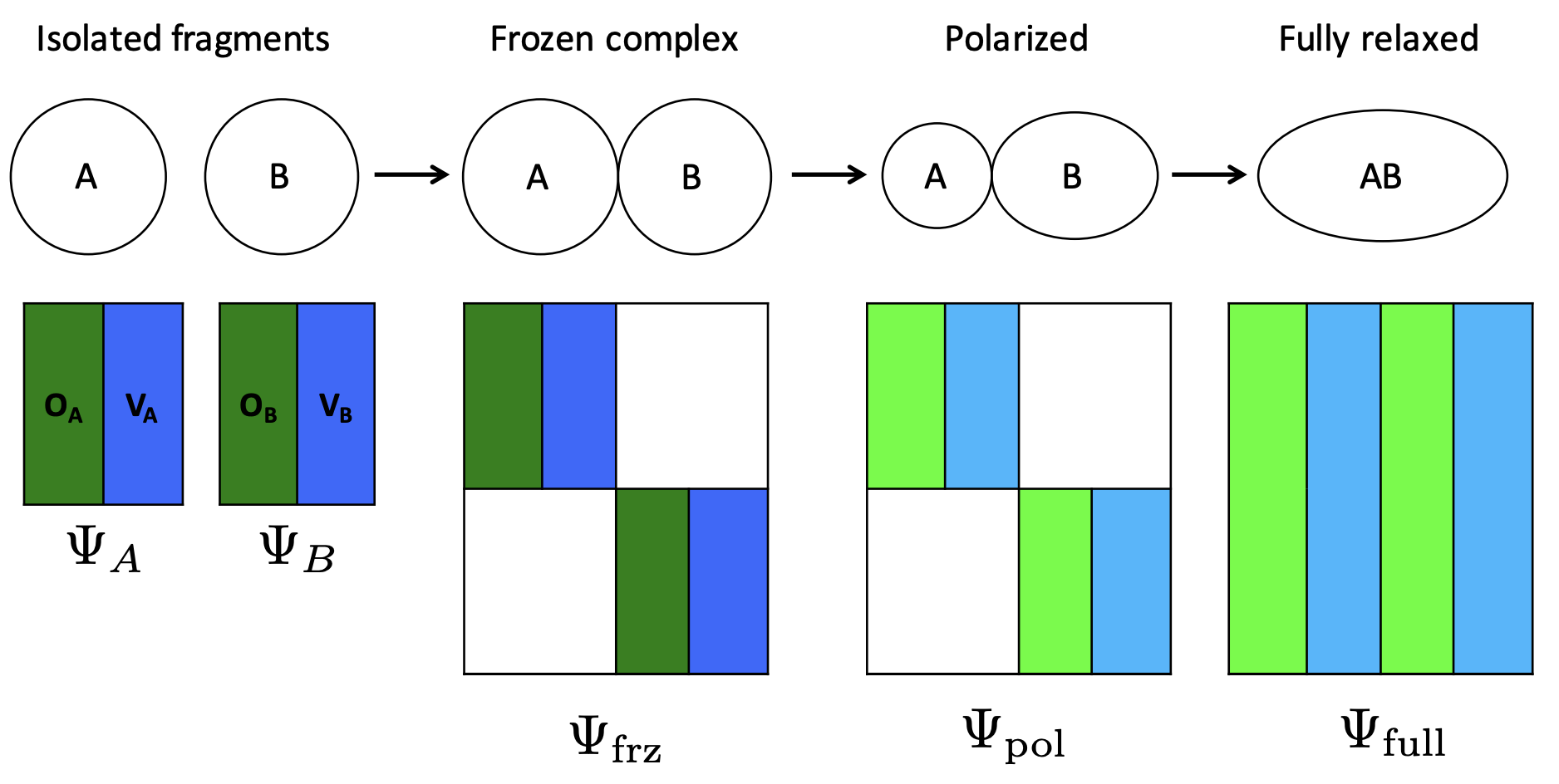}
    \caption{\textbf{ALMO-EDA state definitions used for the $J$-shift analysis.} Adapted from Yuezhi Mao's PhD thesis~\cite{mao2017advances} with permission. Left to right, the schematic shows isolated fragments A and B, the frozen complex $\Psi_\mathrm{frz}$ formed from their direct sum, the polarized state $\Psi_\mathrm{pol}$ obtained by intrafragment relaxation only, and the fully relaxed state $\Psi_\mathrm{full}$ after interfragment mixing is allowed. Green and blue blocks denote occupied (O) and virtual (V) orbital subspaces of each fragment in the molecular-orbital density-matrix representation. The $\Psi_\mathrm{frz}\!\rightarrow\!\Psi_\mathrm{pol}$ density difference corresponds to polarization, whereas $\Psi_\mathrm{pol}\!\rightarrow\!\Psi_\mathrm{full}$ corresponds to charge transfer. The red/blue meshes in \mainfigref[B]{mfig5} correspond to $\Psi_\mathrm{frz}\!\rightarrow\!\Psi_\mathrm{full}$, representing the overall interactions between fragments A and B.}
    \label{fig:ALMO}
\end{figure}

Fig.~\ref{fig:ALMO} schematizes the intermediate states used to construct the electron density difference plots within the ALMO-EDA framework~\cite{horn2016probing}. First, two isolated fragments A and B are calculated separately to obtain their wavefunctions $\Psi_{A}$ and $\Psi_{B}$. The wavefunction of the supersystem frozen state $\Psi_{frz}$ is then formed as a direct sum of fragmental wavefunctions without any interactions, and the electron density of the $\Psi_{\mathrm{frz}}$ state is also obtained from a direct sum of $\Psi_{A}$ and $\Psi_{B}$. For the polarized state $\Psi_{\mathrm{pol}}$, the wavefunction is allowed to relax only within the orbital subspaces of the two fragments. This constraint keeps $\Psi_{\mathrm{pol}}$ block-diagonal, and the electron number of each fragment remains the same as in isolation. The physical interaction that gives rise to the electron-density difference between $\Psi_{\mathrm{frz}}$ and $\Psi_{\mathrm{pol}}$ states is then attributed to polarization. When the block-diagonal constraint is lifted, the total system relaxes fully to $\Psi_{\mathrm{full}}$. Note that $\Psi_{\mathrm{full}}$ is the same full system SCF solution when we do not impose any constraint. The difference between the $\Psi_{\mathrm{pol}}$ and $\Psi_{\mathrm{full}}$ states then reflects charge transfer across the fragments, corresponding to off-diagonal couplings between fragments A and B.

\begin{figure}[ht!]
    \centering
    \subfloat[Electron density difference between $\Psi_{frz}$ and $\Psi_{pol}$ states\label{fig:FA1_EDD_pol}]{%
        \includegraphics[width=0.3\linewidth]{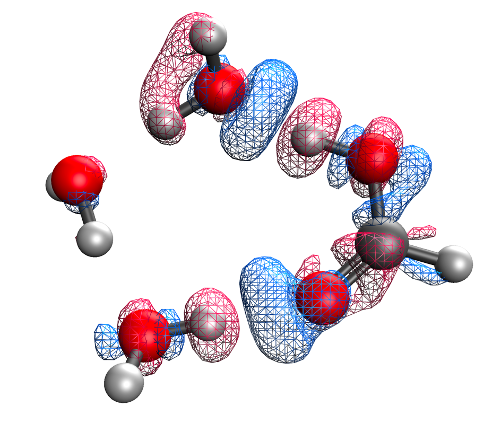}}
    \hfill
    \subfloat[Electron density difference between $\Psi_{pol}$ and $\Psi_{full}$ states\label{fig:FA1_EDD_ct}]{%
        \includegraphics[width=0.3\linewidth]{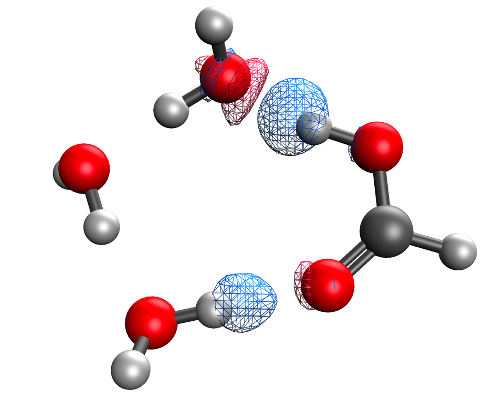}}
    \hfill
    \subfloat[Electron density difference between $\Psi_{frz}$ and $\Psi_{full}$ states\label{fig:FA1_EDD_total}]{%
    \includegraphics[width=0.3\linewidth]{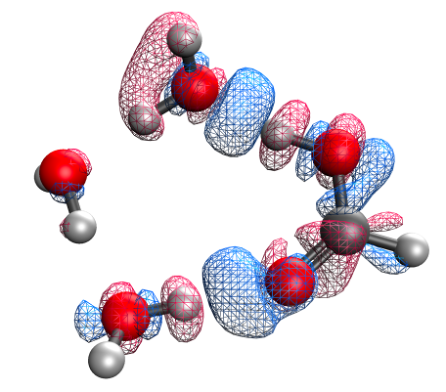}}
    \caption{\textbf{Hydrated formic-acid density differences for the ALMO-EDA analysis.} Structure of the FA1 W3 cluster, corresponding to the intermediate-solvation motif used in \mainfigref[B(ii)]{mfig5}. (a) Electron-density difference between $\Psi_{frz}$ and $\Psi_{pol}$. (b) Electron-density difference between $\Psi_{pol}$ and $\Psi_{full}$. (c) Electron-density difference between $\Psi_{frz}$ and $\Psi_{full}$. Blue mesh denotes density gain and red mesh density loss at the common isovalue used throughout \SIsecref{subsec:e_density_diff}.}
    \label{fig:FA1_EDD}
\end{figure}

Electron-density-difference plots comparing the frozen state $\Psi_{\mathrm{frz}}$ and the fully interacting state $\Psi_{\mathrm{full}}$ states of hydrated formic acid and formate in their global minimum structures are shown in support of \mainfigref[B]{mfig5}. The fragments are chosen as the isolated neutral formic acid molecule or formate anion (fragment A) and the water cluster (fragment B). Blue mesh denotes net electron-density gain and red mesh net electron-density loss. These ALMO-EDA calculations were carried out at the $\omega$B97X-D3/def2-TZVPD level in Q-Chem 6.3, using the same isovalue (0.002 a.u.) for all systems.

\begin{figure}[ht!]
    \centering
    \subfloat[Electron density difference between $\Psi_{frz}$ and $\Psi_{pol}$ states\label{fig:FA-_EDD_pol}]{%
        \includegraphics[width=0.3\linewidth]{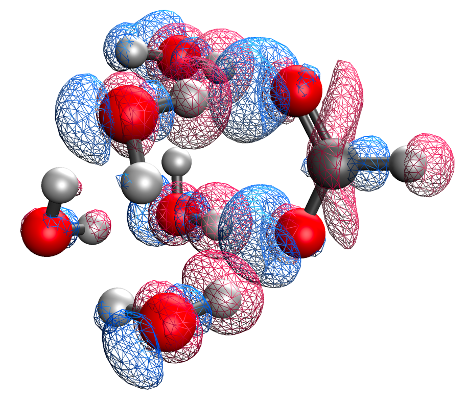}}
    \hfill
    \subfloat[Electron density difference between $\Psi_{pol}$ and $\Psi_{full}$ states\label{fig:FA-_EDD_ct}]{%
        \includegraphics[width=0.3\linewidth]{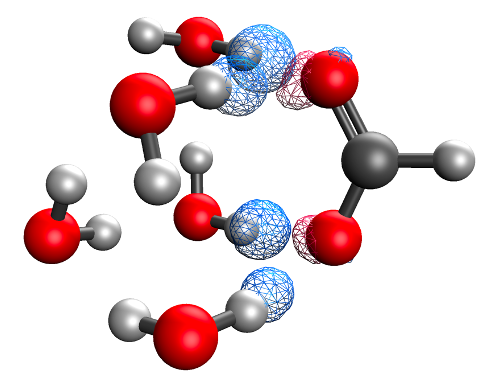}}
    \hfill
    \subfloat[Electron density difference between $\Psi_{frz}$ and $\Psi_{full}$ states\label{fig:FA-_EDD_total}]{%
        \includegraphics[width=0.3\linewidth]{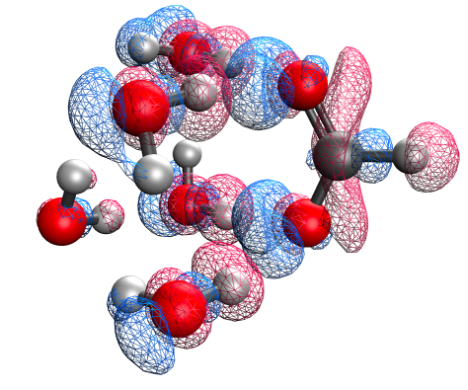}}
    \caption{\textbf{Hydrated formate density differences for the ALMO-EDA analysis.} Structure of the FA$^{-}$ W5 cluster used to interpret the hydrated-formate side of the formic-acid/formate trends in \mainfigref[C--D]{mfig5}. (a) Electron-density difference between $\Psi_{frz}$ and $\Psi_{pol}$. (b) Electron-density difference between $\Psi_{pol}$ and $\Psi_{full}$. Blue mesh denotes density gain and red mesh density loss at the common isovalue used throughout \SIsecref{subsec:e_density_diff}.}
    \label{fig:FA-_EDD}
\end{figure}

Figs.~\ref{fig:FA1_EDD} and \ref{fig:FA-_EDD} show the representative breakdown of the total electron density change due to polarization versus charge transfer effects upon the formation of hydrogen bonds, in hydrated formic acid and formate anion systems respectively. Polarization-driven electron density change (Figs.~\ref{fig:FA1_EDD_pol} and \ref{fig:FA-_EDD_pol}) is more delocalized across the carboxylic $\pi$ system. The charge transfer interactions mostly locate on the hydrogen bond donor and acceptor atoms (Figs.~\ref{fig:FA1_EDD_ct} and \ref{fig:FA-_EDD_ct}), and the charge-transfer-driven density change around the formyl C--H bond is negligible. The overall electron density change upon the formation of hydrogen bonds (Figs.~\ref{fig:FA1_EDD_total} and \ref{fig:FA-_EDD_total}) is very close to the polarization-driven change (Figs.~\ref{fig:FA1_EDD_pol} and \ref{fig:FA-_EDD_pol}). Therefore, the experimentally observed shift in the formyl $^{1} J_{\mathrm{CH}}$ can be mainly attributed to polarization interactions between the solute (formic acid or formate anion) and water solvents. The electron density change in hydrated formate anion (Fig.~\ref{fig:FA-_EDD_total}) is more pronounced along the formyl C--H bond, leading to a much larger shift in formyl $^{1} J_{\mathrm{CH}}$ compared to the neutral formic acid (Fig.~\ref{fig:FA1_EDD_total}).

\subsection{Computational protocol}
\label{subsec:d_comp_workflow}
Unless otherwise stated, all calculations listed here are done using ORCA 6.0.1 with RIJCOSX approximation, where the auxiliary basis set is automatically determined on-the-fly.

For the calculations of isolated molecule in vacuum state, we design the DFT computational workflow as follows:

1) Geometry optimization and subsequent vibrational frequency calculation to ensure minimum structure, at PBE0-D3 /aug-cc-pCVTZ level. For molecules with multiple relevant conformers, an initial conformation search is done with the GOAT module in ORCA using GFN2-xtb. The geometry optimization and vibrational frequency calculations are then done for each conformer separately. All final optimized structures do not have imaginary frequency.

2) Calculating the electronic contribution of J coupling at PBE0-D3/pcJ-3 level using the optimized minimum geometry.

3) Calculating the nuclear vibrational contribution of J coupling using the optimized minimum geometry (cubic force constant at PBE0-D3/aug-cc-pCVTZ level calculated as numerical first derivatives of the harmonic frequency, and $J_{\mathrm{elec}}$'s partial derivative with respect to normal modes also calculated numerically, at PBE0-D3/pcJ-3 level).

4) For the Boltzmann weights of multiple conformers, the electronic energy is refined at the frozen core RHF:RCCSD(T)-F12 /cc-pVDZ-F12 level, with resolution of the identity (RI) approximation applied only for the CCSD(T)-F12 part with cc-pVDZ-F12-CABS near-complete auxiliary basis set \cite{valeev2004improving} and cc-pVTZ auxiliary basis set \cite{weigend2002efficient} for correlation. The final Boltzmann factor at 298K is calculated with CCSD(T)-F12 electronic energy and the PBE0-D3/aug-cc-pCVTZ level zero point vibrational energy. The final $J$ value is a Boltzmann average. 

In the code/data repository (link/DOI to be added at submission), we provide for each conformation of each molecule:
(i) optimized geometries, (ii) the electronic contribution to $J$ couplings, (iii) the nuclear vibrational contribution to $J$ couplings, (iv) the resulting final $J$ couplings, and (v) the predicted ZF spectra from the computed $J$ couplings, together with side-by-side comparisons to experiment.

For the neutral formic acid systems, we follow the aforementioned workflow with the initial micro-solvated structures from reference~\cite{li2024hydrated}, where they optimized each conformation at $\omega$B97X-D/aug-cc-pVDZ after xtb-based conformation search. 

For the hydrated formate anion and hydrated contact ion pair systems, we perform additional global minimum search with the GOAT module in ORCA using GFN2-xtb, with starting micro-solvated structures generated by the SOLVATOR module in ORCA. The conductor-like polarizable continuum model (C-PCM) is used in the $J_{\mathrm{elec}}$ calculation. As ORCA does not have built-in pcJ-3 definitions for Li and K atoms, in the $J_{\mathrm{elec}}$ calculations, the basis set of Li and K atoms are replaced by pc-3. For hydrated HCOOK contact ion pair systems, during the geometry optimization, vibrational frequency, and VPT2 cubic force constant calculation steps, the basis set on K is replaced by aug-pc-2 because aug-cc-pCVTZ does not have a definition on K. Further analysis about the models of hydrated formate anion and hydrated contact ion pair systems will be discussed in the following section.

\subsection{Ion pairing and cation effects in concentrated formate solutions}
\label{subsec:conc_formate_cip}

\begin{figure}[ht!]
    \centering
    \includegraphics[width=0.95\linewidth]{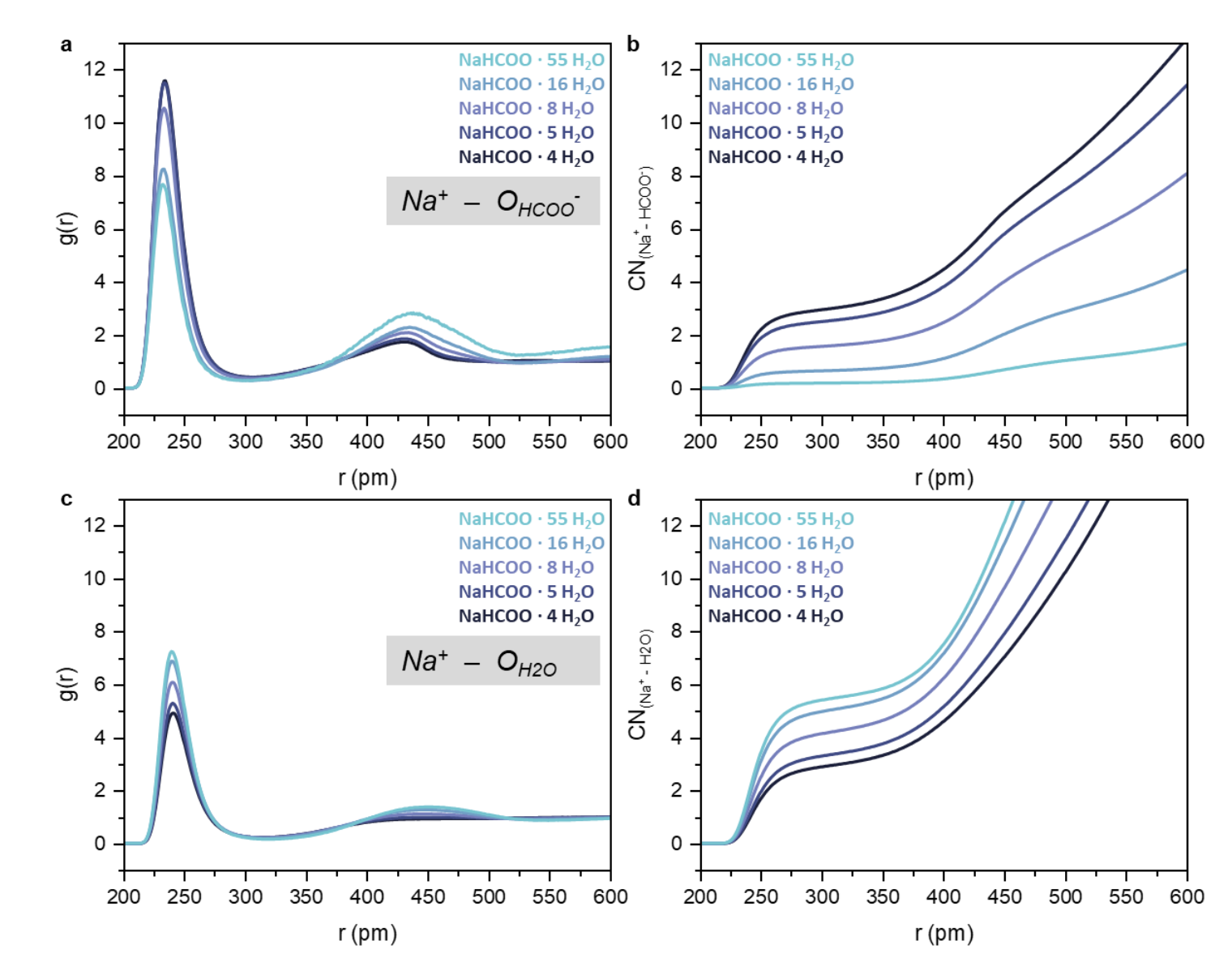}
    \caption{\textbf{Literature MD evidence for ion pairing in concentrated sodium formate.} Adapted from Fig.~S2 of Trapp et al. under CC BY 4.0 license~\cite{trapp2025electrolyte}. (a) Radial distribution function $g(r)$ between Na$^+$ and the formate oxygen, showing the short-range contact-ion-pair feature discussed in connection with \mainfigref[C]{mfig5}. (b) Corresponding coordination number, which grows with concentration and approaches $\sim1$ by \ce{NaHCOO}$\cdot$8\ce{H2O} (7~M). (c) Radial distribution function between Na$^+$ and water oxygen. (d) Corresponding coordination number. Concentrations are labeled by the stoichiometric ratio of solute to water: \ce{NaHCOO}$\cdot$55\ce{H2O} = 1~M, \ce{NaHCOO}$\cdot$16\ce{H2O} = 3.5~M, \ce{NaHCOO}$\cdot$8\ce{H2O} = 7~M, \ce{NaHCOO}$\cdot$5\ce{H2O} = 10.5~M, and \ce{NaHCOO}$\cdot$4\ce{H2O} = 14~M.}
    \label{fig:Na+_RDF}
\end{figure}

In a recent preprint~\cite{trapp2025electrolyte}, ion pairing in concentrated ($>$ 1~M) sodium formate solutions (Fig.~\ref{fig:Na+_RDF}a and Fig.~\ref{fig:Na+_RDF}b) is studied with MD simulations. The radial distribution function (RDF) of Na$^+$ to formate oxygen in Fig.~\ref{fig:Na+_RDF}a shows structural features that correspond to contact ion pair (CIP) and solvent-shared ion pair (SSIP). The average coordination number (CN) between Na$^+$ and the formate oxygen from the CIP clusters is calculated as an integration of radial distribution function over the space within the radius. It is almost 0 at 1~M (\ce{NaHCOO} $\cdot$ 55 \ce{H2O}). The coordination number then eventually increases to $\sim$ 1 at 7~M (\ce{NaHCOO} $\cdot$ 8 \ce{H2O}). 

As discussed in connection with \mainfigref[C]{mfig5}, we experimentally measured a 0.5~M sodium formate solution ($^{13}$C enriched) together with 5~M formate solutions at natural abundance containing lithium, sodium, or potassium counterions. Figure~\ref{fig:Na+_RDF} suggests that contact ion pairs are negligible at 0.5~M sodium formate, whereas contact ion pairs between one formate anion and one cation become dominant in the 5~M formate salt solutions.

Before practically modeling the ion-pair systems, we carried out numerical experiments on $J_{\mathrm{elec}}$ with systems listed in Table \ref{tab:num_exp}. The results demonstrate that ion pairing shifts the formyl $^{1} J_{\mathrm{CH}}$ mostly through local electrostatic interactions between the cation and anion. Such a shift in the calculated formyl $^{1}J_{\mathrm{CH}}$ value is short-ranged: introducing just one explicit water molecule within a polarizable solvent model effectively screens the positive charge, yielding a calculated $^{1}J_{\mathrm{CH}}$ almost identical to the no-point-charge case. Therefore, the experimentally measured 0.5~M sodium formate signal could be mainly attributed to the hydrated anion, while the signals from 5~M solutions include a statistical mixture of both the hydrated anion and the hydrated contact ion pairs with strong local electrostatic interactions.

\begin{table}[ht!]
\centering
\begin{tabular}{l|c} 
\hline
System & Calculated formyl $^{1}J_{\mathrm{CH}}$ \\
&  ($J_{\mathrm{elec}}$ with PBE0-D3/pcJ-3) \\
\hline
Gas phase, \ce{HCOO-} anion  & 154.12 Hz \\
\hline
Gas phase, \ce{Na+ HCOO-} contact ion pair & 191.61 Hz \\
\hline
Gas phase, \ce{HCOO-} in the \ce{Na+ HCOO-} CIP structure & 194.95 Hz \\
 with Na$^{+}$ replaced by a +1 point charge & \\
\hline
C-PCM, \ce{HCOO-} with 1 explicit water & 177.31 Hz \\
\hline
C-PCM, \ce{HCOO-} with 1 explicit water and a +1 point charge & 177.59 Hz \\
 placed at SSIP position (5.5~\AA{}) from formate O & \\
\end{tabular}
\caption{Numerical experiments on formate systems. The structures of \ce{HCOO-} anion, \ce{Na+ HCOO-} contact ion pair, and cluster of \ce{HCOO-} with 1 explicit water are optimized minimum structures without imaginary frequencies. The choice of 5.5~\AA{} to model the SSIP caused shift in ${}^{1}J_{\mathrm{CH}}$ is based on the distance where 1~M sodium formate (\ce{NaHCOO} $\cdot$ 55 \ce{H2O}) system reaches coordination number of 1 in Fig.~\ref{fig:Na+_RDF}b.}
\label{tab:num_exp}
\end{table}

Based on the numerical experiments and the analysis above, we include 1 formate anion and 1 cation in our models for the hydrated contact ion pair clusters. The modeling of the hydrated-anion peak involves one formate anion and several explicit waters, as discussed for the hydration-series analysis in \mainfigref[D]{mfig5}.

\begin{figure}[ht!]
    \centering
    \subfloat[\ce{Li+ HCOO-} contact ion pair\label{fig:Li_CIP}]{%
        \includegraphics[width=0.30\linewidth]{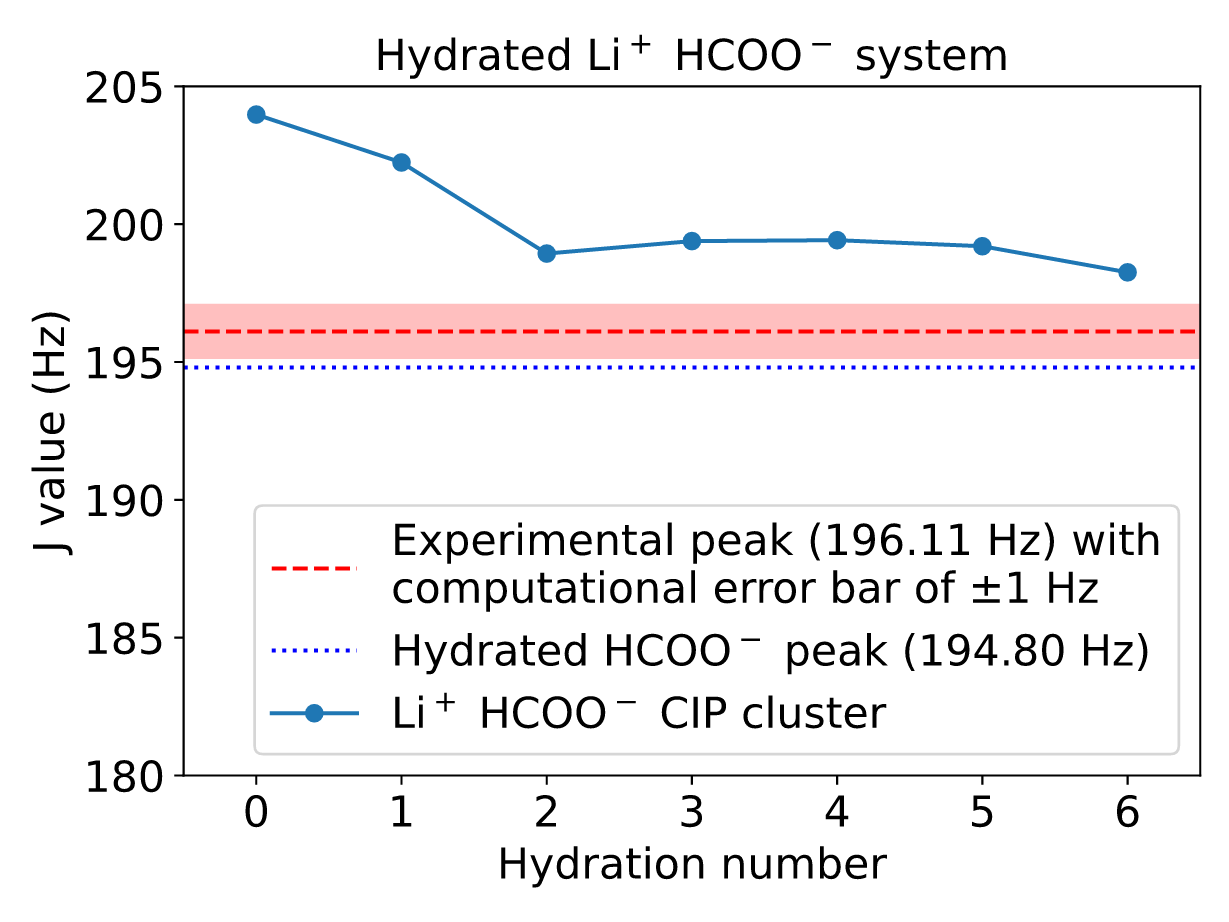}}
    \hfill
    \subfloat[\ce{Na+ HCOO-} contact ion pair\label{fig:Na_CIP}]{%
        \includegraphics[width=0.30\linewidth]{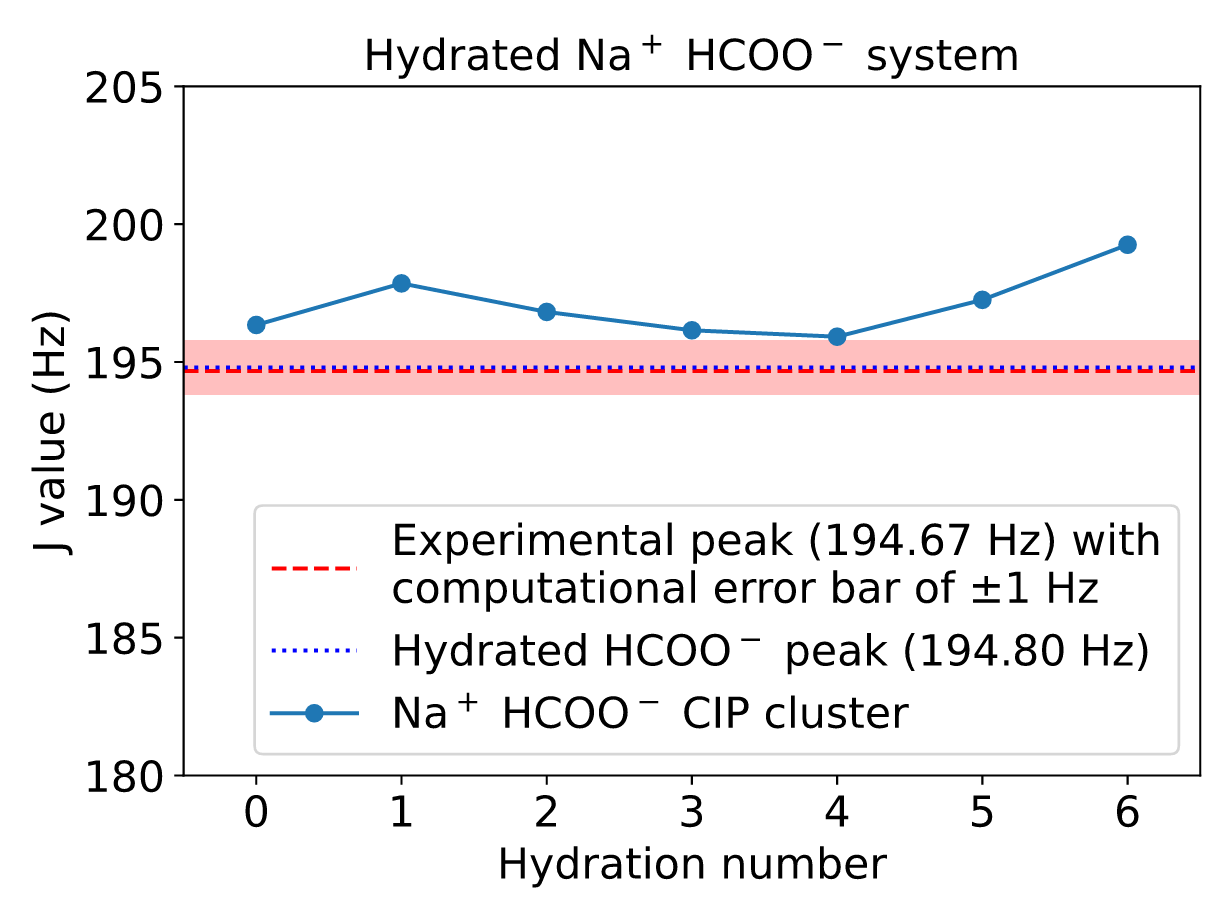}}
    \hfill
    \subfloat[\ce{K+ HCOO-} contact ion pair\label{fig:K_CIP}]{%
        \includegraphics[width=0.30\linewidth]{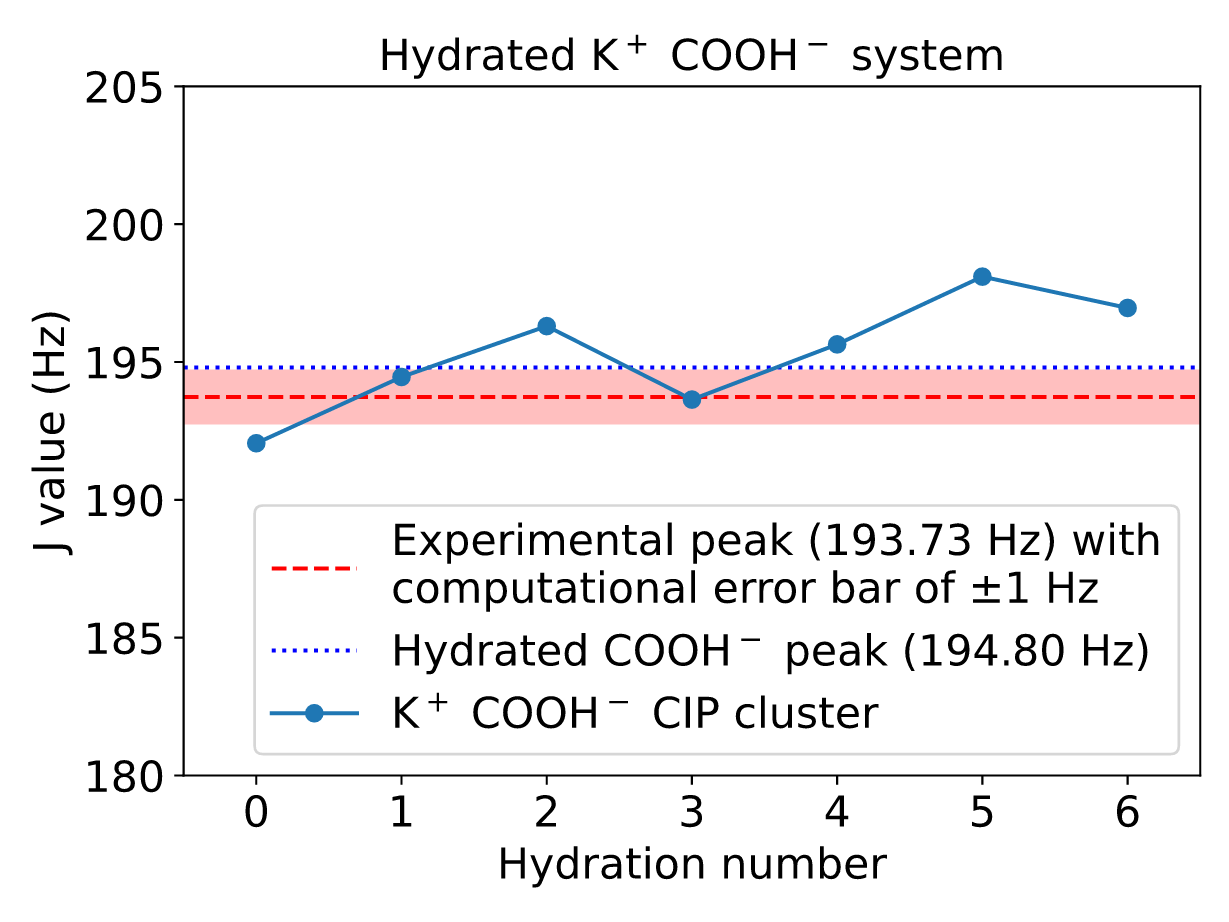}}
    \caption{\textbf{Hydration-number dependence of contact-ion-pair models.} Explicit-solvent calculations corresponding to the counterion-dependent experimental lines in \mainfigref[C]{mfig5}. (a) Li$^+$--formate contact ion pair. (b) Na$^+$--formate contact ion pair. (c) K$^+$--formate contact ion pair. In each panel, the blue points give the calculated $^{1}J_{\mathrm{CH}}$ as the number of explicit waters is varied, the red dashed line marks the experimental 5~M peak for that counterion from \mainfigref[C]{mfig5}, the blue dashed line marks the hydrated-formate reference peak from \mainfigref[C]{mfig5}, and the shaded red band indicates the estimated $\pm1$~Hz DFT uncertainty at fixed geometry.}
    \label{fig:CIP}
\end{figure}

For the contact-ion-pair clusters, the calculated $^{1}J_{\mathrm{CH}}$ values as a function of the number of explicit waters are presented in Fig.~\ref{fig:CIP}. Even without explicit waters (that is, using only the implicit C-PCM), the calculations correctly reproduce the descending periodic trend from lithium to sodium to potassium. This behavior is consistent with cation size: a smaller cation approaches the formate anion more closely and therefore produces a larger upfield shift, so the lithium solution shows the largest shift, followed by sodium and then potassium.

As analyzed earlier, the structure of concentrated salt solution is very dynamic, and the true signal would correspond to a statistical mixture of CIP and SSIP features. We note that for the maximum physical rigor, one could attempt to go beyond using static structures to model the system by applying Boltzmann average from the free energy of the CIP and SSIP states. In a recent benchmark study~\cite{o2024pair}, machine learning assisted ab initio molecular dynamics (AIMD) simulations show that, in 1~M NaCl solution, it requires nano-second long AIMD with MP2/RPA to semi-quantitatively calculate the free energy difference between contact ion pair (CIP) vs solvent-shared ion pair (SSIP) states, yet all tested (non-hybrid) DFT-based AIMD methods still struggle to qualitatively distinguish the relative stability. Our 5~M formate salt solution system is much more complicated and computationally challenging than 1~M NaCl. We consider this to be out of the scope of this work and did not attempt AIMD simulations.

\section{Spin-dynamics simulations and spectral reconstruction}
\label{sec:zf_spin_sim}

This section summarizes the two complementary simulation approaches used in this work: (i) time-domain spin-dynamics simulations (Spinach) and (ii) frequency-domain calculations of transition frequencies and intensities from the ZF Hamiltonian. The latter is particularly convenient for scanning a coupling parameter and evaluating the coupling-response slopes shown in \mainfigref[C]{mfig1}, while the former underlies the simulated spectra in \mainfigref[B]{mfig1}, \mainfigref[A]{mfig3}, and \mainfigref[A--D]{mfig4}.

\subsection{Time-domain simulations in Spinach}
\label{subsec:td_sim}

Time-domain simulations were performed using the Spinach library~\cite{hogbenSpinachSoftwareLibrary2011b}.
For each isotopomer, a spin system is defined by (1) a list of nuclei and (2) an isotropic scalar-coupling network.
The evolution is generated under the ZF Hamiltonian, and the detected signal is the longitudinal magnetization along the OPM sensitive axis,
\begin{equation}
\langle M_z(t)\rangle=\mathrm{Tr}\!\left[\hat{M}_z\,\hat{\rho}(t)\right],
\qquad \hat{M}_z=\sum_i \gamma_i \hat{I}_{z,i}, 
\qquad \hat{\rho}(t) = e^{-i\hat{H}_\mathrm{ZF} t}\, \hat{\rho}_0\, e^{+i\hat{H}_\mathrm{ZF} t}.
\end{equation}
with $\gamma_i$ the gyromagnetic ratio of spin $i$.

A minimal (illustrative) Spinach workflow is:
\begin{verbatim}
% 1) Define nuclei and J-couplings (one isotopomer)
sys.isotopes = {'1H','1H','1H','13C', ...};
inter.coupling.scalar{1,4} = 128.0;   % Hz, example entry
inter.coupling.scalar{2,4} = 128.0;
...

% 2) Build the spin system and run a ZF pulse-acquire experiment
spin_system = create(sys, inter);
parameters = struct(...);   % pulse, acquisition time, linewidth, etc.
fid = liquid(spin_system, @zerofield, parameters);
spec = abs(fftshift(fft(fid)));
\end{verbatim}

The output is an FID with no decoherence, with a specified sampling rate and number of points (care should be taken, as aliasing can occur if sampling rate is too low). The next subsection details how a proper decay profile can be added after-the-fact for comparison with experiment.

\noindent
In this work, the simulation inputs (nuclear identities, coupling lists) are archived in the code/data repository as .csv files
(see the Code and data availability section below), rather than reproduced verbatim here. A comprehensive \href{https://spindynamics.org/wiki/index.php?title=Main_Page}{Spinach wiki} details how to format NMR-active nuclei and couplings, and comes preinstalled with a ZF simulation example file for benzene.

\subsection{Line-shape model: Gaussian and exponential broadening}
\label{subsec:td_sim_proc}

To mimic experimental $T_2^*$ decay, a Gaussian + exponential envelope is applied to the otherwise non-decaying simulated FID defined by the following python function:

\begin{verbatim}
t = np.linspace(0, acq_time_sim, len(fid_sim)) # linear time-series matching simulation
def gaussian_plus_exponential(t, T2e, T2g, translate, A, B):
    return B * np.exp(-((t-translate)/(T2g))**2) + A * np.exp(-t/T2e)
\end{verbatim}

where $\textbf{\texttt{t}}$ is the time axis of the simulated FID, $\textbf{\texttt{T2e}}$ and $\textbf{\texttt{T2g}}$ are the time constants for the exponential and Gaussian decay components respectively (in units matching $\textbf{\texttt{t}}$), $\textbf{\texttt{translate}}$ is a time delay between the decay components (in units matching $\textbf{\texttt{t}}$), and $\textbf{\texttt{A}}$ and $\textbf{\texttt{B}}$ are the relative weights of the exponential and Gaussian decay components respectively. The output is an array with length matching $\textbf{\texttt{t}}$ which is multiplied element-wise with the simulated FID. Table S1 lists these input parameters for every instance in the main text:

\begin{table}[h!]
\centering
\begin{tabular}{cccccc}
\hline
\textbf{instance} & Fig. 1B (ii) & Fig. 4A & Fig. 4B & Fig. 4C & Fig. 4D\\
\hline
$\textbf{\texttt{T2e}}$       & 0.46 & 0.63 & 1.20 & 1.15 & 0.33 \\
$\textbf{\texttt{T2g}}$       & 1.74 & 1.36 & 2.16 & 1.82 & 2.21\\
$\textbf{\texttt{translate}}$ & 0.69 & 0.40 & 0.24 & 0.16 & 0.06\\
$\textbf{\texttt{A}}$         & 1.00 & 1.00 & 1.00 & 1.00 & 1.00\\
$\textbf{\texttt{B}}$         & 1.87 & 3.13 & 7.82 & 5.24 & 7.08\\
\hline
\end{tabular}
\caption{Table detailing parameters used to create Gaussian + exponential decay profile applied to simulation in the time domain via spinach. Parameters can be inputted directly into the python function sketched in this section.}
\label{table:decay_params}
\end{table}

As discussed in the main text, faithful reproduction of experimental ZF multiplets requires sufficiently accurate small $J$ couplings because they determine the internal ``fingerprint'' structure of the spectrum~\cite{theisChemicalAnalysisUsing2013,butlerMultipletsZeroMagnetic2013,blanchardHighResolutionZeroFieldNMR2013b}. To emphasize the accuracy of vibrationally corrected vacuum-state DFT for \mainfigref[A--D]{mfig4}, the dominant $^{1}J_{\mathrm{CH}}$ value for each isotopomer was adjusted manually, because this parameter produces an almost uniform linear shift of the entire multiplet. For example, in a $\ce{^{13}CH}$ group with XA topology, $\partial\nu/\partial {}^{1}J_{\mathrm{CH}}\approx 1$ for all $\nu$, analogous to varying the magnetic-field strength in conventional NMR. This small display-oriented adjustment was applied by eye and could be automated without rerunning the full spin-dynamics simulation. Following the logic of the main text, these residual deviations could in part or in whole be of genuine chemical origin, rather than simply prediction error. Explicit modeling of all relevant interactions in these systems could rectify $^1J_{CH}$ (and other $J$) discrepancies.

These shifts are not represented in the raw simulation inputs provided in the code repository, but are listed here for completeness. No adjustment was made to the symmetric isotopomer of acetone because only a single $J$ coupling defines that system. Table~\ref{table:shifts} reports the applied shifts in $^{1}J_{\mathrm{CH}}$ for the isotopomers shown in \mainfigref[A--D]{mfig4}.

\begin{table}[h!]
\centering
\begin{tabular}{l|c}
Acetone \ce{^{13}CH3}             & -1.935 \\
MF \ce{^{13}CH}                    & -2.244 \\
MF \ce{^{13}CH3}                  & -2.250 \\
DE \ce{^{13}CH2}                  & -1.535 \\
DE \ce{^{13}CH3}                  & -3.210 \\
EA \ce{^{13}CH2}                  & -2.007 \\
EA \ce{^{13}CH3} (adjacent to \ce{C=O})   & -2.271 \\
EA \ce{^{13}CH3} (adjacent to \ce{CH2}) & -3.346 \\

\end{tabular}
\caption{Tabulated uniform frequency shifts applied to $^{1}J_{\R{CH}}$ in the simulated isotopomer spectra in \mainfigref[A--D]{mfig4}. Methyl formate $\rightarrow$ MF, Diethyl Ether $\rightarrow$ DE, Ethyl Acetate $\rightarrow$ EA. In response, multiplets shift linearly, analogous to changing the magnetic field strength in conventional NMR, whereas the other, smaller couplings determine the internal ``fingerprint'' multiplet structure and are not shifted.}
\label{table:shifts}
\end{table}

Finally, each isotopomer spectrum was weighted separately before summation to produce the composite spectra shown in \mainfigref[A--D]{mfig4}. As above, this display-oriented weighting was done by eye and could be automated by normalizing the simulations to natural-abundance statistics.

\subsection{Frequency-domain analysis for \mainfigref[C]{mfig1}}
\label{subsec:fd_sim}
An example Jupyter notebook in the code repository performs the frequency-domain simulation described here. For a time-independent Hamiltonian $\hat{H}$ with eigenstates $\{\lvert n\rangle\}$ and eigenvalues $\{E_n\}$, the density matrix evolves as
$\hat{\rho}(t)=e^{-i\hat{H}t}\hat{\rho}(0)e^{+i\hat{H}t}$.
Inserting a resolution of identity in the energy eigenbasis gives
\begin{equation}
\langle M_z(t)\rangle=\sum_{m>n} (\rho_0)_{mn}\, (M_z)_{mn}\,e^{-i\omega_{mn} t},
\qquad
\omega_{mn}=\frac{E_m-E_n}{\hbar}.
\label{eq:freqdomain_signal}
\end{equation}
in the Hamiltonian eigenbasis between all pairs of eigenstates $n$ and $m$. Thus the spectrum consists of discrete transitions at frequencies $\nu_{mn}=\omega_{mn}/2\pi$ with complex amplitudes proportional to
$\rho_{nm}(0)(M_z)_{mn}$.

The frequency-domain approach is well suited to parameter-response plots because the frequencies and amplitudes are obtained directly. For the $J$-response slopes~\cite{bodenstedtOpticallyDetectedNuclear2024a} shown in \mainfigref[C(ii)]{mfig1}, one can recompute the eigenvalues after shifting a single coupling parameter by a small increment $\Delta J$ and estimate
\begin{equation}
\frac{\partial \nu_{mn}}{\partial J}\approx
\frac{\nu_{mn}(J+\Delta J)-\nu_{mn}(J)}{\Delta J}.
\end{equation}

In contrast, frequencies are not computed directly when simulating via time domain (Spinach) approach, but inferred by manual picking peaks after Fourier transform; in that case, peak positions therefore depend on spectral resolution (FID length, which scales total simulation time) and aliasing must be considered.

\subsection{Machine-readable simulation inputs}
\label{subsec:sim_inputs}

For transparency and reproducibility, the complete lists of nuclei and scalar couplings used for all simulations are provided as machine-readable (.csv) files in the repository described in the Code and data availability section below.
Briefly:
\begin{itemize}
  \item \textbf{\texttt{nuc\_list}} specifies the isotope label for each spin in the simulated isotopomer (e.g., \texttt{['13C','1H','1H',...]}). its length $n$ is the total number of NMR-active nuclei;
  \item \textbf{\texttt{couple\_list}} specifies the nonzero scalar couplings between spins $i$ and $j$ $(J_{ij})$, following indexing consistent with \textbf{\texttt{nuc\_list}}, for each isotopomer. It is therefore an $n \times n$ upper right triangle matrix.
\end{itemize}

\subsection{Double-$^{13}$C acetic acid: extracting $^{1}J_{\mathrm{CC}}$ and simulating \mainfigref[A]{mfig3}}
\label{sec:jcc_acetic}

\begin{figure}[h!]
  \centering
  \includegraphics[width=\SIfigwidth]{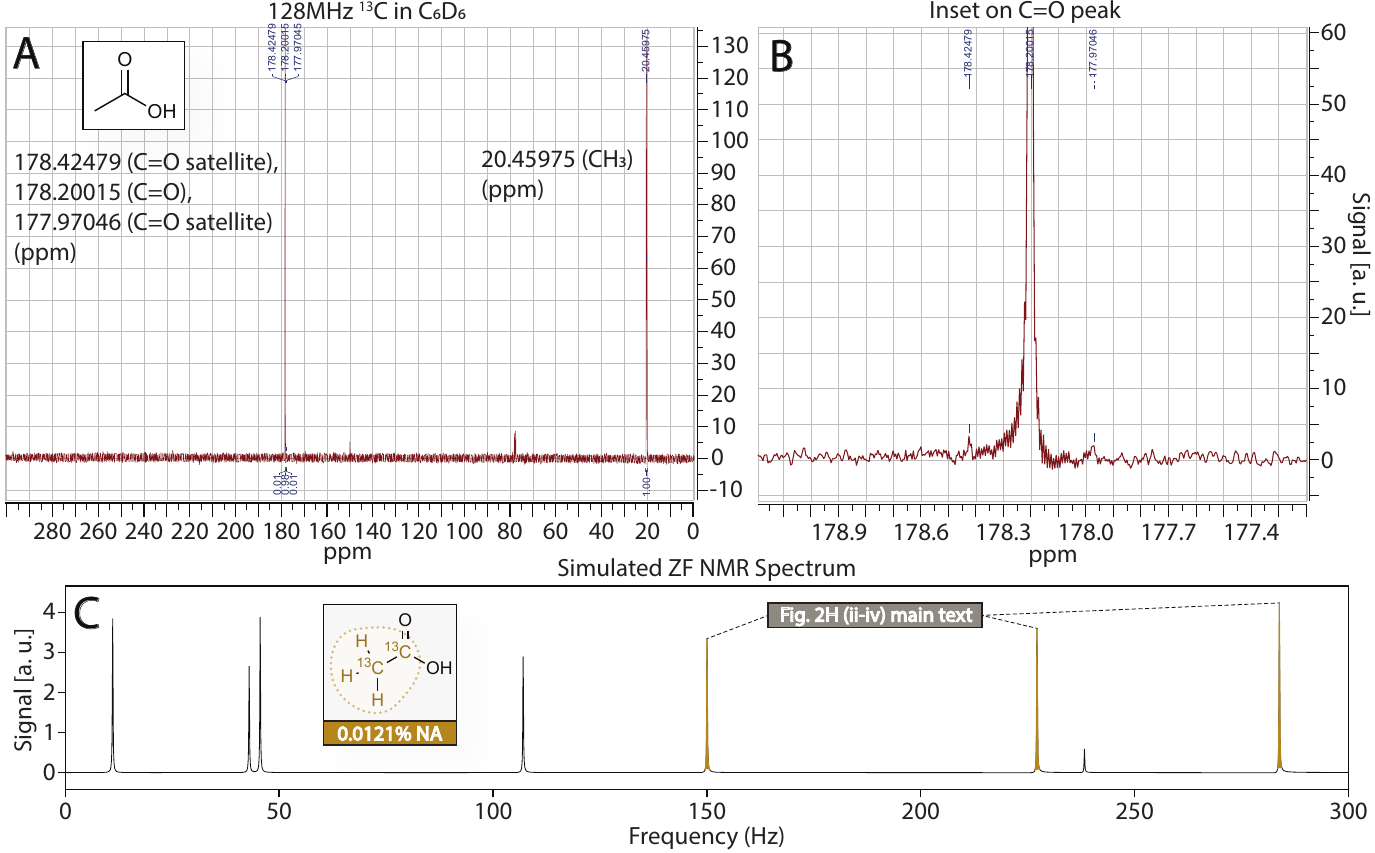}
  \caption{\textbf{Conventional $^{13}$C NMR of acetic acid: $^{1}J_{\mathrm{CC}}$ extraction from satellites.} (A) Full 126~MHz $^{13}$C spectrum of acetic acid in \ce{C6D6}, listing the carbonyl resonance, its two satellite lines, and the methyl resonance. (B) Expanded view of the carbonyl signal from (A), where the weak satellite pair at $p^2\approx0.012\%$ abundance yields $^{1}J_{\mathrm{CC}}$. (C) Simulated ZF spectrum of the doubly $^{13}$C-labelled isotopomer using the extracted $^{1}J_{\mathrm{CC}}$ together with the ZF-derived $^{1,2}J_{\mathrm{CH}}$ values; the shaded representative lines are the ones compared with the experimental peaks in \mainfigref[A(ii--iv)]{mfig3}.}
  \label{fig:Jcc}
\end{figure}

Doubly $^{13}$C-substituted isotopomers of acetic acid occur at natural abundance with probability $p^2\approx 0.012\%$ and are visible in \mainfigref[A(ii--iv)]{mfig3}. Accurate simulation of these features requires inclusion of (i) the $^{13}$C--$^{1}$H scalar couplings associated with each $^{13}$C site (that is, $^{1}J_{\mathrm{CH}}$ and $^{2}J_{\mathrm{CH}}$) and (ii) the additional $^{13}$C--$^{13}$C coupling, $^{1}J_{\mathrm{CC}}$, unique to the doubly labelled isotopomer. We determine $^{1}J_{\mathrm{CC}}$ from a conventional 126~MHz $^{13}$C NMR spectrum of acetic acid in \ce{C6D6} (Fig.~\ref{fig:Jcc}A; satellite inset in Fig.~\ref{fig:Jcc}B). Combining this value with $^{2}J_{\mathrm{CH}}$ and $^{1}J_{\mathrm{CH}}$ extracted from the ZF spectrum in \mainfigref[A(i)]{mfig3} (see also Fig.~\ref{fig:H2Osup}A) specifies all relevant couplings and enables simulation of the ZF spectrum of the $2\times {}^{13}$C isotopomer (Fig.~\ref{fig:Jcc}C).

With the $^{13}$C satellite peak positions in hand, $^1J_{\mathrm{CC}}$ is:
\begin{align}
\mathrm{^1J}_{\mathrm{CC}}
&= (178.424794 - 177.970457)\,\text{(ppm)}
   \times 126\,\text{(MHz)}
= 56.657\,\text{(Hz)}.
\end{align}

This furnishes the simulation inputs used to produce Fig.~\ref{fig:Jcc}C:

{\centering
\begin{verbatim}
nuc_list = ["13C", "13C", "1H", "1H", "1H"]

couple_list = [
[0, 56.657, 129.628, 129.628, 129.628],
[0,      0,    6.74,    6.74,    6.74],
[0,      0,       0,       0,       0],
[0,      0,       0,       0,       0],
[0,      0,       0,       0,       0],
]
\end{verbatim}
\par}

Which is verified in \zfr{mfig3}A(ii--iv) in the main text. In the full simulation, 8 total resonances occur. For comparison, we chose 3 strong, representative lines (shaded and labeled in Fig.~\ref{fig:Jcc}C), as these lines are far away from 1$\times^{13}$C acetic acid resonances, noise artifacts, and the noisier low frequency band (0--50~Hz).

\section{Instrumentation, pulse sequence and experimental workflow}
\label{sec:instrument_exp}

\subsection{Apparatus overview and timing sequence}
\label{subsec:instrument_scheme}

\begin{figure}[h!]
  \centering
  \includegraphics[width=\SIfigwidth]{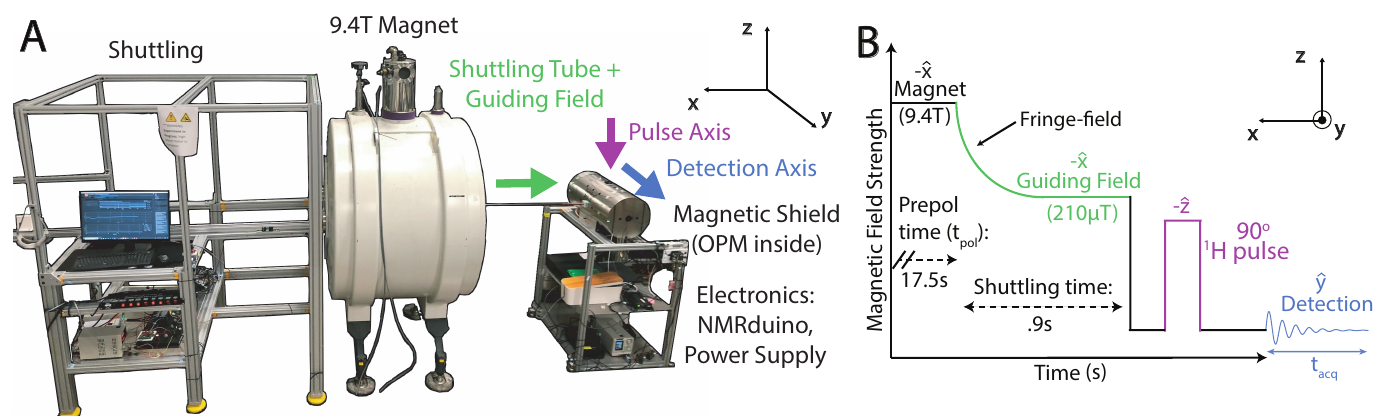}
  \caption{\textbf{Experimental apparatus and pulse sequence schematic.} (A) Photograph of the ZF NMR instrument used for the main-text spectra in \mainfigrange{mfig1}{mfig5}, with the color coding linked to the timing diagram in (B). (B) Pulse-acquire sequence underlying the System and Principle section: pre-polarization in 9.4~T, shuttling under a 210\(\mu\mathrm{T}\) guiding field, switching to zero field, application of a 90$^\circ$ $\Hs$\ pulse, and acquisition of the ZF signal with the OPM.}
  \label{fig:apparatus}
\end{figure}

A schematic of the apparatus and pulse sequence is shown in Fig.~\ref{fig:apparatus}. A previous version of this apparatus is explained in Ref.~\cite{andrewsSensitiveMultichannelZeroto2025}.
Samples are pre-polarized in a 9.4~T inhomogeneous ($\sim$50ppb) magnet, shuttled in $\sim$1~s into a magnetically shielded region under a 210\(\mu\mathrm{T}\) guiding field, and detected with a commercial $^{87}$Rb OPM (QuSpin, Gen 3 dual axis~\cite{shahQZFMGen3}) positioned at a standoff of 2 millimeters from the GCMS sample tube (Agilent 5183-4428).

The detection region uses a multi-layer mu-metal magnetic shield (Twinleaf MS-2, or Twinleaf MS1-LF for the experiments in \mainfigref[B(iii)]{mfig1} and \mainfigref[A,C]{mfig5}) with integrated Helmholtz coils for cancellation of residual fields after initial degaussing. Once cancelled, these residual fields remain stable for long periods~\cite{andrewsSensitiveMultichannelZeroto2025} and need not be readjusted unless the shield is opened (which is unnecessary for routine measurements and only occurs during instrument modification or upgrade).

Samples are mechanically isolated (vibration-damping feet and a flexible sample rod) to reduce motion-induced artifacts during long averaging. The aluminum frame housing the mu-metal magnetic shield is mounted on Sorbothane feet to further mitigate vibration from the environment.
All electronics are powered from an uninterruptible power supply to suppress 60~Hz pickup; a (battery powered) laptop is used to store data and interface with the NMRduino~\cite{taylerNMRduinoModularOpenSource2024}.

\subsection{NMRduino control and pulse program}
\label{subsec:NMRduino}

An NMRduino~\cite{taylerNMRduinoModularOpenSource2024} (Teensy 4.1 microcontroller-based) system is used to synchronize shuttling, guiding-field control, pulsing, and acquisition. The DC pulses applied here, along with low-frequency (0--500~Hz) acquisition, make this ``spectrometer on a chip'' possible with a credit-card-sized footprint.

\begin{figure}[h!]
  \centering
  \includegraphics[width=\SIfigwidthNarrow]{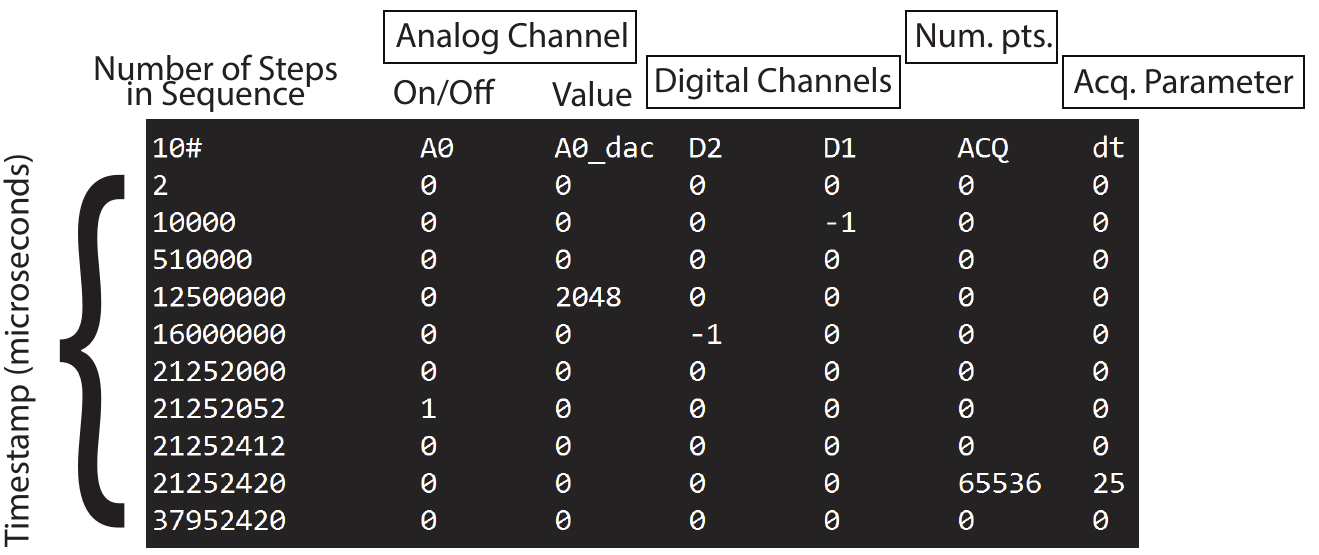}
  \caption{\textbf{Representative NMRduino sequence file format.} Adapted from the NMRduino platform described in Ref.~\cite{taylerNMRduinoModularOpenSource2024}. The annotated columns show, from left to right, the number of steps in the sequence, the timestamp of each step, the analog-channel state, the digital-channel states, the number of acquired points, and the acquisition parameter. In the example shown, the analog channel (A0) applies the 360\(\mu\mathrm{s}\) pulse, D2 controls the 210\(\mu\mathrm{T}\) guiding field, D1 triggers the motorized shuttling sequence, and the ACQ/\texttt{dt} entries define the acquisition block plotted schematically in Fig.~\ref{fig:apparatus}B.}
  \label{fig:nmrduino}
\end{figure}

Figure~\ref{fig:nmrduino} shows the tab-delimited sequence file format used to program the experiment.

In the sequence files used here, the acquisition block records $N_\mathrm{pts}=65536$ points.
The sampling rate is set by an integer parameter \texttt{dt} according to
\begin{equation}
\mathrm{Sampling\ Rate\ (kHz)} = \frac{100}{\texttt{dt}}.
\end{equation}
For example, \texttt{dt}=25 corresponds to a 4~kHz sampling rate and an acquisition time of $65536/4000 \approx 16.384$~s per scan.

For a detailed description of NMRduino operation, we refer the reader to Ref.~\cite{taylerNMRduinoModularOpenSource2024}, which describes the on-chip hardware and provides a tutorial on configuring input and output channels using the text-file interface. 

\subsection{Intrinsic suppression of the water background at zero field}
\label{subsec:water_supp}

\begin{figure}[h!]
  \centering
  \includegraphics[scale=.98]{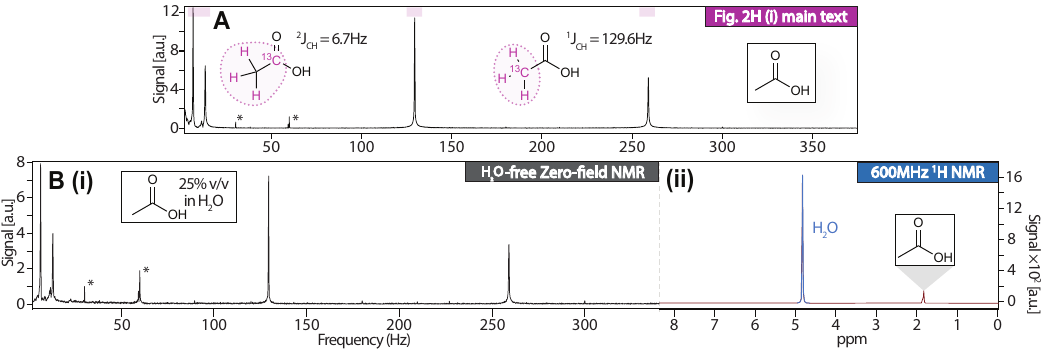}
  \caption{\textbf{Intrinsic water suppression at zero field: acetic-acid demonstration.} (A) Magnitude (unphased) spectrum of neat natural-abundance acetic acid, reproducing the main-text lines of \mainfigref[A(i)]{mfig3} governed by $^{2}J_{\mathrm{CH}}=6.7$~Hz and $^{1}J_{\mathrm{CH}}=129.6$~Hz. (B)(i) Spectrum of 25\% v/v acetic acid in \ce{H2O}, showing that the water-rich sample introduces no additional ZF lines and no detectable degradation of the acetic-acid resonances. (B)(ii) Conventional 600~MHz $^{1}$H spectrum of the same sample, where the water resonance dominates. This contrast illustrates the intrinsic water suppression that also enables the aqueous measurements of \mainfigref[A,C]{mfig5}.}
  \label{fig:H2Osup}
\end{figure}

A feature of ZF NMR is that at least two distinct NMR-active species must be present in order for the spectrum to be observable (in this work, $^{13}$C and $^{1}$H). An interesting consequence is that \ce{H2O} produces no ZF signal, rendering it ``ZF NMR silent". ZF NMR spectroscopy is therefore intrinsically water suppressing. As a simple demonstration, we compare neat acetic acid (Fig.~\ref{fig:H2Osup}A) to 25$\%$ v/v acetic acid (Fig.~\ref{fig:H2Osup}B(i)) in \ce{H2O}, showing no additional lines or degradation of existing resonances despite water constituting the bulk of the sample; this is in contrast to the 600~MHz $^{1}$H NMR of the same sample (Fig.~\ref{fig:H2Osup}B(ii)), where the water resonance (blue) is much stronger than the acetic-acid resonance. This same property underlies the aqueous measurements in \mainfigref[A,C]{mfig5}.

\section{Sensitivity analysis and benchmarking}
\label{sec:sensitivity_budget}

\subsection{Comparison with previous ZF NMR benchmarks}
\label{subsec:prev_sota}

The main text reports a $\sim$3-fold sensitivity gain for the benzene benchmark of \mainfigref[B(i)]{mfig1} (corresponding to $\sim$9-fold less averaging time) over the previous ZF NMR benchmark~\cite{blanchardHighResolutionZeroFieldNMR2013b}, enabled by instrumental refinements. Paired with long-term stability, the present apparatus supports uninterrupted averaging over many days with near-ideal SNR scaling, as illustrated in \mainfigref[B]{mfig3}. In summary, the key experimental differences compared with the previous report are:

\begin{enumerate}
  \item Different pulse sequence and magnetometer sensitive axis.
  \item Natural abundance $^{13}$C “off the shelf” vs. ~100\% 1$\times^{13}$C benzene with paramagnetic oxygen removal (degassing).
  \item Higher prepolarization fields (9.4T vs. 2T).
  \item Extended averaging enabled by long-term stability.
  \item The use of 2mL GCMS vials instead of standard 5mm NMR tubes.
\end{enumerate}

These differences should be considered alongside the general apparatus refinements described in the previous section. We focus below on points 1 and 2, which could feasibly be implemented on our setup, whereas points 3--5 are genuine experimental improvements.

We now quantify this sensitivity enhancement by comparing experimental benzene data, graciously provided by authors of Ref.~\cite{blanchardHighResolutionZeroFieldNMR2013b}. Firstly, the previous report used an OPM situated at the end of the guiding field (shuttling) solenoid, where both the $^{13}$C and $^{1}$H spins were initially polarized along the magnetometer sensitive axis. From there, an orthogonal 4$\pi$ $^{1}$H pulse was applied which effectively flips $^{13}$C spins, leaving $^{1}$H unchanged since $\gamma_H/\gamma_C \app 4$. In our experiment, the OPM is placed outside, perpendicular to the guiding field (shuttling) axis, and thus normal to the initial polarization axis of $^{13}$C and $^{1}$H spins. This alternative scheme was employed for efficient multichannel detection in our previous work~\cite{andrewsSensitiveMultichannelZeroto2025}, with 3 samples and sensors placed longways along the solenoid.  A mutually orthogonal, $\pi$/2 $^{1}$H pulse creates projections of $^{13}$C and $^{1}$H polarization along the magnetometer sensor axis, thus producing signal.

It is straightforward to simulate the difference in signal magnitude between both experiments by modifying $\rho(0)$ and the observable $M_z$, which reveals a 1.37-fold enhancement in the previous case compared with our experimental scheme.

We now compare both experimental data, making adjustments for $^{13}$C enrichment and longer signal lifetimes due to paramagnetic oxygen degassing. To account for longer T2* in Ref.~\cite{blanchardHighResolutionZeroFieldNMR2013b}, we first apply apodization ($e^{-0.67t}$) to match the shorter lifetimes of the benzene data collected here, and subsequently truncate the FID to match total acquisition times (5.00s). From there, SNR was calculated from the strongest peak and the 185--205~Hz noise floor. Lastly, at natural abundance, 1$\times{}^{13}$C benzene occurs at 6.6$\%$ (six equivalent carbon sites, each with a 1.1$\%$ chance), compared with the enriched, ~100$\%$ $^{13}$C sample. Finally, the ~100$\%$ $^{13}$C benzene data provided are an average of 230 scans, so we averaged the first 230 scans of our data accordingly.

With this comparison in place, we find that the previous state of the art produces an SNR of 913.22 in 230 scans, whereas our demonstration has an SNR of 130.52 in 230 scans. Adjusting for (1) and (2), we find:

\begin{align}
\mathrm{SNR}_{\mathrm{prev}}^{(\mathrm{adjusted})}
&= 913.22 
   \times 0.066\,\text{(NA = 6.6\%)}
   \times \frac{1}{1.37}\,\text{(pulse modification)}
= 43.99 . \\
\text{enhancement}
&= \frac{
   130.52\,\text{($\mathrm{SNR}_{\mathrm{current}}$)}
}{
   43.99\,\text{($\mathrm{SNR}_{\mathrm{prev}}^{(\mathrm{adjusted})}$)}
}
= 2.97 .
\end{align}

This corresponds to an 8.82-fold reduction in total averaging time.

\subsection{Simple scaling model for scan-averaged sensitivity}
\label{subsec:sens_model}

For scan-averaged ZF NMR, the signal-to-noise ratio after $N$ scans scales approximately as
\begin{equation}
\mathrm{SNR}(N)\propto \frac{S_1}{\sigma_1}\sqrt{N},
\end{equation}
where $S_1$ and $\sigma_1$ are the single-scan signal amplitude and noise, respectively.
Instrumental changes can improve $S_1$ (e.g., increased sample magnetization at detection, improved pre-polarization, better stand-off geometry)
and/or reduce $\sigma_1$ (e.g., lower magnetic noise at the OPM).

\subsection{Why long-term stability matters: commercial versus home-built OPMs}
\label{subsec:com_vs_home_built}

In practice, the $\sqrt{N}$ scaling requires that the spectrum be stable on the timescale of the full experiment.
Slow drifts (frequency, baseline, gain) degrade coherent averaging and broaden lines, causing SNR to saturate with increasing scan number (total experiment time). The commercial Quspin OPM used here exhibits stable operation over $>$week time scales, enabling narrow (tens-hundreds of mHz) linewidths and near-ideal averaging. This stability, rather than raw single-shot sensitivity alone, is the key enabler for NA measurements and detection of rare isotopomers.

\subsection{Routes to higher sensitivity without sacrificing stability}
\label{subsec:sens_stability}

The main text discusses several hardware routes to further boost sensitivity without resorting to hyperpolarization, including flux concentrators (where fields emanating from the sample are concentrated at the OPM), faster shuttling and/or reduced shuttling path with compact magnets, improved magnetic shielding materials, and reduction of Johnson noise associated with guiding-field and magnetic shielding hardware. These improvements are compatible with the compact, robust OPM detection platform emphasized here. In addition, a larger vapor cell (still sufficiently small to be fully deployed within the magnetic shield) would increase the number of sensing electrons and sensitive volume, allowing for more ZF NMR fields to be captured, potentially increasing SNR.

Much of the stability in our QuSpin OPM~\cite{shahQZFMGen3} lies in iterative engineering improvements across product generations, for instance, precise control loops for magnetic field and temperature stability, and other refinements as a result of commercialization compared to the home-built setup in Ref.~\cite{blanchardHighResolutionZeroFieldNMR2013b}. A marginally bigger cell could in theory be implemented while maintaining stability: Current OPM technologies intentionally use smaller vapor cells; their primary application, magnetoencephalography (MEG) seeks to place $>$100 discrete OPMs across the human scalp for field localization within the brain~\cite{hillOptimisingSensitivityOpticallyPumped2024,zhangMeasuringHumanAuditory2024a,botoMeasuringFunctionalConnectivity2021a}, suggesting that the current vapor cell is not at the size limit. We hope that as ZF NMR methodology improves, OPM manufacturers can provide larger cell devices with optimal SNR for ZF NMR readout in tandem with compact MEG devices. These current, compact MEG-OPMs used here can be arrayed for high-throughput, multichannel ZF NMR detection with potentially $\sim$100 distinct channels~\cite{andrewsSensitiveMultichannelZeroto2025}.

\subsection{Detection-limit benchmark: 100~mM formic acid}
\label{subsec:LOD}

\begin{figure}[h!]
  \centering
  \includegraphics[scale=.98]{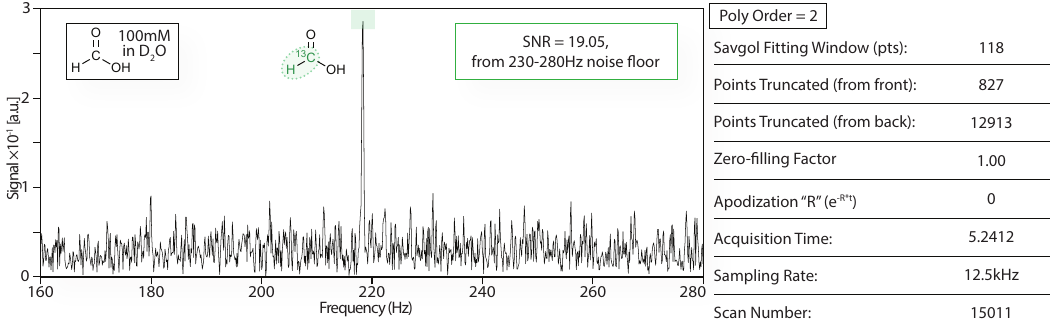}
  \caption{\textbf{Detection-limit benchmark: 100~mM formic acid.} Magnitude (unphased) ZF spectrum of 100~mM natural-abundance formic acid in \ce{D2O}, supporting the statement after \mainfigref[B]{mfig3} that dilute NA formic acid remains detectable. The displayed line has SNR 19.05 after 15{,}011 averages, corresponding to $\mathrm{SNR}_{\mathrm{single-shot}}=0.155$ under ideal $\sqrt{N}$ scaling. The noise floor is taken from 230--280~Hz, and the accompanying parameter panel reports the acquisition and processing settings in the same format as Figs.~\ref{fig:acq_params} and \ref{fig:proc_params}.}
  \label{fig:LOD100mM}
\end{figure}

As noted in the main text after \mainfigref[B]{mfig3}, we present the spectrum of 100~mM natural-abundance formic acid in \ce{D2O} as a detection-limit measurement and SNR benchmark for future comparisons. As shown in Fig.~\ref{fig:LOD100mM}, an SNR of 19.05 is obtained after 15,011 scans, corresponding to a single-shot SNR of 0.155 under ideal $\sqrt{N}$ scaling. Relevant processing parameters are provided, and the magnitude (unphased) spectrum is shown.

\section{Sample preparation and solution compositions}
\label{sec:sample_prep}

\subsection{Commercial ``off-the-shelf'' liquids}
\label{subsec:remarks}

Samples used throughout this work were obtained from synthetic-chemistry research laboratories across the University of California, Berkeley, primarily the laboratories of Richmond Sarpong, Jeffrey Long, and Polly L. Arnold. In this way, the samples were used truly ``off the shelf'' and loaded under ambient atmosphere (that is, without freeze--pump--thaw degassing) into 2~mL glass GC--MS vials compatible with the shuttling geometry. These samples include the compounds highlighted in \mainfigref[A]{mfig1}, \mainfigref[A--G]{mfig2}, \mainfigref[A]{mfig3}, and \mainfigref[A--D]{mfig4}.

\subsection{Formic-acid dilution series}
\label{sec:fa_dilutions}

A formic-acid dilution series in water was prepared volumetrically using micropipettes.

\begin{table}[h!]
\centering
\begin{tabular}{ccc}
\hline
\textbf{$mol frac.$ ($x_\mathrm{FA}$)} & \textbf{$V_\mathrm{FA}$} & \textbf{$V_{\mathrm{H_2O}}$} \\
\hline
1.00 & 2000.00 & 0.00 \\
0.99 & 1990.40 & 9.50 \\
0.98 & 1980.70 & 18.92 \\
0.80 & 1786.70 & 170.64 \\
0.58 & 1486.10 & 513.89 \\
0.30 & 945.96  & 1054.02 \\
0.10 & 377.51  & 1622.45 \\
0.01 & 41.42   & 1958.15 \\
\hline
\end{tabular}
\caption{Table referencing volumes of water and formic acid used ($V$) to produce solutions with formic acid mole fractions ($x_\mathrm{FA}$). Volumes expressed in \(\mu\mathrm{L}\), with a total volume of 2 mL throughout.}
\label{tab:vol_fa}
\end{table}

The mole fraction of formic acid is computed from the micropipette-dispensed volumes $V$ using
\begin{equation}
x_\mathrm{FA}=\frac{n_\mathrm{FA}}{n_\mathrm{FA}+n_{\mathrm{H_2O}}},
\qquad
n=\frac{\rho V}{M},
\end{equation}
with $\rho$ being the density and $M$ the molar mass of the component.
The values used were $\rho_{\mathrm{H_2O}}=1.00$~g/mL, $\rho_{\mathrm{FA}}=1.22$~g/mL, $M_{\mathrm{H_2O}}=18.015$~g/mol, and
$M_{\mathrm{FA}}=46.025$~g/mol.

\subsection{Formate solution compositions}
\label{sec:formate_compisition}
The preparation of the formate solutions used in \mainfigref[C]{mfig5} is summarized here in descending order of concentration. To prepare the 0.5~M $^{13}$C-enriched formate solution, 189\(\mu\mathrm{L}\) of $^{13}$C-enriched formic acid was added to a 10~mL volumetric flask together with 205.7~mg of NaOH, corresponding to a 1.03~equiv. excess of base. $^{13}$C enrichment was used for this sample only as a time-saving measure. To verify complete deprotonation, additional aliquots of base were added with no change in the resonance at 194.80~Hz. In earlier trials, the resonance shifted upward toward the protonated formic-acid resonances (${\app}$218--222~Hz) when insufficient base was added, so care was taken to ensure complete deprotonation.

To prepare the 5~M solutions of reacted formic acid + MOH, we sought to prepare solutions containing 5~M formic acid (formate) and 6M base (MOH), as this creates an excess of counterion while also ensuring complete conversion into formate, both of which encourage contact ion pair formation.

6M-worth of LiOH monohydrate did not fully dissolve in water, so a 5~M solution of Li formate monohydrate was made by adding 3.5401g to a 10~mL volumetric flask. From there, 213.4mg of LiOH monohydrate was added for an intended 6M final concentration, which only partially dissolved and therefore saturated the solution. The solution in \zfr{mfig5}C
therefore represents 5~M formate with somewhere between 5 and 6~M Li${^+}$. We include a $\geq$ next to the stated resonance frequency in \zfr{mfig5}C, as it is unclear if the resonance would have been more upshifted if 6M Li${^+}$ concentration could have been successfully achieved.

For the 5~M solution of formate with 6M Na${^+}$, 1886\(\mu\mathrm{L}\) of formic acid was added to a 10~mL volumetric flask along with 2.4094g of NaOH. Similarly, 1886\(\mu\mathrm{L}\) of formic acid was added to another 10~mL volumetric flask with 3.3894g of KOH to make the 5~M formate, 6M K${^+}$ solution.

\addcontentsline{toc}{section}{Code and data availability}
\section*{Code and data availability}
\label{sec:code_repo}

To keep the SI concise, code listings are not reproduced here.
A public repository (link/DOI to be added at submission) will archive the following materials:
\begin{enumerate}
  \item Scan-averaged time-domain data for each experiment.
  \item Post-processed frequency-domain spectra used in the figures.
  \item A processing notebook documenting the conversion from (1) to (2) using the workflow in \SIsecref{subsec:proc_workflow} and reproducing the display spectra in \mainfigrange{mfig1}{mfig5}.
  \item Reference implementations for (i) frequency-domain (python) simulations and (ii) time-domain Spinach simulations, together with machine-readable
  \texttt{nuc\_list} and \texttt{couple\_list} inputs for all simulated isotopomers.

\end{enumerate}

Additionally, a user-friendly, Python-based zero-field spin-dynamics package developed by Raphael Zumbrunn is openly available at \href{https://github.com/Ajoy-Lab/ZULFPy_release}{ZULFPy\_release on GitHub}. The code adopts an intuitive, beginner-oriented molecular “dictionary” format for specifying J couplings and NMR-active nuclei and is accompanied by clear documentation and informative error messages to support a range of simulation workflows. The repository implements time- and frequency-domain simulations in the zero- and ultralow-field regimes, in one and two dimensions, and includes visualization tools for complex ZF energy-level structure. This software was used extensively during experimental design and troubleshooting, as it contains an in silico virtual replica of our apparatus, which is calibrated to produce a full simulation from an NMRduino sequence text file. It was also used to generate the parameter-response data shown in Fig. 1C.

\end{document}